\documentclass[prx,twocolumn,superscriptaddress,aps,longbibliography,10pt]{revtex4-2}
\usepackage{natbib}
\usepackage{graphicx}
\usepackage{enumitem}
\DeclareGraphicsExtensions{.pdf,.png,.jpg}
\usepackage{array}
\usepackage{amsmath}
\usepackage{amsthm}
\usepackage{amssymb}    
\usepackage{latexsym}
\usepackage{amsfonts}
\usepackage{mathrsfs}
\usepackage{color}
\usepackage{bbold}
\usepackage{kotex}
\usepackage{float}
\usepackage{makecell}
\usepackage{lineno}
\makeatletter
\let\newfloat\newfloat@ltx
\makeatother
\usepackage{algorithm}
\usepackage{algpseudocode}

\usepackage{subfigure}
\usepackage[colorlinks=true, urlcolor=blue,citecolor=blue,linkcolor=blue]{hyperref}
\usepackage{xcolor}

\begin{document}
\newcommand{\bra}[1]{\left< #1\right|}   
\newcommand{\ket}[1]{\left|#1\right>}
\newcommand{\abs}[1]{\left|#1\right|}
\newcommand{\ave}[1]{\left<#1\right>}
\newcommand{\Tr}{\mbox{Tr}}
\renewcommand{\d}[1]{\ensuremath{\operatorname{d}\!{#1}}}
\renewcommand\qedsymbol{$\blacksquare$}
\newcommand{\argmin}{\arg\!\min}
\newcommand{\argmax}{\arg\!\max}

\title{Reinforcement learning to learn quantum states for Heisenberg scaling accuracy}

\author{Jeongwoo Jae}
\email{jeongwoo.jae@samsung.com}
\affiliation{R\&D center, Samsung SDS, Seoul, 05510, Republic of Korea}

\author{Jeonghoon Hong}
\affiliation{R\&D center, Samsung SDS, Seoul, 05510, Republic of Korea}

\author{Jinho Choo}
\affiliation{R\&D center, Samsung SDS, Seoul, 05510, Republic of Korea}

\author{Yeong-Dae Kwon}
\email{y.d.kwon@samsung.com}
\affiliation{R\&D center, Samsung SDS, Seoul, 05510, Republic of Korea}


\begin{abstract}
Learning quantum states is a crucial task for realizing quantum information technology. Recently, neural approaches have emerged as promising methods for learning quantum states. We propose a meta-learning model that utilizes reinforcement learning (RL) to optimize the process of learning quantum states. To improve the data efficiency of the RL, we introduce an action repetition strategy inspired by curriculum learning. The RL agent significantly improves the sample efficiency of learning random quantum states, and achieves infidelity scaling close to the Heisenberg limit. We also show that the RL agent trained using $3$-qubit states can generalize to learning up to $5$-qubit states. These results highlight the utility of RL-driven meta-learning to enhance the efficiency and generalizability of learning quantum states. Our approach can be applied to improve quantum control, quantum optimization, and quantum machine learning.
\end{abstract}
\maketitle

\section{introduction}

Learning quantum states has emerged as a significant challenge in the field of quantum computing and machine learning~\cite{Anshu2024,Gebhart2023}. As a tool for analyzing and benchmarking quantum systems, learning quantum states is essential for advancing quantum information technology~\cite{Vogel1989}. The purpose of learning quantum states is to estimate quantum states with high fidelity based on samples obtained by measurements on multiple copies of a state. Quantum state tomography (QST) is the standard method for the quantum state estimation~\cite{Hradil1997,paris2004}, but this method becomes impractical for large systems since the number of measurements required scales exponentially with the system size. This limitation fuels research directions to devise various strategies for QST using matrix product state~\cite{Cramer2010,Lanyon2017}, permutation invariance of state~\cite{Schwemmer2014}, shadow tomography~\cite{aaronson2018}, and Bayesian estimation~\cite{Lukens2020}.

With the advent of deep learning, there has been a growing body of work using neural network-based models for learning quantum states~\cite{Park2020,Ahmed2021PRR,Ahmed2021PRL,Cha2022,Lange2023,Gaikwad2024,Palmieri2024,Ma2024,QinQin2024}. Generative models provide sample-efficient methods for learning many-body quantum states based on restricted Boltzmann machines~\cite{Torlai2018}, variational autoencoders~\cite{Rocchetto2018}, and recurrent neural networks~\cite{Carrasquilla2019}. These models are typically trained with unsupervised learning. Using supervised learning, a feed-forward neural network model enhances QST by removing errors~\cite{Palmieri2020}. The power of these approaches stems from the {\em expressivity} of neural networks, their ability to represent intricate features of quantum states in terms of a small number of parameters~\cite{Gao2017}. In the framework of quantum machine learning, models based on quantum neural networks have also been employed to learn quantum systems~\cite{Bang2018,Bang2021,Liu2020,Xue2022,Innan2024}. The quantum models can be implemented with parameterized quantum gates~\cite{Wan2017,Beer2020}, making them well suited for learning states of quantum computers. Moreover, a quantum neural network is known to have higher expressive power than classical neural networks of comparable size~\cite{Haghshenas2022,Abbas2021}.

Meanwhile, {\em generalizability}, the ability to generalize to unexperienced tasks, is one of the hallmarks of machine learning models. However, some of the models for learning quantum states are difficult to generalize as each model is tailored to learning a specific quantum state. Learning models with generalizability can be found in attempts to improve a quantum optimization algorithms using deep learning techniques~\cite{farhi2014,verdon2019,Wilson2021,khairy2020}. These works employ a hybrid approach that leverages a classical model to learn a quantum model. Specifically, Verdon et al. propose a model based on a recurrent neural network to initialize the parameters of quantum approximate optimization algorithm (QAOA) circuit~\cite{verdon2019}. Similar work by Wilson et al. use a recurrent neural network model to learn the gradient of objective function in QAOA~\cite{Wilson2021}. Khairy et al. show that a feed-forward neural network model can be used to learn the hyperparameters of QAOA with reinforcement learning~\cite{khairy2020}. They show that the hybrid models can generalize to solving problems of higher dimension than the dimension of problems for training. The deep learning techniques in the aforementioned works can be categorized as meta-learning, also called {\em learning to learn}, which is employed to enable a model to adapt to unexperienced tasks~\cite{li2017learning,chen2017learning}. Meta-learning methods for learning quantum states have also been proposed, where a classical model is used to learn another classical model~\cite{Smith2021} or Bayesian estimation algorithm~\cite{Quek2021}. For the task of learning quantum states, the validity and generalizability of meta-learning models in the hybrid setting remain areas of exploration.

In this work, we propose a meta-learning framework that utilizes reinforcement learning (RL) to learn quantum states. Our meta-learning model is based on quantum and classical neural networks. We adopt a hardware efficient ansatz as our quantum neural network model~\cite{Kandala2017}. The quantum model learns quantum states using a blackbox optimization algorithm called evolution strategy (ES)~\cite{salimans2017}. To enhance the process of learning quantum states, an RL agent consisting of classical neural networks dynamically adjusts the hyperparameters of the ES. To improve the data efficiency of the RL training, we devise {\em action repetition strategy} inspired by curriculum learning, a technique for training a learning model by gradually increasing difficulty of problems~\cite{Narvekar2020,ostaszewski2021rl,patel2024curriculum}. Our results demonstrate that the trained RL agent significantly improves the shot efficiency in quantum state learning, enabling the infidelity $\bar{f}$ to scale according to the Heisenberg limit with respect to the total number of measurement shots ${\cal C}$~\cite{Haah2017}, i.e., $\bar{f}\sim{\cal O}({\cal C}^{-1})$. We also show that our RL agent generalizes to learning quantum states of higher dimensions than those used for training.

The remainder of this paper is organized as follows. We explain the scheme of meta-learning in Section~\ref{sec:meta}. In section~\ref{sec:result}, we present results of meta-learning for random pure states, and introduce the action repetition strategy. We also examine the generalizability of our model and compare our results to the standard QST based on maximum likelihood estimation. Finally, we provide discussion and conclusion in Section~\ref{sec:dis}.

\begin{figure*}[t!]
	\includegraphics[width=0.93\textwidth]{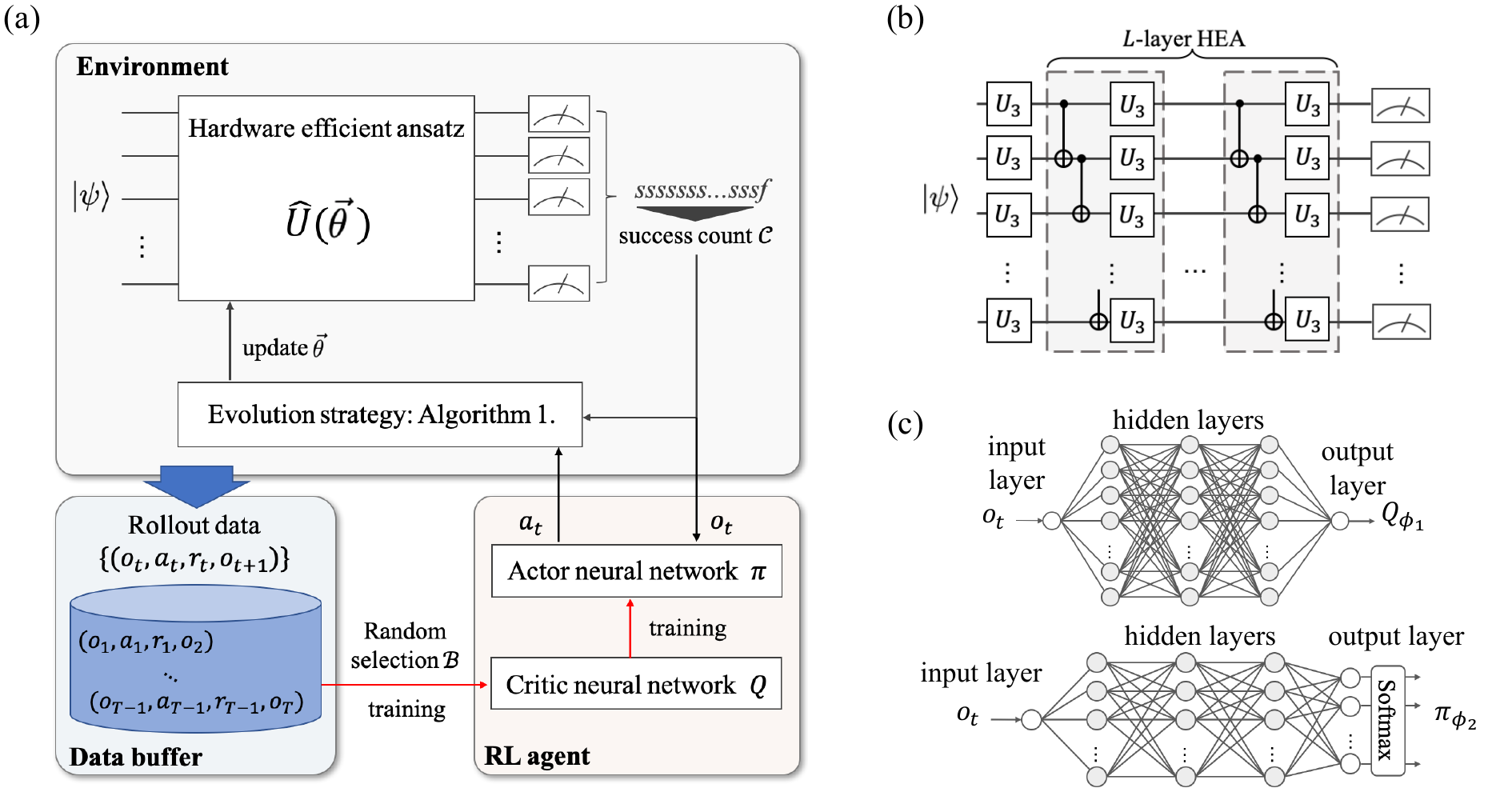}
	\caption{{Schematic of reinforcement learning to learn quantum states.} (a) In the environment, the evolution strategy trains a hardware efficient ansatz (HEA) shown in (b) to learn an input state $\ket{\psi}$. The observation $o_t$ represents the success count $\cal C$, and the HEA is trained to maximize it. The RL agent learns the action $a_t$, the hyperparameters of the evolution strategy. To expedite the training process, we give a penalty $r_t=-1$ to the RL agent at each time step $t$, and no penalty $r_t=0$ for the final time step. The rollout datasets are generated using a policy-guided tree search and stored in the data buffer. We use the Actor-Critic algorithm to train the RL agent with a subdataset $\cal B$ randomly drawn from the buffer. {(b) HEA circuit.} $U_3$ is a generic single-qubit gate $\exp({i\vec{\theta}\cdot \vec{\sigma}/2})$, where $\vec{\theta}=(\theta_x,\theta_y,\theta_z)$ and $\vec{\sigma}$ denotes the vector of Pauli operators $(\hat{\sigma}_x, \hat{\sigma}_y, \hat{\sigma}_z)$. For 1-qubit states, we use a single $U_3$ gate. For $N$-qubit states, we use the HEA of $L$ layers, consisting of $3N(L+1)$ trainable parameters. {(c) Feed-forward neural networks for the RL agent.} Each network adopts the observation $o_t$ as input. $Q_{\phi_1}$ is the observation-value function and $\pi_{\phi_2}$ is the policy of action.
 }
\label{fig:setting}
\end{figure*}

\section{Meta-learning: RL to learn}\label{sec:meta}
A schematic of the meta-learning model is illustrated in Fig.~\ref{fig:setting} (a). The environment consists of an evolution strategy and a quantum system. As depicted in Fig.~\ref{fig:setting} (b), the quantum system comprises a quantum state $\ket{\psi}$, a hardware efficient ansatz (HEA) $\hat{U}(\vec{\theta})$, and a measurement. The evolution strategy optimizes the HEA to learn an input quantum state based on measurement outcomes. The RL agent learns the hyperparameters of the evolution strategy using two classical models in Fig.~\ref{fig:setting} (c), the {\em actor} and {\em critic}.

We formulate our RL scheme using a Markov decision process. The Markov decision process (MDP) is a framework to describe a sequential interaction between an agent and its environment~\cite{sutton2018}. The agent takes an action $a$ based on observation $o$ of the environment. The action alters the state of environment, and the agent receives a reward $r$ in response. The goal of the agent is to learn a policy $\pi$ that maximizes cumulative rewards (see Appendix~\ref{sec:mdp} for details).

In determining approach of agent to formulating and solving a MDP, {\em observability} of the environment is a significant factor to consider. When the elements of an MDP are fully known, that is, assuming full observability of environment, an agent can derive the exact policy based on Bellman's principle of optimality~\cite{bellman1954theory}. However, in many scenarios, an agent cannot completely specify the state of the environment or its dynamics due to limited (or partial) observability. A partially observable MDP (POMDP) extends the standard MDP framework to account for such limited observability~\cite{aastrom1969optimal}. To solve the POMDP, an agent can take a heuristic method through an iterative process. Deep RL is widely used to learn heuristic policies under the limited observability~\cite{hausknecht2015deep}.

The problem we consider is the case that the RL agent lacks full observability of the environment. The quantum state transformed by the HEA, $\hat{U}(\vec{\theta})\ket{\psi}$, represents the state of the environment. However, the agent does not have full knowledge of the exact state of the environment, and quantum measurements can only extract partial information about the quantum state. Thus, we can formulate our problem as a POMDP~\cite{aastrom1969optimal,Barry2014,Sivak2022}, defined by a tuple $\left({\cal S},{\cal A},{\cal T},{\cal R},\Omega,{\cal O}\right)$. The elements of the POMDP are explained in Appendix~\ref{sec:pomdp}.

A major challenge in applying RL to quantum systems is the partial observability due to quantum measurements~\cite{Barry2014}. In applying RL to control of quantum systems, there are attempts to utilize additional information about quantum states such as fidelity~\cite{Bukov2018,Thomas2018,Niu2019,Wang2020,Porotti2022,Metz2023}, but computing this quantity is typically expensive, comparable to learning quantum states by the standard QST methods. In contrast, the RL method proposed by Sivak et al. learns a control policy based on observable quantities~\cite{Sivak2022}. In the same spirit, we train our RL agent using the observation $o$ obtained from the measurement. In the following subsections, we provide a detailed description of the environment, agent, and RL algorithm.

\subsection{Environment}

{\em Learning a quantum state.---}Consider an unknown $N$-qubit state $\ket{\psi}$ as an input of the HEA $\hat{U}(\vec{\theta})$. We measure the quantum state transformed by the HEA, $\hat{U}(\vec{\theta})\ket{\psi}$, with a measurement $M=\{\hat{M}_s, \hat{M}_f\}$ prepared in known bases, where
\begin{eqnarray}
    \hat{M}_s = |s\rangle\langle s|,\quad \hat{M}_f = I-\hat{M}_s.
\end{eqnarray}
The outcome $s$ stands for {\em success} and $f$ does {\em fail}. We perform the measurement until a fail outcome appears, and count the number of consecutive success outcomes before the fail outcome. If a fail outcome appears, we change the parameters of quantum circuit based on the number of success outcomes. Otherwise, we retain the parameters. This measurement scheme is the same as that of single-shot measurement learning (SSML)~\cite{Bang2018,Bang2021}. While the SSML uses a weighted random search algorithm to update the parameters, we use the evolution strategy which will be explained later.

We refer to the number of consecutive success outcomes as {\em success count}. The success count is determined by the probability of success outcome $p_s = \langle\psi|\hat{U}^\dagger(\vec{\theta})\hat{M}_s\hat{U}(\vec{\theta})|\psi\rangle$. Thus, the chance to obtain ${\cal C}$ success count is governed by a geometric distribution
\begin{eqnarray}\label{eq:geo}
    p({\cal C}) = p_s^{\cal C} (1-p_s).
\end{eqnarray}
The average value of success count is then given by $\langle {\cal C} \rangle=\sum_{{\cal C}=0}^\infty {\cal C}p({\cal C}) = p_s/(1-p_s)$. If $p_s \rightarrow 1$, $\langle {\cal C} \rangle \rightarrow \infty$, and, if $p_s= 0$, $\langle {\cal C} \rangle=0$. The success count is the observation $o$, and the geometric distribution determines the transition probability of observation.

We train the HEA until the success count achieves a high value so that the HEA transforms the input quantum state close to the basis of success outcome. We set a target success count ${\cal C}_\text{target}$, and have the training process stop if a success count reaches or exceeds the target value, i.e., ${\cal C}\ge{\cal C}_\text{target}$. The value of target success is related to the accuracy of reconstructed quantum state.

{\em Quantum state reconstruction.---}After the training, we can reconstruct the input quantum state through the trained HEA as
\begin{eqnarray}
\label{eq:estimate}
    \ket{\psi_\text{est}} = \hat{U}^\dagger\left(\vec{\theta}_\text{train}\right)\ket{s}.
\end{eqnarray}
To assess the accuracy of the reconstructed state $\ket{\psi_\text{est}}$, we choose the {\em infidelity} as a figure of merit~\cite{Haah2017} defined by
\begin{equation}
\label{eq:infidelity}
    \bar{f}:=1-|\langle \psi|\psi_\text{est}\rangle|^2
\end{equation}
The infidelity depends on the target success count ${\cal C}_\text{target}$ we initially set; The larger the target success count, the better the accuracy of the estimation.

{\em Optimizer.---}To train the HEA, we employ an evolution strategy (ES)~\cite{Rechenberg1975}. An ES consists of following steps: (i) assuming a probability distribution for sampling, (ii) sampling parameters around a current parameter, (iii) evaluating the parameter samples, and (iv) updating the current parameter based on the evaluations. In our problem, the ES uses a multivariate Gaussian distribution to sample parameters~\cite{salimans2017}, and the evaluation corresponds to the measurement of success count.

The purpose of ES is to find the parameters of HEA $\vec{\theta}$ to achieve the halting condition ${\cal C}(\vec{\theta}) \ge {\cal C}_\text{target}$. For this purpose, we consider the expectation of success count as an objective function of our ES
\begin{eqnarray}
    J(\vec{\theta}) := \frac{1}{{\cal C}_\text{target}}\mathbf{E}_{\vec{\theta}\sim p(\vec{\theta})}\left[ {\cal C}(\vec{\theta}) \right] ,
\end{eqnarray}
where the target success count ${\cal C}_\text{target}$ is introduced in the denominator to normalize the objective function. $p(\vec{\theta})$ is the multivariate Gaussian distribution ${\cal N}(\vec{\theta},\sigma^2 I)$ of mean $\vec{\theta}$ and covariance matrix $\sigma^2 I$, where $\sigma\in(0,\infty)$ is a hyperparameter to determine the {\em range of sampling} over the space of parameters, and $I$ is the identity matrix.

To increase the expected success count, the parameters are updated by following rule
\begin{eqnarray}
    \vec{\theta}\leftarrow\vec{\theta} + \eta \nabla_{\vec{\theta}}\tilde{J}(\vec{\theta}),
\end{eqnarray}
where $\eta\in[0,\infty)$ is the {\em learning rate}, and $\nabla_{\vec{\theta}}\tilde{J}(\vec{\theta})$ is an estimator of gradient. Applying the reparameterization trick~\cite{salimans2017}, the ES estimates the gradient by drawing $\{\vec{\epsilon_i}\}_{i=1}^k$ from a distribution ${\cal N}(\vec{0},I)$, and evaluating $k$ parameter samples $\{\vec{\theta} + \sigma\vec{\epsilon}_i\}_{i=1}^k$. The estimator of gradient is given by
\begin{eqnarray}\label{eq:ESgrad_est}
    \nabla_{\vec{\theta}}\tilde{J}(\vec{\theta}) = \frac{1}{{\cal C}_\text{target}\sigma k }\sum_{i=1}^k {\cal C}(\vec{\theta}+\sigma \vec{\epsilon}_i)\vec{\epsilon}_i.
\end{eqnarray}

\begin{algorithm}[t]
\caption{Evolution strategy with RL agent}
\label{algo:1}
    \begin{algorithmic}[1]
        \State{Set an input quantum state $\ket{\psi_1}$, the HEA $\hat{U}(\vec{\theta}_0)$, the target success count ${\cal C}_\text{target}$, the action repetition time $t_\text{rep}$, the number of samples $k$, and the RL agent $\pi$}
        \For{$t=1,2\cdots,t_\text{max}$}
            \State{Measure ${\cal C}(\vec{\theta}_t)$ and $o_t\leftarrow {\cal C}(\vec{\theta}_t)$}
            \If{$t\%t_\text{rep} = 0$} 
                \State{Draw $\sigma_t$ and $\eta_t$ from the agent, $(\sigma_t,\eta_t)\sim \pi(a_t|o_t)$}
            \EndIf
            \State{Sample $\vec{\epsilon}_1,\cdots, \vec{\epsilon}_k \sim {\cal N}(0,I)$}
            \State{Measure ${\cal C}\left(\vec{\theta}_t + \sigma_t\vec{\epsilon}_i\right)$ for $i=1,\cdots,k$}
            \If{any ${\cal C}\ge {\cal C}_\text{target}$}
                \State{break}
            \EndIf
            \State{$\vec{\theta}_{t+1}\leftarrow\vec{\theta}_t + \eta_t\nabla_{\vec{\theta}}\tilde{J}(\vec{\theta}_t)$}
            \Comment{Eq.~\eqref{eq:ESgrad_est}}
            \State{$\ket{\psi_{t+1}}\leftarrow\hat{U}(\vec{\theta}_{t+1})\ket{\psi_1}$}
        \EndFor
        \State{${\cal C}_\text{total} \leftarrow$ the sum of success counts}
        \State{$t_H \leftarrow$ $t$}
        \State{$\vec{\theta}_\text{train} \leftarrow \vec{\theta}$ s.t. ${\cal C}(\vec{\theta})\ge {\cal C}_\text{target}$}
    \end{algorithmic}
\end{algorithm}

The sampling range, also called step size, is a key hyperparameter that significantly impacts the performance of ES. The sampling range decides the range of search: A larger sampling range induces more exploration on the parameter space. A smaller sampling range focuses on exploitation, searching in a narrower region to enhance solution. In advanced ES variants, such as covariance matrix adaptation ES~\cite{hansen2016cma}, both the sampling range and learning rate are dynamically adjusted. Their interplay determines the covariance matrix, which captures the local geometry of objective function. In our scheme, the sampling range $\sigma$ and learning rate $\eta$ are dynamically changed according to the policy $\pi$.

The pseudocode of ES is shown in Algorithm~\ref{algo:1}. Unlike the two hyperparameters $\sigma$ and $\eta$, we fix the number of samples $k$ to a single value to keep the statistical uncertainty of the gradient estimator at a certain degree throughout the execution of ES. Each action drawn by the agent is applied for $t_\text{rep}$ consecutive time steps, referred to as the action repetition time. The effectiveness of the action repetition will be discussed in the next section. ${\cal C}_\text{total}$ is the total success count used, and the halting time, $t_H$, is the time steps taken until the training is finished. Note that the time here refers to the number of ES iterations required for the success count to reach the halting condition ${\cal C}\ge {\cal C}_\text{target}$. In a real experiment, the total success count is a quantity which is proportional to the physical runtime of the training.

An optimizer typically used to train a quantum neural network is a gradient method~\cite{Mitarai2018,Schuld2019grad}, but it can suffer the gradient vanishing problem, called barren plateau~\cite{McClean2018,Cerezo2021}, making the training intractable. While the gradient vanishing effect cannot be fully alleviated~\cite{Arrasmith2021}, recent studies have shown that ESs can be useful for training some cases of quantum neural networks~\cite{Anand2021,Jianshe2023}. The gradient methods typically require more evaluation points than parameters to obtain directional derivatives~\cite{Mitarai2018}. We will show that the ES can train the HEA with the fewer evaluation points than the parameters of HEA.

\begin{figure*}[t!]
    \centering
    \includegraphics[width=\linewidth]{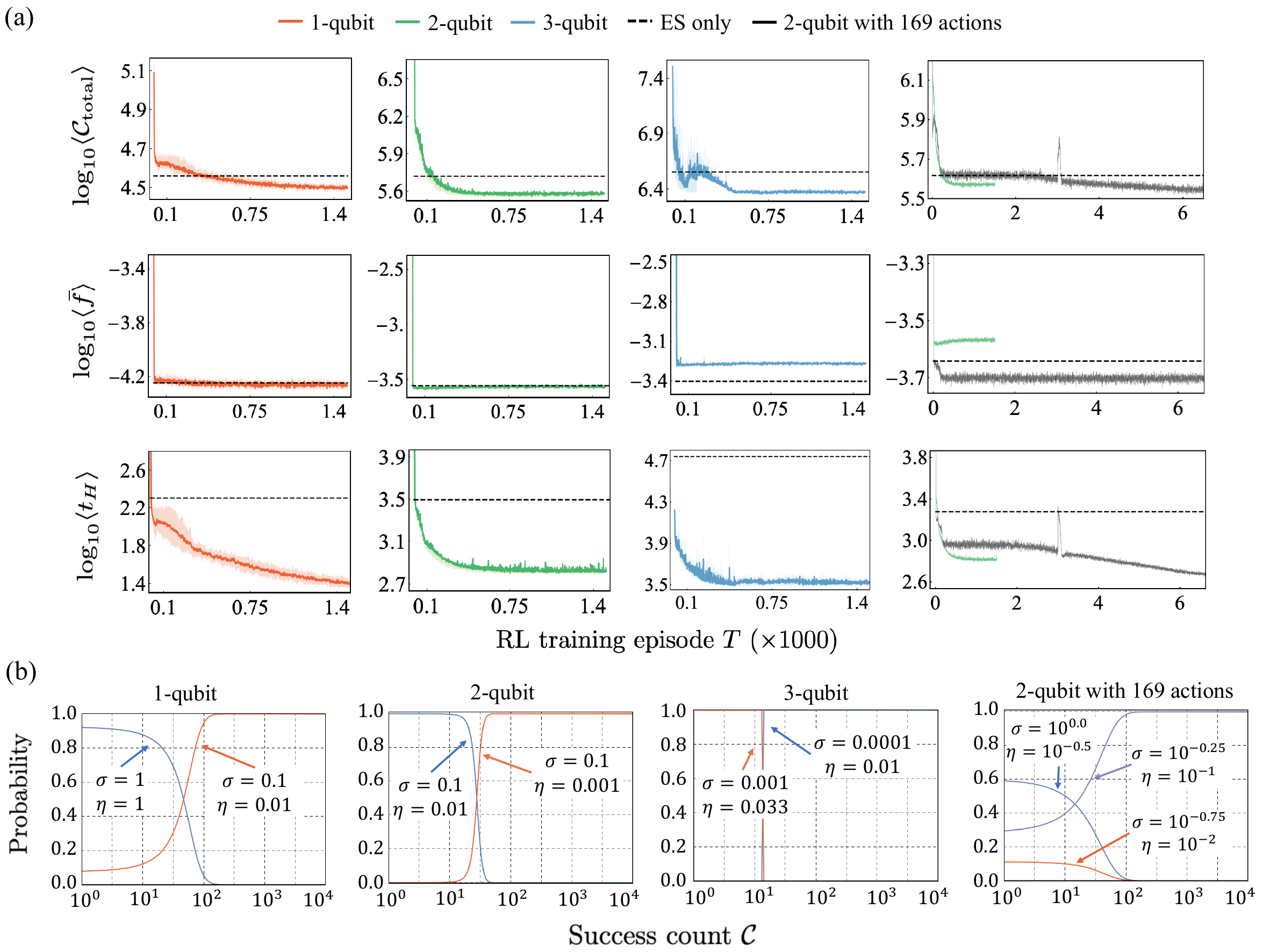}
    \caption{{Results of training RL agent for the target success count ${\cal C}_\text{target} = 10^4$.} For each RL training episode $T$, we perform learning of a thousand random states. $\langle {\cal C}_\text{total}\rangle$ is the average of total success count. $\langle\bar{f}\rangle$ is the average infidelity of learned states. $\langle t_H\rangle$ is the average halting time steps of evolution strategy. (a) The red, green, and blue solid lines are the training curves for learning $1$-, $2$-, and $3$-qubit states with the $16$ actions. The black solid line is obtained by using the extended action space for learning $2$-qubit states. The dashed lines are baseline values. (b) The policy of trained RL agent.}
    \label{fig:RLcurve}
\end{figure*}

\subsection{RL agent}\label{sub:ARL}

The RL agent consists of an actor and a critic, implemented by the feed-forward neural networks [see Fig.~\ref{fig:setting} (c)]. The both networks use three hidden layers. For each hidden layer, the critic has hundred nodes, and the actor has fifty nodes. We use ReLu for activation functions. The actor is to learn a policy of action. We construct the action space ${\cal A}$ by selecting $m$ values from each of the hyperparameter spaces ${\cal A}_\sigma=[0,\infty)$ and ${\cal A}_\eta=[0,\infty)$. So the number of actions the agent can choose from is $m^2=\abs{\cal A}$. The output of actor is the policy $\pi_{\phi_2}=p(\sigma,\eta|o,\phi_2)$, the probability distribution defined over the action space ${\cal A}$. The critic outputs an estimate of observation-value function $Q_{\phi_1}(o)$.

We train the agent to recommend the hyperparmaeters $\sigma$ and $\eta$ that enables the learning of quantum states to be finished quickly. This is accomplished by giving a penalty of $-1$ to the RL agent at each time step, and no penalty for the final time step. Thus, the reward is given by
\begin{eqnarray}
    r_t = \begin{cases}
    ~~0& \text{if}~~t= t_H  \\
    -1& \text{otherwise}.
    \end{cases}
\end{eqnarray}
The empirical cumulative reward is then given by
\begin{eqnarray}
    \label{eq:empcum}
    {\cal R}_t = \sum_t^{{t_H}}r_t  = {t}-{t_H}.
\end{eqnarray}
For the training, we use the Actor-Critic algorithm, and collect training datasets using a tree search guided by the policy. See Appendix~\ref{sec:AC} for details.

\begin{table*}[t]
\caption{\label{tab:setting} Settings used to train the RL agent. The space of action is given by ${\cal A}={\cal A}^4_\sigma\times {\cal A}^4_\eta$. The action repetition time $t_\text{rep}$ is determined by the parameters $(t_l,t_u,T_\text{th})$. For $4$- and $5$-qubit, we perform learning of random pure states with the RL agent trained using the $3$-qubit states.}
\begin{ruledtabular}
\begin{tabular}{cccccc}
 No. qubits $N$&\makecell{Quantum \\ neural network}&\makecell{No. samples $k$}&\makecell{The space of \\ sampling range ${\cal A}_\sigma^4$}&\makecell{The space of \\ learning rate ${\cal A}_\eta^4$}&\makecell{Action repetition \\strategy $(t_l,t_u,T_\text{th})$}\\ \hline
 $1$&$U_3$&5&$\{1.0,0.1,0.01,0.001\}$&$\{1.0,0.1,0.01,0.001\}$&$(1,50,100)$\\
 $2$&HEA, $L=1$&10&$\{1.0,0.1,0.01,0.001\}$&$\{1.0,0.1,0.01,0.001\}$&$(80,800,100)$\\
 $3$&HEA, $L=5$&30&$\{0.1,0.01,0.001,0.0001\}$&$\{1.0, 0.33, 0.01, 0.033\}$&$(300\footnote{We obtain similiar results for $t_l=200$.},2000,500)$\\
 $4$&HEA, $L=10$&100&$\{0.1,0.01,0.001,0.0001\}$&$\{1.0, 0.33, 0.01, 0.033\}$&$t_\text{rep}=300$\\
 $5$&HEA, $L=10$&100&$\{0.1,0.01,0.001,0.0001\}$&$\{1.0, 0.33, 0.01, 0.033\}$&$t_\text{rep}=300$
\end{tabular}
\end{ruledtabular}
\end{table*}

\section{Results}
\label{sec:result}

\subsection{Training RL agent}
We apply our meta-learning scheme to the learning of random pure states. The random states are prepared by applying unitary operators $\hat{V}$ drawn from Haar measure to the initial states $\ket{0}^{\otimes N}$~\cite{mezzadri2007}. To measure these states, we perform measurements in the computational basis, where $|0\rangle\langle0|^{\otimes N}$ corresponds to the basis of the success outcome $|s\rangle\langle s|$. For each RL episode $T$, we prepare a new set of a thousand problem instances, and perform the learning of quantum states to collect datasets for RL. We obtain the following results through simulations implemented by PyTorch~\cite{paszke2019pytorch}, running on four NVIDIA Tesla A$100$ graphic processing units (GPUs). The code is available at Ref.~\cite{code}.

We present the setting of the HEAs, the number of samples, and the space of action in Table~\ref{tab:setting}. While the number of parameters of a unitary operator required to prepare an $N$-qubit state is $4^N$, we show that the HEA can learn the random quantum states with fewer parameters: For the learning $2$- and $3$-qubit states, we use the HEAs which have $12$ and $54$ parameters, respectively. We will also show that HEA of $10$ layers can learn random $4$- and $5$-qubit states. Moreover, to estimate the gradient of HEAs, we use fewer samples than the parameters of HEA circuits. For $2$- and $3$-qubit states, the number of samples is set to $10$ and $30$, respectively. These results show the expressivity of the HEAs.

\begin{figure*}[t!]
    \centering
    \includegraphics[width=0.9\linewidth]{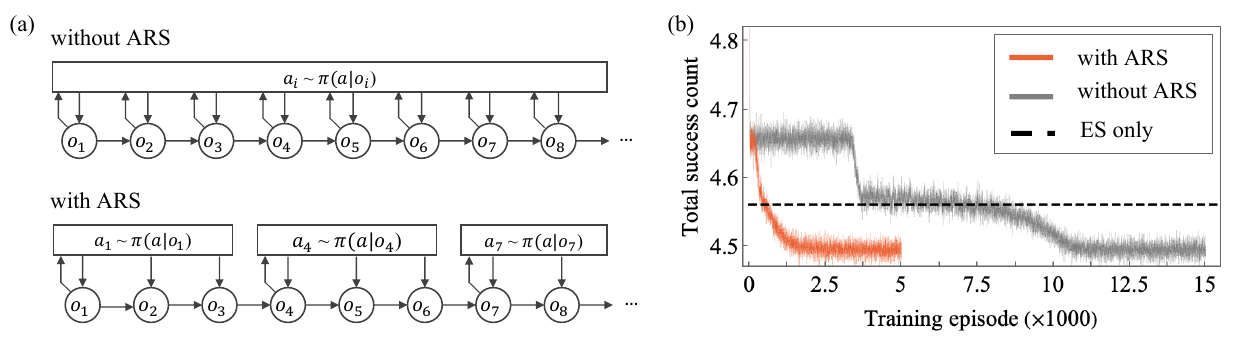}
    \caption{(a) Scheme of action repetition strategy (ARS). Without the ARS ($t_\text{rep}=0$), the agent $\pi$ draws a new action $a_i$ at each time step based on the observation $o_t$. With the ARS ($t_\text{rep}\neq0$), the agent repeats a selected action for $t_\text{rep}$ consecutive time steps. The illustration depicts the cases where $t_\text{rep}=0$ and $t_\text{rep}=3$. (b) {Learning random $1$-qubit states with and without the ARS.} The red solid line represents the result obtained using the ARS with setting $(t_u,t_l,T_\text{th})=(50,1,100)$. The gray solid line corresponds to the case without the ARS. The dashed line indicate the baseline performance (ES only).}
    \label{fig:actrep}
\end{figure*}

The results of training RL agent are illustrated in Fig.~\ref{fig:RLcurve} (a). For these results, we set the target success count to ${\cal C}_\text{target} = 10^4$, and randomly initialize the HEAs. We compare these results to the baseline values, the performance obtained without the RL, i.e., using only a single action for the entire process of learning quantum states. We regard the results obtained from an action which gives the lowest total success count as the baseline values, and train the RL agent until it outperforms the baseline values. The baselines are selected from the data in Appendix~\ref{sec:ESperformance}.

At the end of RL training ($T\approx1500$), the average of total success counts required to learn the $1$-, $2$-, and $3$-qubit states are $3.123\times 10^4$, $3.767\times 10^5$, and $2.363\times 10^6$, which are about $14\%$, $28\%$, and $34\%$ lower than the baseline values, respectively. Using the RL, we obtain the (in)fidelity similar to the baselines, except for the case of $3$-qubit. The average fidelity is given by $0.99994$, $0.99972$, and $0.99946$ for $1$-, $2$-, and $3$-qubit states, respectively. In the early episodes of RL, the average halting time $t_H$ of learning $1$-, $2$-, and $3$-qubit states is about $10^3$, $10^4$, and $1.5\times10^4$, and these are reduced to $19$, $660$, and $3409$ at the end of episode, respectively. For these results, we use the $16$ actions defined in Table~\ref{tab:setting}.

For the $2$-qubit states, we also consider an extended action space defined by ${\cal A}:={\cal A}^{13}_\sigma\times {\cal A}^{13}_\eta$, where
\begin{eqnarray}
\label{eq:169a}
    {\cal A}^{13}_\sigma &=& \left\{\sigma_i ~\bigg|~ \sigma_i=10^{-0.25(i-1)}~\text{for}~i=1,\ldots,13 \right\}\\
    {\cal A}^{13}_\eta &=& \left\{\eta_i~\bigg|~ \eta_i=10^{-0.25(i-1)}~\text{for}~i=1,\ldots,13 \right\}.\nonumber
\end{eqnarray}
The number of actions in the extended space is $169$. Except for the number of actions, we use the same settings for the $2$-qubit case listed in Table.~\ref{tab:setting}. The result is shown in the rightmost column of Fig.~\ref{fig:RLcurve} (a). By extending the action space, we can attain the lower values of average total success count and infidelity than the case of using the $16$ actions; the average total success count is $3.593\times10^5$ and the average fidelity is $0.9998$. To achieve these results, on average, $433$ time steps are required. While the overall performance of learning $2$-qubit states is improved by extending the action space, the agent requires more episodes to be trained.

Fig.~\ref{fig:RLcurve} (b) shows the policy of trained actor. These policies imply that the trained RL agents take a strategy; if the success count is small, the agent recommends the large values for $\sigma$ and $\eta$ to enable the ES to search in a wide range of parameter space; if the success count is large, the agent reduces the values of the recommended hyperparameters to make the learning of quantum states converge.

\begin{figure*}[t!]
    \centering
    \includegraphics[width=\linewidth]{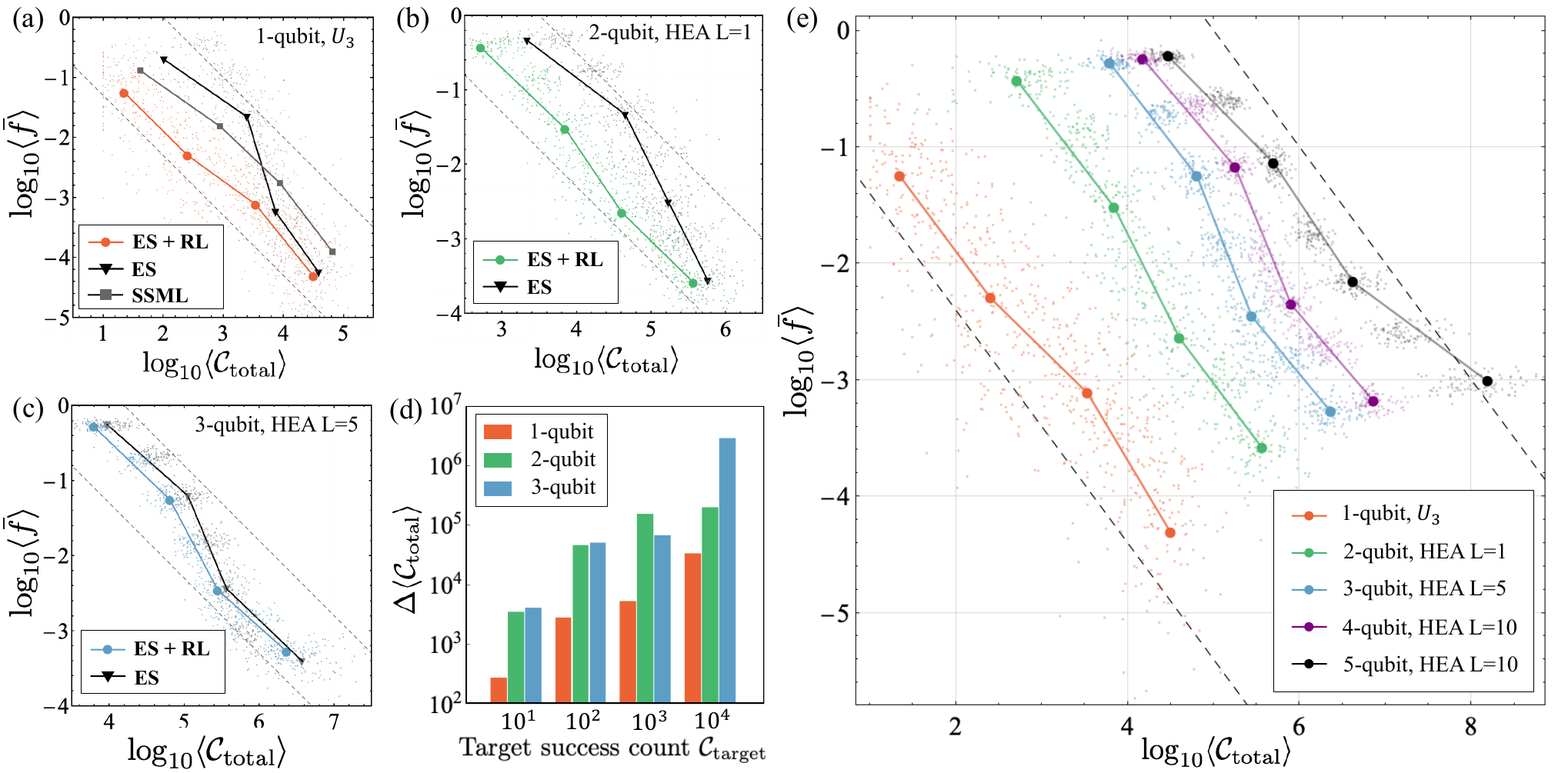}
    \caption{{Learning random pure states with the trained RL agent.} We consider the target success counts from $10^1$ to $10^4$. The dots in (a)--(c) and (e) represent the average total success count $\langle{\cal C}_\text{total}\rangle$ and the average infidelity $\langle\bar{f}\rangle$ for the learning $1$-qubit (red), $2$-qubit (green), and $3$-qubit (blue) states. We denote the results obtained using the trained RL agent, labeled as {\bf ES$+$RL}. The black triangles, labeled as {\bf ES}, are the average values obtained without the RL agent. For the results of {\bf ES}, we use an action $a_\text{base}$ which gives the baseline values. {\bf SSML} represents the results obtained by the single-shot measurement learning. (d) $\Delta\langle {\cal C}_\text{total} \rangle=\langle {\cal C}_\text{total} \rangle_\text{{\bf ES}} - \langle {\cal C}_\text{total}\rangle_\text{{\bf ES+RL}}$ is the amount of average total success counts reduced by using the trained RL agent. (e) The purple and black dots are results of generalizing the RL agent trained using $3$-qubit states to learning random $4$- and $5$-qubit states, respectively. The dashed lines represent the scaling of Heisenberg limit, $\bar{f}\sim{\cal O}({\cal C}^{-1}_\text{total})$.}
    \label{fig:gen}
\end{figure*}

\subsection{Action repetition strategy}
To obtain these results, we use a method, called {\em action repetition strategy} (ARS), that repeats an action selected by the RL agent for consecutive time steps $t_\text{rep}$. The schematic of ARS is illustrated in Fig.~\ref{fig:actrep} (a). For each RL episode $T$, $t_\text{rep}$ is given by
\begin{eqnarray}
\label{eq:ARS}
    t_\text{rep} &=& \max\left(\left\lceil{t_u-\frac{T}{T_\text{th}}\left({t_u - t_l}\right) }\right\rceil, t_l\right),
\end{eqnarray}
where $t_u$ and $t_l$ are the upper and lower value of $t_\text{rep}$, respectively, and $t_u>t_l$. When $T=0$, the action repetition time is $t_u$, and it decreases until the RL episode $T$ reaches $T_\text{th}$. After $T_\text{th}$, $t_\text{rep}$ saturates to $t_l$.

The ARS is a method inspired by {\em curriculum learning}, a technique that trains a machine learning model by gradually increasing the difficulty of problems~\cite{Narvekar2020,ostaszewski2021rl,patel2024curriculum}. Note that the curriculum learning belongs to the category of transfer learning, which reuses a model or knowledge obtained from a task to learn a related task~\cite{Pan2010,Zen2020,Mari2020transfer}. One of factors that determines the difficulty of RL problem is the depth of decision process. In general, the deeper the decision process, the more difficult the problem is considered to be. The depths $t_H$ of decision processes of our problems in the early episodes are about $10^3$, $10^4$, and $1.5\times10^4$ for the learning of $1$-, $2$-, and $3$-qubit states, respectively. With the ARS, we can reduce the depths of decision processes to $\left\lfloor t_H/t_\text{rep}\right\rfloor$, which are shallower than the original depths. Specifically, according to the settings in Table~\ref{tab:setting}, the decision processes can be reduced to tens of steps in the beginnings of the RL trainings. These lead to the significant decrease of difficulties of the problems. As the RL training proceeds, the difficulties of the problems increase again with the decrease of $t_\text{rep}$. Thus, by controlling the action repetition time, the ARS enables the RL agent to experience problems of various difficulty levels during the training process.

For the random $1$-qubit states, we compare the performance obtained with and without the ARS in Fig~\ref{fig:actrep} (b). With the ARS, the RL agents can converge faster than the case without the ARS. For the random $2$- and $3$-qubit states, the training of RL agent without the ARS requires computational resources far exceed our capacity. In some intervals of RL episodes, the difficulties of our problems do not gradually increase since the halting time steps $t_H$ also decrease together with the action repetition times. More sophisticated curriculum for the repetition time may improve the performance of meta-learning, but it is beyond the scope of this work and remains an open question.

\begin{figure*}[t!]
    \centering
    \includegraphics[width=\linewidth]{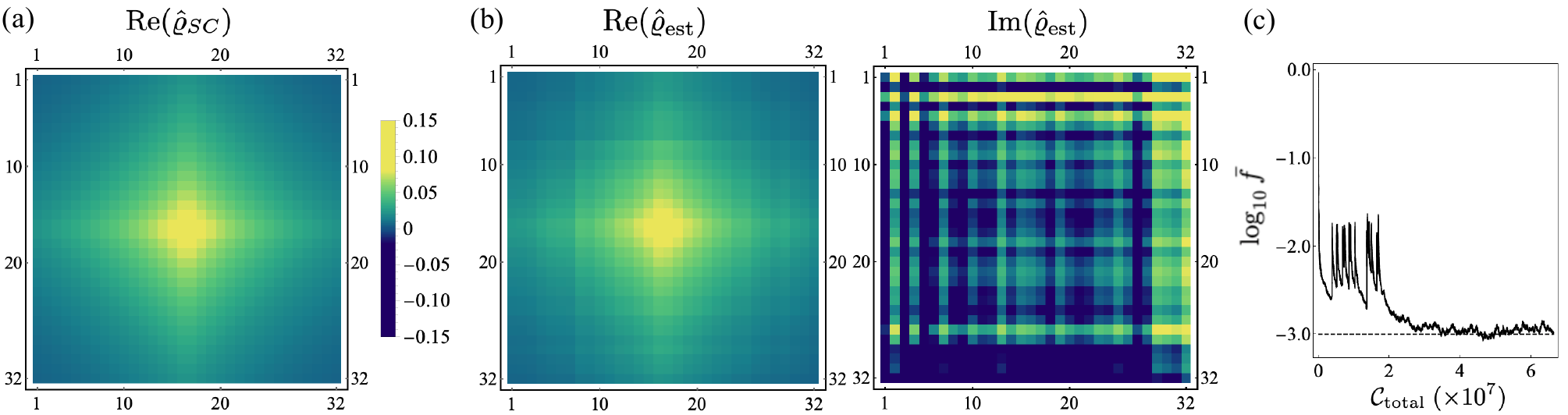}
    \caption{{Learning an entangled $5$-qubit pure state $\hat{\varrho}_{SC}$ with the RL agent trained using the random $3$-qubit states.} {Re}$\left(\cdot\right)$ and {Im}$\left(\cdot\right)$ denote the real and imaginary parts of a density matrix, respectively. (a) The real parts of the density matrix for the $5$-qubit state. The imaginary parts of $\hat{\varrho}_{SC}$ are zero. (b) The reconstructed state $\hat{\varrho}_\text{est}$ for the $5$-qubit state. The reconstructed state achieves fidelity $f\approx0.9989$. (c) The changes of infidelity throughout the learning process.}
    \label{fig:reconstruct}
\end{figure*}

\subsection{The scaling of infidelity}

Using the RL agents trained with the target success count ${\cal C}_\text{target}=10^4$, we perform learning of a hundred random pure states by varying the target success counts from $10^1$ to $10^4$. The results are shown in Fig.~\ref{fig:gen} (a)--(c). We compare these results to the baseline values obtained by an action $a_\text{base}$ (see Appendix~\ref{sec:ESperformance}). Fig.~\ref{fig:gen} (d) shows the amount of total success counts saved by the RL agent according to the increase of the number of qubits $N$ and target success count ${\cal C}_\text{target}$. $\Delta\langle {\cal C}_\text{total} \rangle:=\langle {\cal C}_\text{total} \rangle_\text{{\bf ES}} - \langle {\cal C}_\text{total}\rangle_\text{{\bf ES+RL}}$ quantifies the amount of total success counts saved by the RL agent, where $\langle {\cal C}_\text{total}\rangle_\text{{\bf ES}}$ is obtained using $a_\text{base}$ without the RL agent, and $\langle {\cal C}_\text{total}\rangle_\text{{\bf ES+RL}}$ is the average total success count obtained with the trained RL agent. In the target success count $10^4$, $\Delta\langle {\cal C}_\text{total} \rangle$ is given by $3.274\times10^4$, $1.995\times10^5$, and $2.818\times10^6$ for $1$-, $2$-, and $3$-qubit states, respectively. This leads to enhancement in the scaling of average infidelity $\bar{f}$ with respect to the total success count ${\cal C}_\text{total}$.

The infidelity of our method scales according to the Heisenberg limit~\cite{Haah2017}, $\bar{f}\sim{\cal O}({\cal C}^{-1}_\text{total})$. We curve fit the results of Fig.~\ref{fig:gen} (a)--(c) with $\bar{f}=\alpha{\cal C}_\text{total}^{-\beta}$. The scaling factor is given by $\beta\approx0.948$, $1.161$, and $1.184$ for the learning $1$-, $2$-, and $3$-qubit states, respectively. Our method shows the similar results to previous works on the learning of $1$-qubit states~\cite{Mahler2013,Kravtsov2013,Ferrie2014SGTQT,Bang2018,Bang2021}, where the scaling factors lie in the range $\beta \in [0.9,1.0)$. Typically, the SSML shows the scaling of $\beta\approx0.972$ in our simulation, which is consistent with the results in their works~\cite{Bang2018,Bang2021}. (In the multi-qubit cases, we fail to find the settings of SSML to achieve the optimal scaling.) Our result obtained with the RL agent achieves the similar scaling factor while using fewer total success count than the SSML as shown in Fig.~\ref{fig:gen} (a). The standard QST method cannot achieve Heisenberg-limit scaling without additional information about quantum states, with its scaling limit given by $\beta=0.75$~\cite{Bagan2004}.

\subsection{Generalization of RL agent}

We apply the RL agent trained with the random $3$-qubit states to learning of a hundred random $4$- and $5$-qubit states. The results are illustrated in Fig.~\ref{fig:gen} (e). For the target success count ${\cal C}_\text{target}=10^4$, the average of total success count and infidelity are given by $\langle {\cal C}_\text{total}\rangle \approx 7.261\times 10^6$ and $\bar{f}\approx6.561\times10^{-4}$ for $4$-qubits, and $\langle {\cal C}_\text{total}\rangle\approx1.291\times 10^8$ and $\bar{f}\approx9.311\times10^{-4}$ for $5$-qubits. The scaling factor of infidelity achieves $\beta=-1.189$ and $\beta=-0.829$ for $4$- and $5$-qubit, respectively. For the multi-qubit cases including the $2$- and $3$-qubit results, our method gives the similar results to the self-guided quantum state tomography (SGQT) which shows the scaling factors in the range $\beta\in(0.80, 1.05)$ up to $10$-qubits~\cite{Ferrie2014SGTQT}. In Appendix~\ref{sec:mixed}, we investigate the performance of our scheme for mixed states.

The trained RL agent can generalize to learning of a specific $5$-qubit state. As illustrated in Fig.~\ref{fig:reconstruct} (a), the density matrix of the $5$-qubit state $\hat{\varrho}_{SC}$ is given by a Shen-Castan matrix, where each element is determined by a real-valued function $\exp[-(\abs{m-16.5}+\abs{n-16.5}+1)/10]/K$ for $m,n=1,\ldots,32$ and $K=0.00366$. (A Shen-Castan matrix is a filer used for edge detection in image processing~\cite{SHEN1992112}.) This quantum state is pure, i.e., $\Tr\hat{\varrho}_{SC}^2=1$, and the five qubits are entangled. For a reduced state of $i$-th subsystem $\hat{\varrho}_i$, Von Neumann entropy ${ S}_i=-\Tr\hat{\varrho}_i\log_2 \hat{\varrho}_i$ has nonzero value: ${ S}_1=0.640$, ${ S}_2=0.547$, ${ S}_3=0.230$, ${ S}_4=0.080$, and ${ S}_5=0.025$. (Note that subsystems of quantum states sampled from Haar measure exhibit nearly maximal entropy~\cite{Page1993}.) In the learning of the $5$-qubit state, we achieve fidelity ${f}\approx0.9989$ by using $6.62\times 10^7$ success count. The result of state reconstruction is illustrated in Fig.~\ref{fig:reconstruct} (b). Fig.~\ref{fig:reconstruct} (c) shows the changes of infidelity throughout the learning process. The policy of the RL agent induces exploration in the early time steps, and, as the ES approaches the optimal point, the agent focuses on exploitation to improve the quality of solution.

\begin{figure*}[t!]
    \centering
    \includegraphics[width=0.85\linewidth]{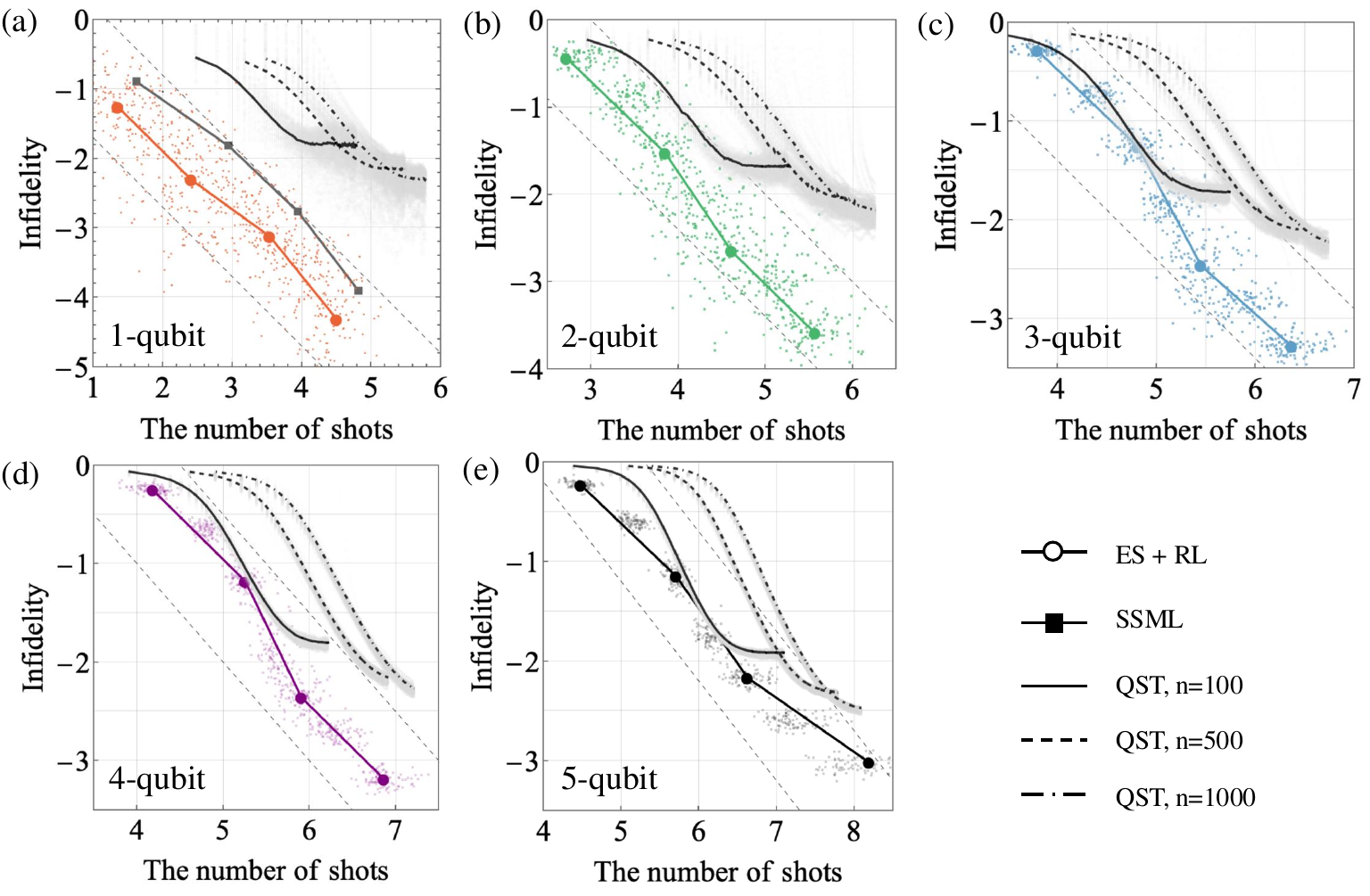}
    \caption{{Comparison between our meta-learning method and the quantum state tomography (QST).} The results of learning a hundred random pure states. The infidelity and the number of shots are represented on a logarithmic scale in base $10$. The red, green, blue, purple, and black dots represent data obtained using our meta-learning. These are equivalent to the data labeled as ES+RL in Fig.~\ref{fig:gen}. The square dots indicate results from SSML. The solid, dashed, and dot-dashed lines depict the results of QST based on the maximum likelihood estimation. For estimating random $N$-qubit states, the QST considers the $4^N-1$ measurement settings, where each setting uses $n$ copies of the states (or $n$ shots).}
    \label{fig:QST}
\end{figure*}

\subsection{Comparison with the quantum state tomography}
We compare our results with the performance of standard method, quantum state tomography (QST), based on maximum likelihood estimation~\cite{Hradil1997,paris2004}. Fig.~\ref{fig:QST} shows the results of QST, demonstrating that the RL-driven meta-learning scheme learns the random quantum states more efficiently in terms of resource usage than the standard method.

To estimate a hundred random $N$-qubit states with QST, we assume a parametric model $\hat{\varrho}(\vec{r})=\vec{r}\cdot{\cal P}_N/2^N$, where $\vec{r}$ is a generalized Bloch vector of $4^N$ dimension, and ${\cal P}_N$ is a string of Pauli operators, $\{I,\hat{\sigma}_x,\hat{\sigma}_y,\hat{\sigma}_z\}^{\otimes N}$. The first element of parameter vector is fixed to one, and the $4^N-1$ elements of the initial parameter vector $\vec{r}_0$ are randomly chosen. We consider $4^N-1$ measurements of which outcomes are associated with projectors given by $\{M_I,M_{\pm X},M_{\pm Y},M_{\pm Z}\}^{\otimes N}/M^{\otimes N}_I$, where $M_I=I/2$, $M_{\pm X} = (I\pm \hat{\sigma}_x)/2$, $M_{\pm Y} = (I\pm \hat{\sigma}_y)/2$, and $M_{\pm Z}=(I \pm \hat{\sigma}_z)/2$. Each measurement uses $n$ copies of unknown states (or equivalently $n$ shots). See Appendix~\ref{sec:QST} for a detailed description of the QST method.

\section{Discussion \& Conclusion}
\label{sec:dis}

The single-shot measurement scheme used in this work is proposed by the single-shot measurement learning (SSML)~\cite{Bang2018,Bang2021}. Notably, the SSML addresses following problems of the standard approach: (i) The way to reconstruct a learned state is not known. (ii) The learned state can have negative eigenvalues~\cite{paris2004}. (iii) The scaling of infidelity is limited to ${\cal O}(n^{-3/4})$ for $n$ copies of states~\cite{Bagan2004}. Our method also addresses these limitations: The trained HEA immediately provides a way to reconstruct the estimated state, ensuring that the reconstructed state always satisfies the conditions of a valid quantum state, as it is represented by a quantum circuit. Also, the average infidelity nearly achieves the scaling of statistical limit~\cite{Haah2017}. While the SSML shows that the optimal scaling in the learning up to $6$-dimensional states~\cite{Bang2021}, our results show the optimal scaling for higher dimensional states. Implementing the SSML on a quantum computer is challenging for high-dimensional states as it uses a parameterized special unitary gate and compiling the parameterized circuit typically incurs a high computational cost. In constrast, our approach employs the hardware efficient ansatzs of which structure can be compatible with quantum computers of low connectivity such as a one-dimensional qubit array with nearest-neighbor interactions.

A strategy of action repetition has been used in a reinforcement learning scheme for games~\cite{sharma2017learning}. In this context, the motivation behind action repetition is to imitate human behaviors in playing games. In contrast, we use the action repetition strategy to control the depth of the decision process through the curriculum so that the agent can experience the problems of various difficulties. Reducing the depth and breadth of the decision process is a key issue in successfully applying reinforcement learning to large-scale problems~\cite{Silver2016}. (Note that in our problem, discretizing the action space reduces the breadth of the decision process.) The effectiveness of curriculum learning has been shown in meta-learning schemes for gradient descent~\cite{chen2017learning} and the quantum optimization problems~\cite{Wilson2021}. In our setting, the curriculum of the previous works can be realized by gradually increasing the maximum time steps of evolution strategy from a small value to a large value. However, we do not apply this curriculum to our method as controlling the maximum time step directly impacts on the estimation of the observation-value function. If the maximum time step is set to a small value, the learning process may not complete, leading to inaccurate cumulative reward estimation. This inaccuracy can bias the agent to solve only easier problem instances, which are the quantum states that can be learned within a short time step.

For the task of learning quantum states, we propose the meta-learning scheme based on the hybrid model of quantum and classical neural networks. Our approach trains a hardware-efficient ansatz (HEA) with low connectivity using an evolution strategy to learn Haar-random quantum states. To enhance the learning of quantum states, we use the reinforcement learning (RL), where an RL agent optimizes the hyperparameters of the evolution strategy. The trained RL agent enables the reconstructed states to achieve infidelity scaling close to the Heisenberg limit~\cite{Haah2017}. Our meta-learning model demonstrates generalizability, as the RL agent can be applied to learning quantum states of higher dimensions than those used for training. The generalization is possible since the structure of RL agent is independent of the dimension of quantum system. To obtain these results, we introduce the action repetition strategy (ARS), which facilitates RL training. We expect that our meta-learning scheme including ARS can be leveraged to enhance existing classical and quantum algorithms for quantum state learning, quantum control~\cite{Porotti2023}, quantum machine learning~\cite{Maria2015Intro}, and quantum optimization~\cite{farhi2014,Abbas2024}. Moreover, meta-learning for hyperparameter optimization of blackbox optimization algorithms has broad applications across multiple disciplines~\cite{Wilson2021,chen17e,tv2019meta,Shala2020,lange2023discovering}.

\begin{acknowledgments}
Authors thank Minyoung Lee and Taehee Lee for their support. J.J. thanks Sang Min Lee, Jeongho Bang, and Akash Kundu for discussions.
\end{acknowledgments}

\appendix
\setcounter{equation}{0}
\setcounter{section}{0}
\setcounter{table}{0}
\setcounter{figure}{0}
\renewcommand{\thetable}{A\arabic{table}}
\renewcommand{\d}[1]{\ensuremath{\operatorname{d}\!{#1}}}
\renewcommand{\thesubsection}{A\arabic{subsection}}
\renewcommand{\theequation}{A\arabic{equation}}
\renewcommand{\thefigure}{A\arabic{figure}}

\begin{figure*}[t]
    \centering
    \includegraphics[width=\linewidth]{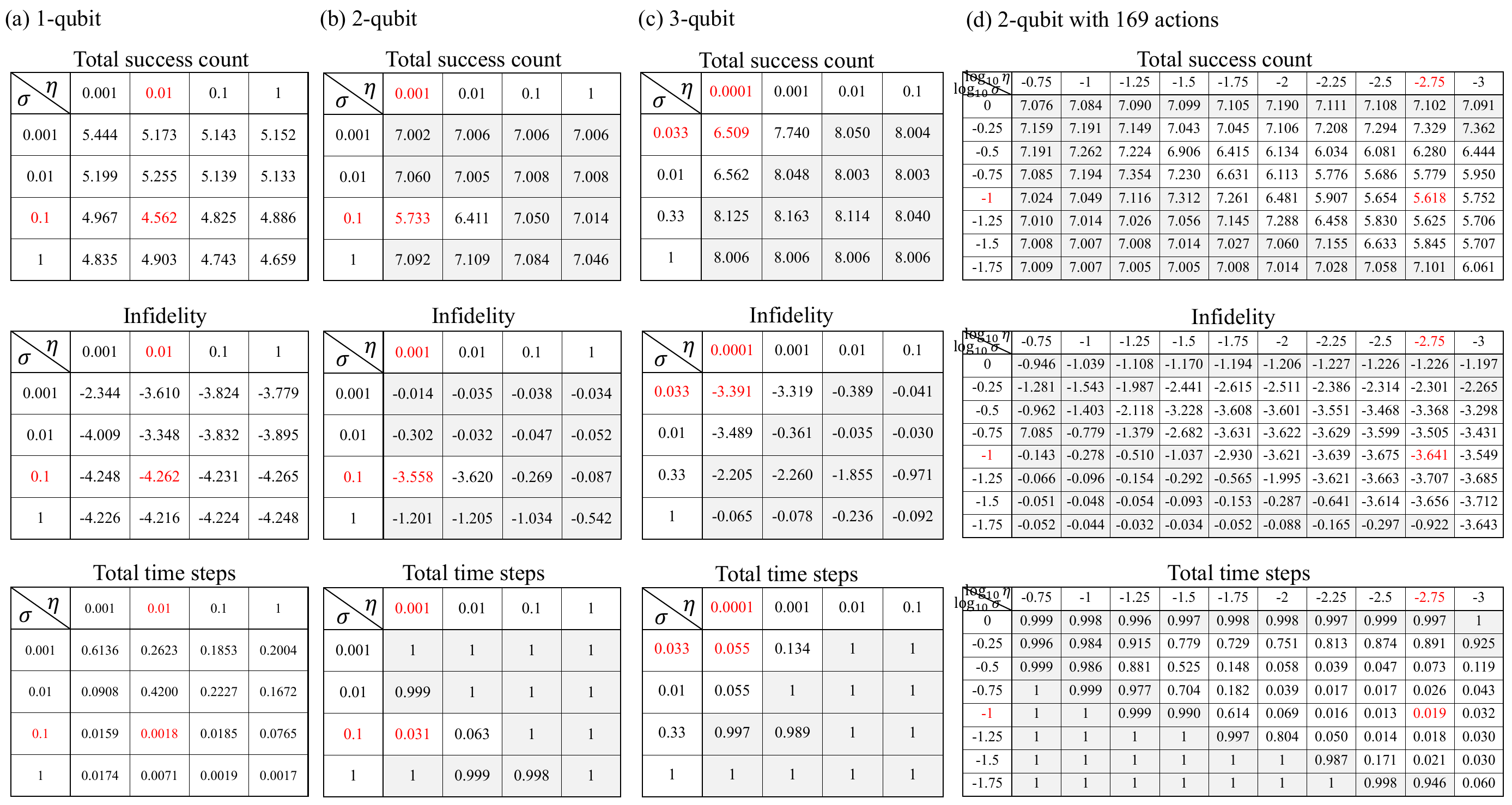}
    \caption{Results of learning a thousand of random pure states obtained without the RL agent. The target success count is set to ${\cal C}_\text{target}=10^4$. We express the average of total success count and infidelity as logarithmic values in base $10$. The total time steps are represented by ratio $t_H/t_\text{max}$, where $t_\text{max}=10^4$, $10^5$, and $10^6$ for $1$-, $2$-, and $3$-qubit state, respectively. The shaded cells are where the evolution strategy fails to learn quantum states within the maximum time step. The tables in the columns (a)--(c) are the results obtained by learning of $1$-, $2$-, and $3$-qubit states with the $16$ actions, respectively. The column (d) is the results of learning $2$-qubit states with the $169$ actions. The red fonts represent the baseline values and respective actions $a_\text{base}$.}
    \label{fig:optpara}
\end{figure*}

\section{Markov decision process}\label{sec:mdp}

For finite time steps $t=1,2,\ldots,T$, a markov decision process (MDP) is represented by a tuple $\left( {\cal S},{\cal A},{\cal T},{\cal R} \right)$, where ${\cal S}$ is the space of state $s_t$, ${\cal A}$ is the space of action $a_t$, ${\cal T}$ is the space of transition probability $p(s_{t+1}|s_t, a_t)$ from a state $s_{t}$ to $s_{t+1}$ by an action $a_t$, and ${\cal R}$ is the space of reward $r_t$. In the description of MDP, a trajectory of decision process $\tau$, e.g., $s_1,a_1,r_1,\ldots,s_{T-1},a_{T-1},s_T,r_T$, is deemed to satisfy Markov condition, which assumes that a current state and reward only depend on a previous state and action. Specifically, the assumption implies that a probability distribution which governs the trajectory $\tau$ can be written as
\begin{eqnarray}
    p(\tau)&=&p(s_1,a_1,r_1,s_2,a_2,r_2\ldots,s_{T-1},a_{T-1},s_T,r_T) \nonumber\\
    &=& p(s_1)\prod_{t=1}^{T-1} p(s_{t+1},r_{t+1}|s_t,a_t)p(a_t|s_t),
\end{eqnarray}
where the probability $p(a_t|s_t)$ is called {\em policy}. The purpose of agent is to learn the optimal policy so as to maximize the expected cumulative reward $G_t$ at each time step $t$, where
\begin{eqnarray}
    G_t := {\mathbf{E}}_{\tau \sim p(\tau)}\left[\sum_t^{T}r_t \right].
\end{eqnarray}
Finally, the RL problem that the agent solves is an optimization to find a policy $\tilde{\pi}$ such that
\begin{eqnarray}
    \tilde{\pi} = \mathop{\mathrm{argmax}}_{\{p(a_t|s_t)\}} G_0.
\end{eqnarray}

\section{Elements of POMDP}\label{sec:pomdp}

Our RL problem can be formulated with a POMDP defined by a tuple $\left({\cal S},{\cal A},{\cal T},{\cal R},\Omega,{\cal O}\right)$:
\begin{enumerate}
    
    \item ${\cal S}$ is the Hilbert space of $N$-qubit states. The state $s_t$ is a quantum state $\hat{U}(\vec{\theta}_t)\ket{\psi}$ transformed by the HEA in Fig.~\ref{fig:setting} (b).

    \item ${\cal A}$ is the space of action $a_t=(\sigma_t,\eta_t)$, where $\sigma$ and $\eta$ are hyperprameters of evolution strategy which determine {\em sampling range} and {\em learning rate}, respectively~\cite{salimans2017}.

    \item ${\cal T}$ is the space of transition probability defined by $p(s_{t+1}|s_t,a_t): {\cal S}\times{\cal A}\times{\cal S} \rightarrow [0,1]$.

    \item ${\cal R}$ is the space of reward $r_t$. The reward $r_t=0$ if the quantum state learning is finished at time step $t$. Otherwise, $r_t=-1$.

    \item $\Omega$ is the space of observation $o_t$. To observe the state, we use a measurement of binary outcome; {\em success} and {\em fail}. We perform the measurement until a fail outcome appears. The observation is defined as the number of consecutive success outcomes before the fail outcome.

    \item ${\cal O}$ is the space of the transition probability of observation defined by $p(o_{t}|s_t): {\cal O}\times{\cal S}\rightarrow[0,1]$. 
  
\end{enumerate}

\section{Actor-Critic algorithm}\label{sec:AC}
Actor-Critic algorithm proceeds with iteration of generating training datasets, value evaluation, and policy improvement~\cite{konda1999actor,wang2016sample}. We generate the rollout datasets $\{(o_t,a_t,r_t,o_{t+1})\}_{t=1}^{t_H-1}$ by using a tree search guided by the policy and store them in the data buffer. We randomly sample a datasets $\cal B$ from the data buffer~\cite{Mnih2015}, and, based on the sampled datasets $\cal B$, the critic network evaluates the observation-value function $Q_{\phi_1}$. We take the empirical cumulative reward~\eqref{eq:empcum} as an estimator of the observation-value function:
\begin{eqnarray}
\label{eq:ovalue}
    \tilde{Q}(o_t) = \frac{{\cal R}_t}{{t}_\text{max}},
\end{eqnarray}
where ${t}_\text{max}$ is the maximum time step of ES which is set to $3\times10^3$, $10^4$, and $2\times 10^4$ for $1$-, $2$-, and $3$-qubit, respectively. We introduce ${t}_\text{max}$ in the denominator to normalize the empirical cumulative reward, and its value is empirically chosen. The loss function of critic network is given by the difference between the output of critic network and the estimator of observation-value function, i.e.,
\begin{eqnarray}
    loss_\text{critic}(\phi_1) := \frac{1}{2}\left(Q_{\phi_1} (o_t) - \tilde{Q}(o_t) \right)^2
\end{eqnarray}
We update the parameters of critic network $\phi_1$ by using a gradient descent method. Based on the critic network, we improve the policy $\pi_{\phi_2}$ by training the actor network. The loss function of actor network is given by
\begin{eqnarray}
    loss_\text{actor}(\phi_2) := {A(o_t)} \log \pi_{\phi_2} (a_t|o_t),
\end{eqnarray}
where $A(o_t) := Q_{\phi_1} (o_t) - \tilde{Q}(o_t)$ is called advantage. The parameters of actor network $\phi_2$ is updated by a gradient ascent.

In the typical algorithm of Advantage Actor-Critic (A$2$C)~\cite{pmlr-v48-mniha16}, the value function and advantage are estimated by using the time difference (TD) error. In our case, we estimate these quantities based on the empirical value function obtained by a policy-guided tree search, similar to RL using Monte-Carlo tree search~\cite{Silver2016,Kocsis2006}. We use the ADAM optimizer for the gradient methods~\cite{kingma2014adam}. The learning rate of ADAM optimizer is set to $10^{-4}$ for learning $1$-qubit states, and $3\times 10^{-5}$ for learning $2$- and $3$-qubit states.

\section{Learning quantum states without RL agent}
\label{sec:ESperformance}

In Fig.~\ref{fig:optpara}, we investigate the performance of learning a thousand random pure states without using the RL agent. We choose the performance given by an action $a_\text{base}$ which gives the lowest total success count as the baseline. For $1$-qubit, $a_\text{base}=(\sigma= 0.1 , \eta=0.01)$. In case of $2$-qubit, $a_\text{base}=(\sigma= 0.1 , \eta=0.001)$ for the $16$ actions, and $a_\text{base}=(\sigma=10^{-1}, \eta=10^{-2.75})$ for the $169$ actions. For $3$-qubit, $a_\text{base}=(\sigma=0.033, \eta=0.0001)$. The respective baseline values are represented by the red fonts in Fig.~\ref{fig:optpara}, and the dashed lines in Fig.~\ref{fig:RLcurve}. The shaded cells are where the evolution strategy fails to learn quantum states within the maximum time step $t_\text{max}$. These results imply that, without the RL agent, finding an action which can successfully complete the quantum state learning requires much amount of computational resources. For example, without the trained RL agent, learning a thousand $3$-qubit states requires more than a million time steps for most of the actions considered.

\begin{figure}[t!]
    \centering
    \includegraphics[width=\linewidth]{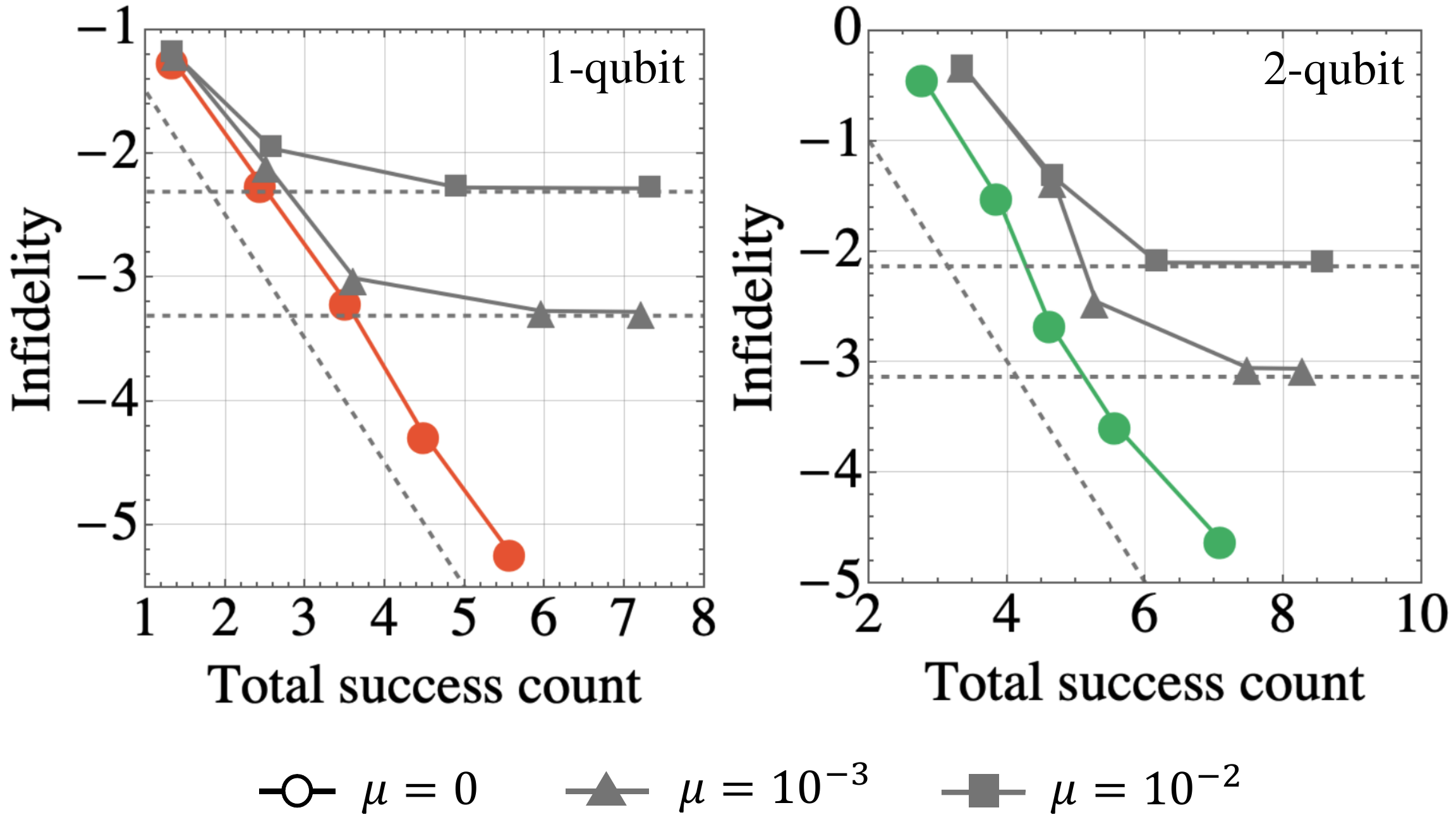}
    \caption{Results of learning $1$- and $2$-qubit mixed states. The average infidelity $\langle\bar{f}\rangle$ and the average total success count $\langle{\cal C}_\text{total}\rangle$ are represented on a logarithmic scale in base $10$. $\mu$ represents the amount of noise in the mixed states. The target success counts ${\cal C}_\text{target}$ are set to values from $10^1$ to $10^{5}$. The horizontal dashed lines represent the lower bound of infidelity $\bar{f}_l=\mu/2$. The diagonal dashed lines indicate the scaling of Heisenberg limit.}
    \label{fig:noise}
\end{figure}

\section{Learning mixed states}
\label{sec:mixed}
We investigate the performance of our meta-learning scheme for a hundred random mixed states $\hat{\varrho}$. A random $N$-qubit mixed state is given by
\begin{eqnarray}
    \hat{\varrho} = (1-\mu)|\psi\rangle\langle\psi| + \mu\frac{I}{2^N},
\end{eqnarray}
where $\ket{\psi}$ is a random pure state, and $\mu$ quantifies how much noise is in the state $\hat{\varrho}$. If $\mu=0$, the state is pure, and, if $\mu=1$, the state is the maximally mixed state. Fig.~\ref{fig:noise} shows the performance of our meta-learning scheme for the random mixed states of $\mu=0$, $\mu=10^{-3}$ and $\mu=10^{-2}$. The noise degrades the performance of our scheme. The lower bound of infidelity is given by $\bar{f}_l=1-\langle\psi|\hat{\varrho}|\psi\rangle=\mu/2$. Our method nearly achieves the lower bound.

\section{Quantum state tomography}
\label{sec:QST}

We briefly review the process of quantum state tomography (QST). Let $\hat{\varrho}_\text{true}$ be an unknown state to be estimated with the QST, and let $\hat{\varrho}(\vec{r})$ be a parametric model of the unknown state. To estimate $\hat{\varrho}_\text{true}$, measurements $M$s are performed on the multiple copies of unknown state, and the frequency of each outcome $i$ is recorded as $f_i$. The QST concerns how much the probabilities of model $p_i(\vec{r}) = \Tr\hat{\varrho}(\vec{r})\hat{M}_i$ differ from the frequencies. The difference can be quantified by the Kullback–Leibler (KL) divergence, $\sum_i f_i\log(f_i/p_i(\vec{r}))$. Maximum likelihood estimation finds the parameters of the minimum KL divergence, $\vec{r}_\text{ML}$, by maximizing the log-likelihood function ${\cal L}(\vec{r}) := \sum_{i}f_i\log p_i(\vec{r})$. In other words, the optimal parameters are obtained by
\begin{eqnarray}
    \vec{r}_\text{ML} = \argmax_{\vec{r}}{\cal L}(\vec{r}).
\end{eqnarray}
Thus, in QST, the state estimation problem amounts to the optimization problem of the log-likelihood function.

For the optimization, we employ a heuristic algorithm, so-called $R\varrho R$ method~\cite{Jaroslav2007}, which updates the model $\hat{\varrho}(\vec{r})$ by following rule
\begin{eqnarray}
    \hat{\varrho}(\vec{r}_{t+1}) \leftarrow \alpha\hat{\varrho}(\vec{r}_{t}) + (1-\alpha)\frac{\hat{R}_t\hat{\varrho}(\vec{r}_{t}) \hat{R}_t}{\Tr\left(\hat{R}_t\hat{\varrho}(\vec{r}_{t}) \hat{R}_t\right)},
\end{eqnarray}
where $\hat{R}_t := \frac{1}{m}\sum_i (f_i/p_i(\vec{r}_t)) \hat{\Pi}_i$, and $m$ is the number of measurement bases used for the estimation. The learning rate is set to $\alpha=0.5$. This method is based on the expectation maximization algorithm, and the fact that the optimal state $\hat{\varrho}(\vec{r}_\text{ML})$ is an eigenstate of the operator $R$.


\begin{thebibliography}{96}%
\makeatletter
\providecommand \@ifxundefined [1]{%
 \@ifx{#1\undefined}
}%
\providecommand \@ifnum [1]{%
 \ifnum #1\expandafter \@firstoftwo
 \else \expandafter \@secondoftwo
 \fi
}%
\providecommand \@ifx [1]{%
 \ifx #1\expandafter \@firstoftwo
 \else \expandafter \@secondoftwo
 \fi
}%
\providecommand \natexlab [1]{#1}%
\providecommand \enquote  [1]{``#1''}%
\providecommand \bibnamefont  [1]{#1}%
\providecommand \bibfnamefont [1]{#1}%
\providecommand \citenamefont [1]{#1}%
\providecommand \href@noop [0]{\@secondoftwo}%
\providecommand \href [0]{\begingroup \@sanitize@url \@href}%
\providecommand \@href[1]{\@@startlink{#1}\@@href}%
\providecommand \@@href[1]{\endgroup#1\@@endlink}%
\providecommand \@sanitize@url [0]{\catcode `\\12\catcode `\$12\catcode `\&12\catcode `\#12\catcode `\^12\catcode `\_12\catcode `\%12\relax}%
\providecommand \@@startlink[1]{}%
\providecommand \@@endlink[0]{}%
\providecommand \url  [0]{\begingroup\@sanitize@url \@url }%
\providecommand \@url [1]{\endgroup\@href {#1}{\urlprefix }}%
\providecommand \urlprefix  [0]{URL }%
\providecommand \Eprint [0]{\href }%
\providecommand \doibase [0]{https://doi.org/}%
\providecommand \selectlanguage [0]{\@gobble}%
\providecommand \bibinfo  [0]{\@secondoftwo}%
\providecommand \bibfield  [0]{\@secondoftwo}%
\providecommand \translation [1]{[#1]}%
\providecommand \BibitemOpen [0]{}%
\providecommand \bibitemStop [0]{}%
\providecommand \bibitemNoStop [0]{.\EOS\space}%
\providecommand \EOS [0]{\spacefactor3000\relax}%
\providecommand \BibitemShut  [1]{\csname bibitem#1\endcsname}%
\let\auto@bib@innerbib\@empty
\bibitem [{\citenamefont {Anshu}\ and\ \citenamefont {Arunachalam}(2024)}]{Anshu2024}%
  \BibitemOpen
  \bibfield  {author} {\bibinfo {author} {\bibfnamefont {A.}~\bibnamefont {Anshu}}\ and\ \bibinfo {author} {\bibfnamefont {S.}~\bibnamefont {Arunachalam}},\ }\bibfield  {title} {\bibinfo {title} {A survey on the complexity of learning quantum states},\ }\href {https://doi.org/10.1038/s42254-023-00662-4} {\bibfield  {journal} {\bibinfo  {journal} {Nature Reviews Physics}\ }\textbf {\bibinfo {volume} {6}},\ \bibinfo {pages} {59} (\bibinfo {year} {2024})}\BibitemShut {NoStop}%
\bibitem [{\citenamefont {Gebhart}\ \emph {et~al.}(2023)\citenamefont {Gebhart}, \citenamefont {Santagati}, \citenamefont {Gentile}, \citenamefont {Gauger}, \citenamefont {Craig}, \citenamefont {Ares}, \citenamefont {Banchi}, \citenamefont {Marquardt}, \citenamefont {Pezz{\`e}},\ and\ \citenamefont {Bonato}}]{Gebhart2023}%
  \BibitemOpen
  \bibfield  {author} {\bibinfo {author} {\bibfnamefont {V.}~\bibnamefont {Gebhart}}, \bibinfo {author} {\bibfnamefont {R.}~\bibnamefont {Santagati}}, \bibinfo {author} {\bibfnamefont {A.~A.}\ \bibnamefont {Gentile}}, \bibinfo {author} {\bibfnamefont {E.~M.}\ \bibnamefont {Gauger}}, \bibinfo {author} {\bibfnamefont {D.}~\bibnamefont {Craig}}, \bibinfo {author} {\bibfnamefont {N.}~\bibnamefont {Ares}}, \bibinfo {author} {\bibfnamefont {L.}~\bibnamefont {Banchi}}, \bibinfo {author} {\bibfnamefont {F.}~\bibnamefont {Marquardt}}, \bibinfo {author} {\bibfnamefont {L.}~\bibnamefont {Pezz{\`e}}},\ and\ \bibinfo {author} {\bibfnamefont {C.}~\bibnamefont {Bonato}},\ }\bibfield  {title} {\bibinfo {title} {Learning quantum systems},\ }\href {https://doi.org/10.1038/s42254-022-00552-1} {\bibfield  {journal} {\bibinfo  {journal} {Nature Reviews Physics}\ }\textbf {\bibinfo {volume} {5}},\ \bibinfo {pages} {141} (\bibinfo {year} {2023})}\BibitemShut {NoStop}%
\bibitem [{\citenamefont {Vogel}\ and\ \citenamefont {Risken}(1989)}]{Vogel1989}%
  \BibitemOpen
  \bibfield  {author} {\bibinfo {author} {\bibfnamefont {K.}~\bibnamefont {Vogel}}\ and\ \bibinfo {author} {\bibfnamefont {H.}~\bibnamefont {Risken}},\ }\bibfield  {title} {\bibinfo {title} {Determination of quasiprobability distributions in terms of probability distributions for the rotated quadrature phase},\ }\href {https://doi.org/10.1103/PhysRevA.40.2847} {\bibfield  {journal} {\bibinfo  {journal} {Phys. Rev. A}\ }\textbf {\bibinfo {volume} {40}},\ \bibinfo {pages} {2847} (\bibinfo {year} {1989})}\BibitemShut {NoStop}%
\bibitem [{\citenamefont {Hradil}(1997)}]{Hradil1997}%
  \BibitemOpen
  \bibfield  {author} {\bibinfo {author} {\bibfnamefont {Z.}~\bibnamefont {Hradil}},\ }\bibfield  {title} {\bibinfo {title} {Quantum-state estimation},\ }\href {https://doi.org/10.1103/PhysRevA.55.R1561} {\bibfield  {journal} {\bibinfo  {journal} {Phys. Rev. A}\ }\textbf {\bibinfo {volume} {55}},\ \bibinfo {pages} {R1561} (\bibinfo {year} {1997})}\BibitemShut {NoStop}%
\bibitem [{\citenamefont {Paris}\ and\ \citenamefont {Rehacek}(2004)}]{paris2004}%
  \BibitemOpen
  \bibfield  {author} {\bibinfo {author} {\bibfnamefont {M.}~\bibnamefont {Paris}}\ and\ \bibinfo {author} {\bibfnamefont {J.}~\bibnamefont {Rehacek}},\ }\href@noop {} {\emph {\bibinfo {title} {Quantum state estimation}}},\ Vol.\ \bibinfo {volume} {649}\ (\bibinfo  {publisher} {Springer Science \& Business Media},\ \bibinfo {year} {2004})\BibitemShut {NoStop}%
\bibitem [{\citenamefont {Cramer}\ \emph {et~al.}(2010)\citenamefont {Cramer}, \citenamefont {Plenio}, \citenamefont {Flammia}, \citenamefont {Somma}, \citenamefont {Gross}, \citenamefont {Bartlett}, \citenamefont {Landon-Cardinal}, \citenamefont {Poulin},\ and\ \citenamefont {Liu}}]{Cramer2010}%
  \BibitemOpen
  \bibfield  {author} {\bibinfo {author} {\bibfnamefont {M.}~\bibnamefont {Cramer}}, \bibinfo {author} {\bibfnamefont {M.~B.}\ \bibnamefont {Plenio}}, \bibinfo {author} {\bibfnamefont {S.~T.}\ \bibnamefont {Flammia}}, \bibinfo {author} {\bibfnamefont {R.}~\bibnamefont {Somma}}, \bibinfo {author} {\bibfnamefont {D.}~\bibnamefont {Gross}}, \bibinfo {author} {\bibfnamefont {S.~D.}\ \bibnamefont {Bartlett}}, \bibinfo {author} {\bibfnamefont {O.}~\bibnamefont {Landon-Cardinal}}, \bibinfo {author} {\bibfnamefont {D.}~\bibnamefont {Poulin}},\ and\ \bibinfo {author} {\bibfnamefont {Y.-K.}\ \bibnamefont {Liu}},\ }\bibfield  {title} {\bibinfo {title} {Efficient quantum state tomography},\ }\href {https://doi.org/10.1038/ncomms1147} {\bibfield  {journal} {\bibinfo  {journal} {Nature Communications}\ }\textbf {\bibinfo {volume} {1}},\ \bibinfo {pages} {149} (\bibinfo {year} {2010})}\BibitemShut {NoStop}%
\bibitem [{\citenamefont {Lanyon}\ \emph {et~al.}(2017)\citenamefont {Lanyon}, \citenamefont {Maier}, \citenamefont {Holz{\"a}pfel}, \citenamefont {Baumgratz}, \citenamefont {Hempel}, \citenamefont {Jurcevic}, \citenamefont {Dhand}, \citenamefont {Buyskikh}, \citenamefont {Daley}, \citenamefont {Cramer}, \citenamefont {Plenio}, \citenamefont {Blatt},\ and\ \citenamefont {Roos}}]{Lanyon2017}%
  \BibitemOpen
  \bibfield  {author} {\bibinfo {author} {\bibfnamefont {B.~P.}\ \bibnamefont {Lanyon}}, \bibinfo {author} {\bibfnamefont {C.}~\bibnamefont {Maier}}, \bibinfo {author} {\bibfnamefont {M.}~\bibnamefont {Holz{\"a}pfel}}, \bibinfo {author} {\bibfnamefont {T.}~\bibnamefont {Baumgratz}}, \bibinfo {author} {\bibfnamefont {C.}~\bibnamefont {Hempel}}, \bibinfo {author} {\bibfnamefont {P.}~\bibnamefont {Jurcevic}}, \bibinfo {author} {\bibfnamefont {I.}~\bibnamefont {Dhand}}, \bibinfo {author} {\bibfnamefont {A.~S.}\ \bibnamefont {Buyskikh}}, \bibinfo {author} {\bibfnamefont {A.~J.}\ \bibnamefont {Daley}}, \bibinfo {author} {\bibfnamefont {M.}~\bibnamefont {Cramer}}, \bibinfo {author} {\bibfnamefont {M.~B.}\ \bibnamefont {Plenio}}, \bibinfo {author} {\bibfnamefont {R.}~\bibnamefont {Blatt}},\ and\ \bibinfo {author} {\bibfnamefont {C.~F.}\ \bibnamefont {Roos}},\ }\bibfield  {title} {\bibinfo {title} {Efficient tomography of a quantum many-body system},\ }\href {https://doi.org/10.1038/nphys4244} {\bibfield  {journal}
  {\bibinfo  {journal} {Nature Physics}\ }\textbf {\bibinfo {volume} {13}},\ \bibinfo {pages} {1158} (\bibinfo {year} {2017})}\BibitemShut {NoStop}%
\bibitem [{\citenamefont {Schwemmer}\ \emph {et~al.}(2014)\citenamefont {Schwemmer}, \citenamefont {T\'oth}, \citenamefont {Niggebaum}, \citenamefont {Moroder}, \citenamefont {Gross}, \citenamefont {G\"uhne},\ and\ \citenamefont {Weinfurter}}]{Schwemmer2014}%
  \BibitemOpen
  \bibfield  {author} {\bibinfo {author} {\bibfnamefont {C.}~\bibnamefont {Schwemmer}}, \bibinfo {author} {\bibfnamefont {G.}~\bibnamefont {T\'oth}}, \bibinfo {author} {\bibfnamefont {A.}~\bibnamefont {Niggebaum}}, \bibinfo {author} {\bibfnamefont {T.}~\bibnamefont {Moroder}}, \bibinfo {author} {\bibfnamefont {D.}~\bibnamefont {Gross}}, \bibinfo {author} {\bibfnamefont {O.}~\bibnamefont {G\"uhne}},\ and\ \bibinfo {author} {\bibfnamefont {H.}~\bibnamefont {Weinfurter}},\ }\bibfield  {title} {\bibinfo {title} {Experimental comparison of efficient tomography schemes for a six-qubit state},\ }\href {https://doi.org/10.1103/PhysRevLett.113.040503} {\bibfield  {journal} {\bibinfo  {journal} {Phys. Rev. Lett.}\ }\textbf {\bibinfo {volume} {113}},\ \bibinfo {pages} {040503} (\bibinfo {year} {2014})}\BibitemShut {NoStop}%
\bibitem [{\citenamefont {Aaronson}(2018)}]{aaronson2018}%
  \BibitemOpen
  \bibfield  {author} {\bibinfo {author} {\bibfnamefont {S.}~\bibnamefont {Aaronson}},\ }\bibfield  {title} {\bibinfo {title} {Shadow tomography of quantum states},\ }in\ \href@noop {} {\emph {\bibinfo {booktitle} {Proceedings of the 50th annual ACM SIGACT symposium on theory of computing}}}\ (\bibinfo {year} {2018})\ pp.\ \bibinfo {pages} {325--338}\BibitemShut {NoStop}%
\bibitem [{\citenamefont {Lukens}\ \emph {et~al.}(2020)\citenamefont {Lukens}, \citenamefont {Law}, \citenamefont {Jasra},\ and\ \citenamefont {Lougovski}}]{Lukens2020}%
  \BibitemOpen
  \bibfield  {author} {\bibinfo {author} {\bibfnamefont {J.~M.}\ \bibnamefont {Lukens}}, \bibinfo {author} {\bibfnamefont {K.~J.~H.}\ \bibnamefont {Law}}, \bibinfo {author} {\bibfnamefont {A.}~\bibnamefont {Jasra}},\ and\ \bibinfo {author} {\bibfnamefont {P.}~\bibnamefont {Lougovski}},\ }\bibfield  {title} {\bibinfo {title} {A practical and efficient approach for bayesian quantum state estimation},\ }\href {https://doi.org/10.1088/1367-2630/ab8efa} {\bibfield  {journal} {\bibinfo  {journal} {New Journal of Physics}\ }\textbf {\bibinfo {volume} {22}},\ \bibinfo {pages} {063038} (\bibinfo {year} {2020})}\BibitemShut {NoStop}%
\bibitem [{\citenamefont {Park}\ and\ \citenamefont {Kastoryano}(2020)}]{Park2020}%
  \BibitemOpen
  \bibfield  {author} {\bibinfo {author} {\bibfnamefont {C.-Y.}\ \bibnamefont {Park}}\ and\ \bibinfo {author} {\bibfnamefont {M.~J.}\ \bibnamefont {Kastoryano}},\ }\bibfield  {title} {\bibinfo {title} {Geometry of learning neural quantum states},\ }\href {https://doi.org/10.1103/PhysRevResearch.2.023232} {\bibfield  {journal} {\bibinfo  {journal} {Phys. Rev. Res.}\ }\textbf {\bibinfo {volume} {2}},\ \bibinfo {pages} {023232} (\bibinfo {year} {2020})}\BibitemShut {NoStop}%
\bibitem [{\citenamefont {Ahmed}\ \emph {et~al.}(2021{\natexlab{a}})\citenamefont {Ahmed}, \citenamefont {S\'anchez Mu\~noz}, \citenamefont {Nori},\ and\ \citenamefont {Kockum}}]{Ahmed2021PRR}%
  \BibitemOpen
  \bibfield  {author} {\bibinfo {author} {\bibfnamefont {S.}~\bibnamefont {Ahmed}}, \bibinfo {author} {\bibfnamefont {C.}~\bibnamefont {S\'anchez Mu\~noz}}, \bibinfo {author} {\bibfnamefont {F.}~\bibnamefont {Nori}},\ and\ \bibinfo {author} {\bibfnamefont {A.~F.}\ \bibnamefont {Kockum}},\ }\bibfield  {title} {\bibinfo {title} {Classification and reconstruction of optical quantum states with deep neural networks},\ }\href {https://doi.org/10.1103/PhysRevResearch.3.033278} {\bibfield  {journal} {\bibinfo  {journal} {Phys. Rev. Res.}\ }\textbf {\bibinfo {volume} {3}},\ \bibinfo {pages} {033278} (\bibinfo {year} {2021}{\natexlab{a}})}\BibitemShut {NoStop}%
\bibitem [{\citenamefont {Ahmed}\ \emph {et~al.}(2021{\natexlab{b}})\citenamefont {Ahmed}, \citenamefont {S\'anchez Mu\~noz}, \citenamefont {Nori},\ and\ \citenamefont {Kockum}}]{Ahmed2021PRL}%
  \BibitemOpen
  \bibfield  {author} {\bibinfo {author} {\bibfnamefont {S.}~\bibnamefont {Ahmed}}, \bibinfo {author} {\bibfnamefont {C.}~\bibnamefont {S\'anchez Mu\~noz}}, \bibinfo {author} {\bibfnamefont {F.}~\bibnamefont {Nori}},\ and\ \bibinfo {author} {\bibfnamefont {A.~F.}\ \bibnamefont {Kockum}},\ }\bibfield  {title} {\bibinfo {title} {Quantum state tomography with conditional generative adversarial networks},\ }\href {https://doi.org/10.1103/PhysRevLett.127.140502} {\bibfield  {journal} {\bibinfo  {journal} {Phys. Rev. Lett.}\ }\textbf {\bibinfo {volume} {127}},\ \bibinfo {pages} {140502} (\bibinfo {year} {2021}{\natexlab{b}})}\BibitemShut {NoStop}%
\bibitem [{\citenamefont {Cha}\ \emph {et~al.}(2021)\citenamefont {Cha}, \citenamefont {Ginsparg}, \citenamefont {Wu}, \citenamefont {Carrasquilla}, \citenamefont {McMahon},\ and\ \citenamefont {Kim}}]{Cha2022}%
  \BibitemOpen
  \bibfield  {author} {\bibinfo {author} {\bibfnamefont {P.}~\bibnamefont {Cha}}, \bibinfo {author} {\bibfnamefont {P.}~\bibnamefont {Ginsparg}}, \bibinfo {author} {\bibfnamefont {F.}~\bibnamefont {Wu}}, \bibinfo {author} {\bibfnamefont {J.}~\bibnamefont {Carrasquilla}}, \bibinfo {author} {\bibfnamefont {P.~L.}\ \bibnamefont {McMahon}},\ and\ \bibinfo {author} {\bibfnamefont {E.-A.}\ \bibnamefont {Kim}},\ }\bibfield  {title} {\bibinfo {title} {Attention-based quantum tomography},\ }\href {https://doi.org/10.1088/2632-2153/ac362b} {\bibfield  {journal} {\bibinfo  {journal} {Machine Learning: Science and Technology}\ }\textbf {\bibinfo {volume} {3}},\ \bibinfo {pages} {01LT01} (\bibinfo {year} {2021})}\BibitemShut {NoStop}%
\bibitem [{\citenamefont {Lange}\ \emph {et~al.}(2023{\natexlab{a}})\citenamefont {Lange}, \citenamefont {Kebri{\v{c}}}, \citenamefont {Buser}, \citenamefont {Schollw{\"{o}}ck}, \citenamefont {Grusdt},\ and\ \citenamefont {Bohrdt}}]{Lange2023}%
  \BibitemOpen
  \bibfield  {author} {\bibinfo {author} {\bibfnamefont {H.}~\bibnamefont {Lange}}, \bibinfo {author} {\bibfnamefont {M.}~\bibnamefont {Kebri{\v{c}}}}, \bibinfo {author} {\bibfnamefont {M.}~\bibnamefont {Buser}}, \bibinfo {author} {\bibfnamefont {U.}~\bibnamefont {Schollw{\"{o}}ck}}, \bibinfo {author} {\bibfnamefont {F.}~\bibnamefont {Grusdt}},\ and\ \bibinfo {author} {\bibfnamefont {A.}~\bibnamefont {Bohrdt}},\ }\bibfield  {title} {\bibinfo {title} {Adaptive {Q}uantum {S}tate {T}omography with {A}ctive {L}earning},\ }\href {https://doi.org/10.22331/q-2023-10-09-1129} {\bibfield  {journal} {\bibinfo  {journal} {{Quantum}}\ }\textbf {\bibinfo {volume} {7}},\ \bibinfo {pages} {1129} (\bibinfo {year} {2023}{\natexlab{a}})}\BibitemShut {NoStop}%
\bibitem [{\citenamefont {Gaikwad}\ \emph {et~al.}(2024)\citenamefont {Gaikwad}, \citenamefont {Bihani}, \citenamefont {Arvind},\ and\ \citenamefont {Dorai}}]{Gaikwad2024}%
  \BibitemOpen
  \bibfield  {author} {\bibinfo {author} {\bibfnamefont {A.}~\bibnamefont {Gaikwad}}, \bibinfo {author} {\bibfnamefont {O.}~\bibnamefont {Bihani}}, \bibinfo {author} {\bibnamefont {Arvind}},\ and\ \bibinfo {author} {\bibfnamefont {K.}~\bibnamefont {Dorai}},\ }\bibfield  {title} {\bibinfo {title} {Neural-network-assisted quantum state and process tomography using limited data sets},\ }\href {https://doi.org/10.1103/PhysRevA.109.012402} {\bibfield  {journal} {\bibinfo  {journal} {Phys. Rev. A}\ }\textbf {\bibinfo {volume} {109}},\ \bibinfo {pages} {012402} (\bibinfo {year} {2024})}\BibitemShut {NoStop}%
\bibitem [{\citenamefont {Palmieri}\ \emph {et~al.}(2024)\citenamefont {Palmieri}, \citenamefont {M\"uller-Rigat}, \citenamefont {Srivastava}, \citenamefont {Lewenstein}, \citenamefont {Rajchel-Mieldzio\ifmmode~\acute{c}\else \'{c}\fi{}},\ and\ \citenamefont {P\l{}odzie\ifmmode~\acute{n}\else \'{n}\fi{}}}]{Palmieri2024}%
  \BibitemOpen
  \bibfield  {author} {\bibinfo {author} {\bibfnamefont {A.~M.}\ \bibnamefont {Palmieri}}, \bibinfo {author} {\bibfnamefont {G.}~\bibnamefont {M\"uller-Rigat}}, \bibinfo {author} {\bibfnamefont {A.~K.}\ \bibnamefont {Srivastava}}, \bibinfo {author} {\bibfnamefont {M.}~\bibnamefont {Lewenstein}}, \bibinfo {author} {\bibfnamefont {G.}~\bibnamefont {Rajchel-Mieldzio\ifmmode~\acute{c}\else \'{c}\fi{}}},\ and\ \bibinfo {author} {\bibfnamefont {M.}~\bibnamefont {P\l{}odzie\ifmmode~\acute{n}\else \'{n}\fi{}}},\ }\bibfield  {title} {\bibinfo {title} {Enhancing quantum state tomography via resource-efficient attention-based neural networks},\ }\href {https://doi.org/10.1103/PhysRevResearch.6.033248} {\bibfield  {journal} {\bibinfo  {journal} {Phys. Rev. Res.}\ }\textbf {\bibinfo {volume} {6}},\ \bibinfo {pages} {033248} (\bibinfo {year} {2024})}\BibitemShut {NoStop}%
\bibitem [{\citenamefont {Ma}\ \emph {et~al.}(2024)\citenamefont {Ma}, \citenamefont {Dong}, \citenamefont {Petersen}, \citenamefont {Huang},\ and\ \citenamefont {Xiang}}]{Ma2024}%
  \BibitemOpen
  \bibfield  {author} {\bibinfo {author} {\bibfnamefont {H.}~\bibnamefont {Ma}}, \bibinfo {author} {\bibfnamefont {D.}~\bibnamefont {Dong}}, \bibinfo {author} {\bibfnamefont {I.~R.}\ \bibnamefont {Petersen}}, \bibinfo {author} {\bibfnamefont {C.-J.}\ \bibnamefont {Huang}},\ and\ \bibinfo {author} {\bibfnamefont {G.-Y.}\ \bibnamefont {Xiang}},\ }\bibfield  {title} {\bibinfo {title} {Neural networks for quantum state tomography with constrained measurements},\ }\href {https://doi.org/10.1007/s11128-024-04522-7} {\bibfield  {journal} {\bibinfo  {journal} {Quantum Information Processing}\ }\textbf {\bibinfo {volume} {23}},\ \bibinfo {pages} {317} (\bibinfo {year} {2024})}\BibitemShut {NoStop}%
\bibitem [{\citenamefont {Wang}\ \emph {et~al.}(2024)\citenamefont {Wang}, \citenamefont {Dong}, \citenamefont {Li}, \citenamefont {Xu}, \citenamefont {Wang}, \citenamefont {Han}, \citenamefont {Yung}, \citenamefont {Han}, \citenamefont {Li},\ and\ \citenamefont {Guo}}]{QinQin2024}%
  \BibitemOpen
  \bibfield  {author} {\bibinfo {author} {\bibfnamefont {Q.-Q.}\ \bibnamefont {Wang}}, \bibinfo {author} {\bibfnamefont {S.}~\bibnamefont {Dong}}, \bibinfo {author} {\bibfnamefont {X.-W.}\ \bibnamefont {Li}}, \bibinfo {author} {\bibfnamefont {X.-Y.}\ \bibnamefont {Xu}}, \bibinfo {author} {\bibfnamefont {C.}~\bibnamefont {Wang}}, \bibinfo {author} {\bibfnamefont {S.}~\bibnamefont {Han}}, \bibinfo {author} {\bibfnamefont {M.-H.}\ \bibnamefont {Yung}}, \bibinfo {author} {\bibfnamefont {Y.-J.}\ \bibnamefont {Han}}, \bibinfo {author} {\bibfnamefont {C.-F.}\ \bibnamefont {Li}},\ and\ \bibinfo {author} {\bibfnamefont {G.-C.}\ \bibnamefont {Guo}},\ }\bibfield  {title} {\bibinfo {title} {Efficient learning of mixed-state tomography for photonic quantum walk},\ }\href {https://doi.org/10.1126/sciadv.adl4871} {\bibfield  {journal} {\bibinfo  {journal} {Science Advances}\ }\textbf {\bibinfo {volume} {10}},\ \bibinfo {pages} {eadl4871} (\bibinfo {year} {2024})}\BibitemShut {NoStop}%
\bibitem [{\citenamefont {Torlai}\ \emph {et~al.}(2018)\citenamefont {Torlai}, \citenamefont {Mazzola}, \citenamefont {Carrasquilla}, \citenamefont {Troyer}, \citenamefont {Melko},\ and\ \citenamefont {Carleo}}]{Torlai2018}%
  \BibitemOpen
  \bibfield  {author} {\bibinfo {author} {\bibfnamefont {G.}~\bibnamefont {Torlai}}, \bibinfo {author} {\bibfnamefont {G.}~\bibnamefont {Mazzola}}, \bibinfo {author} {\bibfnamefont {J.}~\bibnamefont {Carrasquilla}}, \bibinfo {author} {\bibfnamefont {M.}~\bibnamefont {Troyer}}, \bibinfo {author} {\bibfnamefont {R.}~\bibnamefont {Melko}},\ and\ \bibinfo {author} {\bibfnamefont {G.}~\bibnamefont {Carleo}},\ }\bibfield  {title} {\bibinfo {title} {Neural-network quantum state tomography},\ }\href {https://doi.org/10.1038/s41567-018-0048-5} {\bibfield  {journal} {\bibinfo  {journal} {Nature Physics}\ }\textbf {\bibinfo {volume} {14}},\ \bibinfo {pages} {447} (\bibinfo {year} {2018})}\BibitemShut {NoStop}%
\bibitem [{\citenamefont {Rocchetto}\ \emph {et~al.}(2018)\citenamefont {Rocchetto}, \citenamefont {Grant}, \citenamefont {Strelchuk}, \citenamefont {Carleo},\ and\ \citenamefont {Severini}}]{Rocchetto2018}%
  \BibitemOpen
  \bibfield  {author} {\bibinfo {author} {\bibfnamefont {A.}~\bibnamefont {Rocchetto}}, \bibinfo {author} {\bibfnamefont {E.}~\bibnamefont {Grant}}, \bibinfo {author} {\bibfnamefont {S.}~\bibnamefont {Strelchuk}}, \bibinfo {author} {\bibfnamefont {G.}~\bibnamefont {Carleo}},\ and\ \bibinfo {author} {\bibfnamefont {S.}~\bibnamefont {Severini}},\ }\bibfield  {title} {\bibinfo {title} {Learning hard quantum distributions with variational autoencoders},\ }\href {https://doi.org/10.1038/s41534-018-0077-z} {\bibfield  {journal} {\bibinfo  {journal} {npj Quantum Information}\ }\textbf {\bibinfo {volume} {4}},\ \bibinfo {pages} {28} (\bibinfo {year} {2018})}\BibitemShut {NoStop}%
\bibitem [{\citenamefont {Carrasquilla}\ \emph {et~al.}(2019)\citenamefont {Carrasquilla}, \citenamefont {Torlai}, \citenamefont {Melko},\ and\ \citenamefont {Aolita}}]{Carrasquilla2019}%
  \BibitemOpen
  \bibfield  {author} {\bibinfo {author} {\bibfnamefont {J.}~\bibnamefont {Carrasquilla}}, \bibinfo {author} {\bibfnamefont {G.}~\bibnamefont {Torlai}}, \bibinfo {author} {\bibfnamefont {R.~G.}\ \bibnamefont {Melko}},\ and\ \bibinfo {author} {\bibfnamefont {L.}~\bibnamefont {Aolita}},\ }\bibfield  {title} {\bibinfo {title} {Reconstructing quantum states with generative models},\ }\href {https://doi.org/10.1038/s42256-019-0028-1} {\bibfield  {journal} {\bibinfo  {journal} {Nature Machine Intelligence}\ }\textbf {\bibinfo {volume} {1}},\ \bibinfo {pages} {155} (\bibinfo {year} {2019})}\BibitemShut {NoStop}%
\bibitem [{\citenamefont {Palmieri}\ \emph {et~al.}(2020)\citenamefont {Palmieri}, \citenamefont {Kovlakov}, \citenamefont {Bianchi}, \citenamefont {Yudin}, \citenamefont {Straupe}, \citenamefont {Biamonte},\ and\ \citenamefont {Kulik}}]{Palmieri2020}%
  \BibitemOpen
  \bibfield  {author} {\bibinfo {author} {\bibfnamefont {A.~M.}\ \bibnamefont {Palmieri}}, \bibinfo {author} {\bibfnamefont {E.}~\bibnamefont {Kovlakov}}, \bibinfo {author} {\bibfnamefont {F.}~\bibnamefont {Bianchi}}, \bibinfo {author} {\bibfnamefont {D.}~\bibnamefont {Yudin}}, \bibinfo {author} {\bibfnamefont {S.}~\bibnamefont {Straupe}}, \bibinfo {author} {\bibfnamefont {J.~D.}\ \bibnamefont {Biamonte}},\ and\ \bibinfo {author} {\bibfnamefont {S.}~\bibnamefont {Kulik}},\ }\bibfield  {title} {\bibinfo {title} {Experimental neural network enhanced quantum tomography},\ }\href {https://doi.org/10.1038/s41534-020-0248-6} {\bibfield  {journal} {\bibinfo  {journal} {npj Quantum Information}\ }\textbf {\bibinfo {volume} {6}},\ \bibinfo {pages} {20} (\bibinfo {year} {2020})}\BibitemShut {NoStop}%
\bibitem [{\citenamefont {Gao}\ and\ \citenamefont {Duan}(2017)}]{Gao2017}%
  \BibitemOpen
  \bibfield  {author} {\bibinfo {author} {\bibfnamefont {X.}~\bibnamefont {Gao}}\ and\ \bibinfo {author} {\bibfnamefont {L.-M.}\ \bibnamefont {Duan}},\ }\bibfield  {title} {\bibinfo {title} {Efficient representation of quantum many-body states with deep neural networks},\ }\href {https://doi.org/10.1038/s41467-017-00705-2} {\bibfield  {journal} {\bibinfo  {journal} {Nature Communications}\ }\textbf {\bibinfo {volume} {8}},\ \bibinfo {pages} {662} (\bibinfo {year} {2017})}\BibitemShut {NoStop}%
\bibitem [{\citenamefont {Lee}\ \emph {et~al.}(2018)\citenamefont {Lee}, \citenamefont {Lee},\ and\ \citenamefont {Bang}}]{Bang2018}%
  \BibitemOpen
  \bibfield  {author} {\bibinfo {author} {\bibfnamefont {S.~M.}\ \bibnamefont {Lee}}, \bibinfo {author} {\bibfnamefont {J.}~\bibnamefont {Lee}},\ and\ \bibinfo {author} {\bibfnamefont {J.}~\bibnamefont {Bang}},\ }\bibfield  {title} {\bibinfo {title} {Learning unknown pure quantum states},\ }\href {https://doi.org/10.1103/PhysRevA.98.052302} {\bibfield  {journal} {\bibinfo  {journal} {Phys. Rev. A}\ }\textbf {\bibinfo {volume} {98}},\ \bibinfo {pages} {052302} (\bibinfo {year} {2018})}\BibitemShut {NoStop}%
\bibitem [{\citenamefont {Lee}\ \emph {et~al.}(2021)\citenamefont {Lee}, \citenamefont {Park}, \citenamefont {Lee}, \citenamefont {Kim},\ and\ \citenamefont {Bang}}]{Bang2021}%
  \BibitemOpen
  \bibfield  {author} {\bibinfo {author} {\bibfnamefont {S.~M.}\ \bibnamefont {Lee}}, \bibinfo {author} {\bibfnamefont {H.~S.}\ \bibnamefont {Park}}, \bibinfo {author} {\bibfnamefont {J.}~\bibnamefont {Lee}}, \bibinfo {author} {\bibfnamefont {J.}~\bibnamefont {Kim}},\ and\ \bibinfo {author} {\bibfnamefont {J.}~\bibnamefont {Bang}},\ }\bibfield  {title} {\bibinfo {title} {Quantum state learning via single-shot measurements},\ }\href {https://doi.org/10.1103/PhysRevLett.126.170504} {\bibfield  {journal} {\bibinfo  {journal} {Phys. Rev. Lett.}\ }\textbf {\bibinfo {volume} {126}},\ \bibinfo {pages} {170504} (\bibinfo {year} {2021})}\BibitemShut {NoStop}%
\bibitem [{\citenamefont {Liu}\ \emph {et~al.}(2020)\citenamefont {Liu}, \citenamefont {Wang}, \citenamefont {Xue}, \citenamefont {Huang}, \citenamefont {Fu}, \citenamefont {Qiang}, \citenamefont {Xu}, \citenamefont {Huang}, \citenamefont {Deng}, \citenamefont {Guo}, \citenamefont {Yang},\ and\ \citenamefont {Wu}}]{Liu2020}%
  \BibitemOpen
  \bibfield  {author} {\bibinfo {author} {\bibfnamefont {Y.}~\bibnamefont {Liu}}, \bibinfo {author} {\bibfnamefont {D.}~\bibnamefont {Wang}}, \bibinfo {author} {\bibfnamefont {S.}~\bibnamefont {Xue}}, \bibinfo {author} {\bibfnamefont {A.}~\bibnamefont {Huang}}, \bibinfo {author} {\bibfnamefont {X.}~\bibnamefont {Fu}}, \bibinfo {author} {\bibfnamefont {X.}~\bibnamefont {Qiang}}, \bibinfo {author} {\bibfnamefont {P.}~\bibnamefont {Xu}}, \bibinfo {author} {\bibfnamefont {H.-L.}\ \bibnamefont {Huang}}, \bibinfo {author} {\bibfnamefont {M.}~\bibnamefont {Deng}}, \bibinfo {author} {\bibfnamefont {C.}~\bibnamefont {Guo}}, \bibinfo {author} {\bibfnamefont {X.}~\bibnamefont {Yang}},\ and\ \bibinfo {author} {\bibfnamefont {J.}~\bibnamefont {Wu}},\ }\bibfield  {title} {\bibinfo {title} {Variational quantum circuits for quantum state tomography},\ }\href {https://doi.org/10.1103/PhysRevA.101.052316} {\bibfield  {journal} {\bibinfo  {journal} {Phys. Rev. A}\ }\textbf {\bibinfo {volume} {101}},\ \bibinfo {pages} {052316}
  (\bibinfo {year} {2020})}\BibitemShut {NoStop}%
\bibitem [{\citenamefont {Xue}\ \emph {et~al.}(2022)\citenamefont {Xue}, \citenamefont {Liu}, \citenamefont {Wang}, \citenamefont {Zhu}, \citenamefont {Guo},\ and\ \citenamefont {Wu}}]{Xue2022}%
  \BibitemOpen
  \bibfield  {author} {\bibinfo {author} {\bibfnamefont {S.}~\bibnamefont {Xue}}, \bibinfo {author} {\bibfnamefont {Y.}~\bibnamefont {Liu}}, \bibinfo {author} {\bibfnamefont {Y.}~\bibnamefont {Wang}}, \bibinfo {author} {\bibfnamefont {P.}~\bibnamefont {Zhu}}, \bibinfo {author} {\bibfnamefont {C.}~\bibnamefont {Guo}},\ and\ \bibinfo {author} {\bibfnamefont {J.}~\bibnamefont {Wu}},\ }\bibfield  {title} {\bibinfo {title} {Variational quantum process tomography of unitaries},\ }\href {https://doi.org/10.1103/PhysRevA.105.032427} {\bibfield  {journal} {\bibinfo  {journal} {Phys. Rev. A}\ }\textbf {\bibinfo {volume} {105}},\ \bibinfo {pages} {032427} (\bibinfo {year} {2022})}\BibitemShut {NoStop}%
\bibitem [{\citenamefont {Innan}\ \emph {et~al.}(2024)\citenamefont {Innan}, \citenamefont {Siddiqui}, \citenamefont {Arora}, \citenamefont {Ghosh}, \citenamefont {Ko{\c{c}}ak}, \citenamefont {Paragas}, \citenamefont {Galib}, \citenamefont {Khan},\ and\ \citenamefont {Bennai}}]{Innan2024}%
  \BibitemOpen
  \bibfield  {author} {\bibinfo {author} {\bibfnamefont {N.}~\bibnamefont {Innan}}, \bibinfo {author} {\bibfnamefont {O.~I.}\ \bibnamefont {Siddiqui}}, \bibinfo {author} {\bibfnamefont {S.}~\bibnamefont {Arora}}, \bibinfo {author} {\bibfnamefont {T.}~\bibnamefont {Ghosh}}, \bibinfo {author} {\bibfnamefont {Y.~P.}\ \bibnamefont {Ko{\c{c}}ak}}, \bibinfo {author} {\bibfnamefont {D.}~\bibnamefont {Paragas}}, \bibinfo {author} {\bibfnamefont {A.~A.~O.}\ \bibnamefont {Galib}}, \bibinfo {author} {\bibfnamefont {M.~A.-Z.}\ \bibnamefont {Khan}},\ and\ \bibinfo {author} {\bibfnamefont {M.}~\bibnamefont {Bennai}},\ }\bibfield  {title} {\bibinfo {title} {Quantum state tomography using quantum machine learning},\ }\href {https://doi.org/10.1007/s42484-024-00162-3} {\bibfield  {journal} {\bibinfo  {journal} {Quantum Machine Intelligence}\ }\textbf {\bibinfo {volume} {6}},\ \bibinfo {pages} {28} (\bibinfo {year} {2024})}\BibitemShut {NoStop}%
\bibitem [{\citenamefont {Wan}\ \emph {et~al.}(2017)\citenamefont {Wan}, \citenamefont {Dahlsten}, \citenamefont {Kristj{\'a}nsson}, \citenamefont {Gardner},\ and\ \citenamefont {Kim}}]{Wan2017}%
  \BibitemOpen
  \bibfield  {author} {\bibinfo {author} {\bibfnamefont {K.~H.}\ \bibnamefont {Wan}}, \bibinfo {author} {\bibfnamefont {O.}~\bibnamefont {Dahlsten}}, \bibinfo {author} {\bibfnamefont {H.}~\bibnamefont {Kristj{\'a}nsson}}, \bibinfo {author} {\bibfnamefont {R.}~\bibnamefont {Gardner}},\ and\ \bibinfo {author} {\bibfnamefont {M.~S.}\ \bibnamefont {Kim}},\ }\bibfield  {title} {\bibinfo {title} {Quantum generalisation of feedforward neural networks},\ }\href {https://doi.org/10.1038/s41534-017-0032-4} {\bibfield  {journal} {\bibinfo  {journal} {npj Quantum Information}\ }\textbf {\bibinfo {volume} {3}},\ \bibinfo {pages} {36} (\bibinfo {year} {2017})}\BibitemShut {NoStop}%
\bibitem [{\citenamefont {Beer}\ \emph {et~al.}(2020)\citenamefont {Beer}, \citenamefont {Bondarenko}, \citenamefont {Farrelly}, \citenamefont {Osborne}, \citenamefont {Salzmann}, \citenamefont {Scheiermann},\ and\ \citenamefont {Wolf}}]{Beer2020}%
  \BibitemOpen
  \bibfield  {author} {\bibinfo {author} {\bibfnamefont {K.}~\bibnamefont {Beer}}, \bibinfo {author} {\bibfnamefont {D.}~\bibnamefont {Bondarenko}}, \bibinfo {author} {\bibfnamefont {T.}~\bibnamefont {Farrelly}}, \bibinfo {author} {\bibfnamefont {T.~J.}\ \bibnamefont {Osborne}}, \bibinfo {author} {\bibfnamefont {R.}~\bibnamefont {Salzmann}}, \bibinfo {author} {\bibfnamefont {D.}~\bibnamefont {Scheiermann}},\ and\ \bibinfo {author} {\bibfnamefont {R.}~\bibnamefont {Wolf}},\ }\bibfield  {title} {\bibinfo {title} {Training deep quantum neural networks},\ }\href {https://doi.org/10.1038/s41467-020-14454-2} {\bibfield  {journal} {\bibinfo  {journal} {Nature Communications}\ }\textbf {\bibinfo {volume} {11}},\ \bibinfo {pages} {808} (\bibinfo {year} {2020})}\BibitemShut {NoStop}%
\bibitem [{\citenamefont {Haghshenas}\ \emph {et~al.}(2022)\citenamefont {Haghshenas}, \citenamefont {Gray}, \citenamefont {Potter},\ and\ \citenamefont {Chan}}]{Haghshenas2022}%
  \BibitemOpen
  \bibfield  {author} {\bibinfo {author} {\bibfnamefont {R.}~\bibnamefont {Haghshenas}}, \bibinfo {author} {\bibfnamefont {J.}~\bibnamefont {Gray}}, \bibinfo {author} {\bibfnamefont {A.~C.}\ \bibnamefont {Potter}},\ and\ \bibinfo {author} {\bibfnamefont {G.~K.-L.}\ \bibnamefont {Chan}},\ }\bibfield  {title} {\bibinfo {title} {Variational power of quantum circuit tensor networks},\ }\href {https://doi.org/10.1103/PhysRevX.12.011047} {\bibfield  {journal} {\bibinfo  {journal} {Phys. Rev. X}\ }\textbf {\bibinfo {volume} {12}},\ \bibinfo {pages} {011047} (\bibinfo {year} {2022})}\BibitemShut {NoStop}%
\bibitem [{\citenamefont {Abbas}\ \emph {et~al.}(2021)\citenamefont {Abbas}, \citenamefont {Sutter}, \citenamefont {Zoufal}, \citenamefont {Lucchi}, \citenamefont {Figalli},\ and\ \citenamefont {Woerner}}]{Abbas2021}%
  \BibitemOpen
  \bibfield  {author} {\bibinfo {author} {\bibfnamefont {A.}~\bibnamefont {Abbas}}, \bibinfo {author} {\bibfnamefont {D.}~\bibnamefont {Sutter}}, \bibinfo {author} {\bibfnamefont {C.}~\bibnamefont {Zoufal}}, \bibinfo {author} {\bibfnamefont {A.}~\bibnamefont {Lucchi}}, \bibinfo {author} {\bibfnamefont {A.}~\bibnamefont {Figalli}},\ and\ \bibinfo {author} {\bibfnamefont {S.}~\bibnamefont {Woerner}},\ }\bibfield  {title} {\bibinfo {title} {The power of quantum neural networks},\ }\href {https://doi.org/10.1038/s43588-021-00084-1} {\bibfield  {journal} {\bibinfo  {journal} {Nature Computational Science}\ }\textbf {\bibinfo {volume} {1}},\ \bibinfo {pages} {403} (\bibinfo {year} {2021})}\BibitemShut {NoStop}%
\bibitem [{\citenamefont {Farhi}\ \emph {et~al.}(2014)\citenamefont {Farhi}, \citenamefont {Goldstone},\ and\ \citenamefont {Gutmann}}]{farhi2014}%
  \BibitemOpen
  \bibfield  {author} {\bibinfo {author} {\bibfnamefont {E.}~\bibnamefont {Farhi}}, \bibinfo {author} {\bibfnamefont {J.}~\bibnamefont {Goldstone}},\ and\ \bibinfo {author} {\bibfnamefont {S.}~\bibnamefont {Gutmann}},\ }\bibfield  {title} {\bibinfo {title} {A quantum approximate optimization algorithm},\ }\href@noop {} {\bibfield  {journal} {\bibinfo  {journal} {arXiv preprint arXiv:1411.4028}\ } (\bibinfo {year} {2014})}\BibitemShut {NoStop}%
\bibitem [{\citenamefont {Verdon}\ \emph {et~al.}(2019)\citenamefont {Verdon}, \citenamefont {Broughton}, \citenamefont {McClean}, \citenamefont {Sung}, \citenamefont {Babbush}, \citenamefont {Jiang}, \citenamefont {Neven},\ and\ \citenamefont {Mohseni}}]{verdon2019}%
  \BibitemOpen
  \bibfield  {author} {\bibinfo {author} {\bibfnamefont {G.}~\bibnamefont {Verdon}}, \bibinfo {author} {\bibfnamefont {M.}~\bibnamefont {Broughton}}, \bibinfo {author} {\bibfnamefont {J.~R.}\ \bibnamefont {McClean}}, \bibinfo {author} {\bibfnamefont {K.~J.}\ \bibnamefont {Sung}}, \bibinfo {author} {\bibfnamefont {R.}~\bibnamefont {Babbush}}, \bibinfo {author} {\bibfnamefont {Z.}~\bibnamefont {Jiang}}, \bibinfo {author} {\bibfnamefont {H.}~\bibnamefont {Neven}},\ and\ \bibinfo {author} {\bibfnamefont {M.}~\bibnamefont {Mohseni}},\ }\href {https://arxiv.org/abs/1907.05415} {\bibinfo {title} {Learning to learn with quantum neural networks via classical neural networks}} (\bibinfo {year} {2019}),\ \Eprint {https://arxiv.org/abs/1907.05415} {arXiv:1907.05415 [quant-ph]} \BibitemShut {NoStop}%
\bibitem [{\citenamefont {Wilson}\ \emph {et~al.}(2021)\citenamefont {Wilson}, \citenamefont {Stromswold}, \citenamefont {Wudarski}, \citenamefont {Hadfield}, \citenamefont {Tubman},\ and\ \citenamefont {Rieffel}}]{Wilson2021}%
  \BibitemOpen
  \bibfield  {author} {\bibinfo {author} {\bibfnamefont {M.}~\bibnamefont {Wilson}}, \bibinfo {author} {\bibfnamefont {R.}~\bibnamefont {Stromswold}}, \bibinfo {author} {\bibfnamefont {F.}~\bibnamefont {Wudarski}}, \bibinfo {author} {\bibfnamefont {S.}~\bibnamefont {Hadfield}}, \bibinfo {author} {\bibfnamefont {N.~M.}\ \bibnamefont {Tubman}},\ and\ \bibinfo {author} {\bibfnamefont {E.~G.}\ \bibnamefont {Rieffel}},\ }\bibfield  {title} {\bibinfo {title} {Optimizing quantum heuristics with meta-learning},\ }\href {https://doi.org/10.1007/s42484-020-00022-w} {\bibfield  {journal} {\bibinfo  {journal} {Quantum Machine Intelligence}\ }\textbf {\bibinfo {volume} {3}},\ \bibinfo {pages} {13} (\bibinfo {year} {2021})}\BibitemShut {NoStop}%
\bibitem [{\citenamefont {Khairy}\ \emph {et~al.}(2020)\citenamefont {Khairy}, \citenamefont {Shaydulin}, \citenamefont {Cincio}, \citenamefont {Alexeev},\ and\ \citenamefont {Balaprakash}}]{khairy2020}%
  \BibitemOpen
  \bibfield  {author} {\bibinfo {author} {\bibfnamefont {S.}~\bibnamefont {Khairy}}, \bibinfo {author} {\bibfnamefont {R.}~\bibnamefont {Shaydulin}}, \bibinfo {author} {\bibfnamefont {L.}~\bibnamefont {Cincio}}, \bibinfo {author} {\bibfnamefont {Y.}~\bibnamefont {Alexeev}},\ and\ \bibinfo {author} {\bibfnamefont {P.}~\bibnamefont {Balaprakash}},\ }\bibfield  {title} {\bibinfo {title} {Learning to optimize variational quantum circuits to solve combinatorial problems},\ }in\ \href@noop {} {\emph {\bibinfo {booktitle} {Proceedings of the AAAI conference on artificial intelligence}}},\ Vol.~\bibinfo {volume} {34}\ (\bibinfo {year} {2020})\ pp.\ \bibinfo {pages} {2367--2375}\BibitemShut {NoStop}%
\bibitem [{\citenamefont {Li}\ and\ \citenamefont {Malik}(2017)}]{li2017learning}%
  \BibitemOpen
  \bibfield  {author} {\bibinfo {author} {\bibfnamefont {K.}~\bibnamefont {Li}}\ and\ \bibinfo {author} {\bibfnamefont {J.}~\bibnamefont {Malik}},\ }\bibfield  {title} {\bibinfo {title} {Learning to optimize},\ }in\ \href {https://openreview.net/forum?id=ry4Vrt5gl} {\emph {\bibinfo {booktitle} {International Conference on Learning Representations}}}\ (\bibinfo {year} {2017})\BibitemShut {NoStop}%
\bibitem [{\citenamefont {Chen}\ \emph {et~al.}(2017{\natexlab{a}})\citenamefont {Chen}, \citenamefont {Hoffman}, \citenamefont {Colmenarejo}, \citenamefont {Denil}, \citenamefont {Lillicrap}, \citenamefont {Botvinick},\ and\ \citenamefont {Freitas}}]{chen2017learning}%
  \BibitemOpen
  \bibfield  {author} {\bibinfo {author} {\bibfnamefont {Y.}~\bibnamefont {Chen}}, \bibinfo {author} {\bibfnamefont {M.~W.}\ \bibnamefont {Hoffman}}, \bibinfo {author} {\bibfnamefont {S.~G.}\ \bibnamefont {Colmenarejo}}, \bibinfo {author} {\bibfnamefont {M.}~\bibnamefont {Denil}}, \bibinfo {author} {\bibfnamefont {T.~P.}\ \bibnamefont {Lillicrap}}, \bibinfo {author} {\bibfnamefont {M.}~\bibnamefont {Botvinick}},\ and\ \bibinfo {author} {\bibfnamefont {N.}~\bibnamefont {Freitas}},\ }\bibfield  {title} {\bibinfo {title} {Learning to learn without gradient descent by gradient descent},\ }in\ \href@noop {} {\emph {\bibinfo {booktitle} {International Conference on Machine Learning}}}\ (\bibinfo {organization} {PMLR},\ \bibinfo {year} {2017})\ pp.\ \bibinfo {pages} {748--756}\BibitemShut {NoStop}%
\bibitem [{\citenamefont {Smith}\ \emph {et~al.}(2021)\citenamefont {Smith}, \citenamefont {Gray},\ and\ \citenamefont {Kim}}]{Smith2021}%
  \BibitemOpen
  \bibfield  {author} {\bibinfo {author} {\bibfnamefont {A.~W.~R.}\ \bibnamefont {Smith}}, \bibinfo {author} {\bibfnamefont {J.}~\bibnamefont {Gray}},\ and\ \bibinfo {author} {\bibfnamefont {M.~S.}\ \bibnamefont {Kim}},\ }\bibfield  {title} {\bibinfo {title} {Efficient quantum state sample tomography with basis-dependent neural networks},\ }\href {https://doi.org/10.1103/PRXQuantum.2.020348} {\bibfield  {journal} {\bibinfo  {journal} {PRX Quantum}\ }\textbf {\bibinfo {volume} {2}},\ \bibinfo {pages} {020348} (\bibinfo {year} {2021})}\BibitemShut {NoStop}%
\bibitem [{\citenamefont {Quek}\ \emph {et~al.}(2021)\citenamefont {Quek}, \citenamefont {Fort},\ and\ \citenamefont {Ng}}]{Quek2021}%
  \BibitemOpen
  \bibfield  {author} {\bibinfo {author} {\bibfnamefont {Y.}~\bibnamefont {Quek}}, \bibinfo {author} {\bibfnamefont {S.}~\bibnamefont {Fort}},\ and\ \bibinfo {author} {\bibfnamefont {H.~K.}\ \bibnamefont {Ng}},\ }\bibfield  {title} {\bibinfo {title} {Adaptive quantum state tomography with neural networks},\ }\href {https://doi.org/10.1038/s41534-021-00436-9} {\bibfield  {journal} {\bibinfo  {journal} {npj Quantum Information}\ }\textbf {\bibinfo {volume} {7}},\ \bibinfo {pages} {105} (\bibinfo {year} {2021})}\BibitemShut {NoStop}%
\bibitem [{\citenamefont {Kandala}\ \emph {et~al.}(2017)\citenamefont {Kandala}, \citenamefont {Mezzacapo}, \citenamefont {Temme}, \citenamefont {Takita}, \citenamefont {Brink}, \citenamefont {Chow},\ and\ \citenamefont {Gambetta}}]{Kandala2017}%
  \BibitemOpen
  \bibfield  {author} {\bibinfo {author} {\bibfnamefont {A.}~\bibnamefont {Kandala}}, \bibinfo {author} {\bibfnamefont {A.}~\bibnamefont {Mezzacapo}}, \bibinfo {author} {\bibfnamefont {K.}~\bibnamefont {Temme}}, \bibinfo {author} {\bibfnamefont {M.}~\bibnamefont {Takita}}, \bibinfo {author} {\bibfnamefont {M.}~\bibnamefont {Brink}}, \bibinfo {author} {\bibfnamefont {J.~M.}\ \bibnamefont {Chow}},\ and\ \bibinfo {author} {\bibfnamefont {J.~M.}\ \bibnamefont {Gambetta}},\ }\bibfield  {title} {\bibinfo {title} {Hardware-efficient variational quantum eigensolver for small molecules and quantum magnets},\ }\href {https://doi.org/10.1038/nature23879} {\bibfield  {journal} {\bibinfo  {journal} {Nature}\ }\textbf {\bibinfo {volume} {549}},\ \bibinfo {pages} {242} (\bibinfo {year} {2017})}\BibitemShut {NoStop}%
\bibitem [{\citenamefont {Salimans}\ \emph {et~al.}(2017)\citenamefont {Salimans}, \citenamefont {Ho}, \citenamefont {Chen}, \citenamefont {Sidor},\ and\ \citenamefont {Sutskever}}]{salimans2017}%
  \BibitemOpen
  \bibfield  {author} {\bibinfo {author} {\bibfnamefont {T.}~\bibnamefont {Salimans}}, \bibinfo {author} {\bibfnamefont {J.}~\bibnamefont {Ho}}, \bibinfo {author} {\bibfnamefont {X.}~\bibnamefont {Chen}}, \bibinfo {author} {\bibfnamefont {S.}~\bibnamefont {Sidor}},\ and\ \bibinfo {author} {\bibfnamefont {I.}~\bibnamefont {Sutskever}},\ }\bibfield  {title} {\bibinfo {title} {Evolution strategies as a scalable alternative to reinforcement learning},\ }\href@noop {} {\bibfield  {journal} {\bibinfo  {journal} {arXiv preprint arXiv:1703.03864}\ } (\bibinfo {year} {2017})}\BibitemShut {NoStop}%
\bibitem [{\citenamefont {Narvekar}\ \emph {et~al.}(2020)\citenamefont {Narvekar}, \citenamefont {Peng}, \citenamefont {Leonetti}, \citenamefont {Sinapov}, \citenamefont {Taylor},\ and\ \citenamefont {Stone}}]{Narvekar2020}%
  \BibitemOpen
  \bibfield  {author} {\bibinfo {author} {\bibfnamefont {S.}~\bibnamefont {Narvekar}}, \bibinfo {author} {\bibfnamefont {B.}~\bibnamefont {Peng}}, \bibinfo {author} {\bibfnamefont {M.}~\bibnamefont {Leonetti}}, \bibinfo {author} {\bibfnamefont {J.}~\bibnamefont {Sinapov}}, \bibinfo {author} {\bibfnamefont {M.~E.}\ \bibnamefont {Taylor}},\ and\ \bibinfo {author} {\bibfnamefont {P.}~\bibnamefont {Stone}},\ }\bibfield  {title} {\bibinfo {title} {Curriculum learning for reinforcement learning domains: a framework and survey},\ }\href@noop {} {\bibfield  {journal} {\bibinfo  {journal} {J. Mach. Learn. Res.}\ }\textbf {\bibinfo {volume} {21}} (\bibinfo {year} {2020})}\BibitemShut {NoStop}%
\bibitem [{\citenamefont {Ostaszewski}\ \emph {et~al.}(2021)\citenamefont {Ostaszewski}, \citenamefont {Trenkwalder}, \citenamefont {Masarczyk}, \citenamefont {Scerri},\ and\ \citenamefont {Dunjko}}]{ostaszewski2021rl}%
  \BibitemOpen
  \bibfield  {author} {\bibinfo {author} {\bibfnamefont {M.}~\bibnamefont {Ostaszewski}}, \bibinfo {author} {\bibfnamefont {L.~M.}\ \bibnamefont {Trenkwalder}}, \bibinfo {author} {\bibfnamefont {W.}~\bibnamefont {Masarczyk}}, \bibinfo {author} {\bibfnamefont {E.}~\bibnamefont {Scerri}},\ and\ \bibinfo {author} {\bibfnamefont {V.}~\bibnamefont {Dunjko}},\ }\bibfield  {title} {\bibinfo {title} {Reinforcement learning for optimization of variational quantum circuit architectures},\ }\href@noop {} {\bibfield  {journal} {\bibinfo  {journal} {Advances in Neural Information Processing Systems}\ }\textbf {\bibinfo {volume} {34}},\ \bibinfo {pages} {18182} (\bibinfo {year} {2021})}\BibitemShut {NoStop}%
\bibitem [{\citenamefont {Patel}\ \emph {et~al.}(2024)\citenamefont {Patel}, \citenamefont {Kundu}, \citenamefont {Ostaszewski}, \citenamefont {Bonet-Monroig}, \citenamefont {Dunjko},\ and\ \citenamefont {Danaci}}]{patel2024curriculum}%
  \BibitemOpen
  \bibfield  {author} {\bibinfo {author} {\bibfnamefont {Y.~J.}\ \bibnamefont {Patel}}, \bibinfo {author} {\bibfnamefont {A.}~\bibnamefont {Kundu}}, \bibinfo {author} {\bibfnamefont {M.}~\bibnamefont {Ostaszewski}}, \bibinfo {author} {\bibfnamefont {X.}~\bibnamefont {Bonet-Monroig}}, \bibinfo {author} {\bibfnamefont {V.}~\bibnamefont {Dunjko}},\ and\ \bibinfo {author} {\bibfnamefont {O.}~\bibnamefont {Danaci}},\ }\bibfield  {title} {\bibinfo {title} {Curriculum reinforcement learning for quantum architecture search under hardware errors},\ }in\ \href {https://openreview.net/forum?id=rINBD8jPoP} {\emph {\bibinfo {booktitle} {The Twelfth International Conference on Learning Representations}}}\ (\bibinfo {year} {2024})\BibitemShut {NoStop}%
\bibitem [{\citenamefont {Haah}\ \emph {et~al.}(2017)\citenamefont {Haah}, \citenamefont {Harrow}, \citenamefont {Ji}, \citenamefont {Wu},\ and\ \citenamefont {Yu}}]{Haah2017}%
  \BibitemOpen
  \bibfield  {author} {\bibinfo {author} {\bibfnamefont {J.}~\bibnamefont {Haah}}, \bibinfo {author} {\bibfnamefont {A.~W.}\ \bibnamefont {Harrow}}, \bibinfo {author} {\bibfnamefont {Z.}~\bibnamefont {Ji}}, \bibinfo {author} {\bibfnamefont {X.}~\bibnamefont {Wu}},\ and\ \bibinfo {author} {\bibfnamefont {N.}~\bibnamefont {Yu}},\ }\bibfield  {title} {\bibinfo {title} {Sample-optimal tomography of quantum states},\ }\href {https://doi.org/10.1109/TIT.2017.2719044} {\bibfield  {journal} {\bibinfo  {journal} {IEEE Transactions on Information Theory}\ }\textbf {\bibinfo {volume} {63}},\ \bibinfo {pages} {5628} (\bibinfo {year} {2017})}\BibitemShut {NoStop}%
\bibitem [{\citenamefont {Sutton}\ and\ \citenamefont {Barto}(2018)}]{sutton2018}%
  \BibitemOpen
  \bibfield  {author} {\bibinfo {author} {\bibfnamefont {R.~S.}\ \bibnamefont {Sutton}}\ and\ \bibinfo {author} {\bibfnamefont {A.~G.}\ \bibnamefont {Barto}},\ }\href@noop {} {\emph {\bibinfo {title} {Reinforcement learning: An introduction}}}\ (\bibinfo  {publisher} {MIT press},\ \bibinfo {year} {2018})\BibitemShut {NoStop}%
\bibitem [{\citenamefont {Bellman}(1954)}]{bellman1954theory}%
  \BibitemOpen
  \bibfield  {author} {\bibinfo {author} {\bibfnamefont {R.}~\bibnamefont {Bellman}},\ }\bibfield  {title} {\bibinfo {title} {The theory of dynamic programming},\ }\href@noop {} {\bibfield  {journal} {\bibinfo  {journal} {Bulletin of the American Mathematical Society}\ }\textbf {\bibinfo {volume} {60}},\ \bibinfo {pages} {503} (\bibinfo {year} {1954})}\BibitemShut {NoStop}%
\bibitem [{\citenamefont {{\AA}str{\"o}m}(1969)}]{aastrom1969optimal}%
  \BibitemOpen
  \bibfield  {author} {\bibinfo {author} {\bibfnamefont {K.~J.}\ \bibnamefont {{\AA}str{\"o}m}},\ }\bibfield  {title} {\bibinfo {title} {Optimal control of markov processes with incomplete state information ii: The convexity of the lossfunction},\ }\href@noop {} {\bibfield  {journal} {\bibinfo  {journal} {Journal of Mathematical Analysis and Applications}\ }\textbf {\bibinfo {volume} {26}},\ \bibinfo {pages} {403} (\bibinfo {year} {1969})}\BibitemShut {NoStop}%
\bibitem [{\citenamefont {Hausknecht}\ and\ \citenamefont {Stone}(2015)}]{hausknecht2015deep}%
  \BibitemOpen
  \bibfield  {author} {\bibinfo {author} {\bibfnamefont {M.}~\bibnamefont {Hausknecht}}\ and\ \bibinfo {author} {\bibfnamefont {P.}~\bibnamefont {Stone}},\ }\bibfield  {title} {\bibinfo {title} {Deep recurrent q-learning for partially observable mdps},\ }in\ \href@noop {} {\emph {\bibinfo {booktitle} {2015 aaai fall symposium series}}}\ (\bibinfo {year} {2015})\BibitemShut {NoStop}%
\bibitem [{\citenamefont {Barry}\ \emph {et~al.}(2014)\citenamefont {Barry}, \citenamefont {Barry},\ and\ \citenamefont {Aaronson}}]{Barry2014}%
  \BibitemOpen
  \bibfield  {author} {\bibinfo {author} {\bibfnamefont {J.}~\bibnamefont {Barry}}, \bibinfo {author} {\bibfnamefont {D.~T.}\ \bibnamefont {Barry}},\ and\ \bibinfo {author} {\bibfnamefont {S.}~\bibnamefont {Aaronson}},\ }\bibfield  {title} {\bibinfo {title} {Quantum partially observable markov decision processes},\ }\href {https://doi.org/10.1103/PhysRevA.90.032311} {\bibfield  {journal} {\bibinfo  {journal} {Phys. Rev. A}\ }\textbf {\bibinfo {volume} {90}},\ \bibinfo {pages} {032311} (\bibinfo {year} {2014})}\BibitemShut {NoStop}%
\bibitem [{\citenamefont {Sivak}\ \emph {et~al.}(2022)\citenamefont {Sivak}, \citenamefont {Eickbusch}, \citenamefont {Liu}, \citenamefont {Royer}, \citenamefont {Tsioutsios},\ and\ \citenamefont {Devoret}}]{Sivak2022}%
  \BibitemOpen
  \bibfield  {author} {\bibinfo {author} {\bibfnamefont {V.~V.}\ \bibnamefont {Sivak}}, \bibinfo {author} {\bibfnamefont {A.}~\bibnamefont {Eickbusch}}, \bibinfo {author} {\bibfnamefont {H.}~\bibnamefont {Liu}}, \bibinfo {author} {\bibfnamefont {B.}~\bibnamefont {Royer}}, \bibinfo {author} {\bibfnamefont {I.}~\bibnamefont {Tsioutsios}},\ and\ \bibinfo {author} {\bibfnamefont {M.~H.}\ \bibnamefont {Devoret}},\ }\bibfield  {title} {\bibinfo {title} {Model-free quantum control with reinforcement learning},\ }\href {https://doi.org/10.1103/PhysRevX.12.011059} {\bibfield  {journal} {\bibinfo  {journal} {Phys. Rev. X}\ }\textbf {\bibinfo {volume} {12}},\ \bibinfo {pages} {011059} (\bibinfo {year} {2022})}\BibitemShut {NoStop}%
\bibitem [{\citenamefont {Bukov}\ \emph {et~al.}(2018)\citenamefont {Bukov}, \citenamefont {Day}, \citenamefont {Sels}, \citenamefont {Weinberg}, \citenamefont {Polkovnikov},\ and\ \citenamefont {Mehta}}]{Bukov2018}%
  \BibitemOpen
  \bibfield  {author} {\bibinfo {author} {\bibfnamefont {M.}~\bibnamefont {Bukov}}, \bibinfo {author} {\bibfnamefont {A.~G.~R.}\ \bibnamefont {Day}}, \bibinfo {author} {\bibfnamefont {D.}~\bibnamefont {Sels}}, \bibinfo {author} {\bibfnamefont {P.}~\bibnamefont {Weinberg}}, \bibinfo {author} {\bibfnamefont {A.}~\bibnamefont {Polkovnikov}},\ and\ \bibinfo {author} {\bibfnamefont {P.}~\bibnamefont {Mehta}},\ }\bibfield  {title} {\bibinfo {title} {Reinforcement learning in different phases of quantum control},\ }\href {https://doi.org/10.1103/PhysRevX.8.031086} {\bibfield  {journal} {\bibinfo  {journal} {Phys. Rev. X}\ }\textbf {\bibinfo {volume} {8}},\ \bibinfo {pages} {031086} (\bibinfo {year} {2018})}\BibitemShut {NoStop}%
\bibitem [{\citenamefont {F\"osel}\ \emph {et~al.}(2018)\citenamefont {F\"osel}, \citenamefont {Tighineanu}, \citenamefont {Weiss},\ and\ \citenamefont {Marquardt}}]{Thomas2018}%
  \BibitemOpen
  \bibfield  {author} {\bibinfo {author} {\bibfnamefont {T.}~\bibnamefont {F\"osel}}, \bibinfo {author} {\bibfnamefont {P.}~\bibnamefont {Tighineanu}}, \bibinfo {author} {\bibfnamefont {T.}~\bibnamefont {Weiss}},\ and\ \bibinfo {author} {\bibfnamefont {F.}~\bibnamefont {Marquardt}},\ }\bibfield  {title} {\bibinfo {title} {Reinforcement learning with neural networks for quantum feedback},\ }\href {https://doi.org/10.1103/PhysRevX.8.031084} {\bibfield  {journal} {\bibinfo  {journal} {Phys. Rev. X}\ }\textbf {\bibinfo {volume} {8}},\ \bibinfo {pages} {031084} (\bibinfo {year} {2018})}\BibitemShut {NoStop}%
\bibitem [{\citenamefont {Niu}\ \emph {et~al.}(2019)\citenamefont {Niu}, \citenamefont {Boixo}, \citenamefont {Smelyanskiy},\ and\ \citenamefont {Neven}}]{Niu2019}%
  \BibitemOpen
  \bibfield  {author} {\bibinfo {author} {\bibfnamefont {M.~Y.}\ \bibnamefont {Niu}}, \bibinfo {author} {\bibfnamefont {S.}~\bibnamefont {Boixo}}, \bibinfo {author} {\bibfnamefont {V.~N.}\ \bibnamefont {Smelyanskiy}},\ and\ \bibinfo {author} {\bibfnamefont {H.}~\bibnamefont {Neven}},\ }\bibfield  {title} {\bibinfo {title} {Universal quantum control through deep reinforcement learning},\ }\href {https://doi.org/10.1038/s41534-019-0141-3} {\bibfield  {journal} {\bibinfo  {journal} {npj Quantum Information}\ }\textbf {\bibinfo {volume} {5}},\ \bibinfo {pages} {33} (\bibinfo {year} {2019})}\BibitemShut {NoStop}%
\bibitem [{\citenamefont {Wang}\ \emph {et~al.}(2020)\citenamefont {Wang}, \citenamefont {Ashida},\ and\ \citenamefont {Ueda}}]{Wang2020}%
  \BibitemOpen
  \bibfield  {author} {\bibinfo {author} {\bibfnamefont {Z.~T.}\ \bibnamefont {Wang}}, \bibinfo {author} {\bibfnamefont {Y.}~\bibnamefont {Ashida}},\ and\ \bibinfo {author} {\bibfnamefont {M.}~\bibnamefont {Ueda}},\ }\bibfield  {title} {\bibinfo {title} {Deep reinforcement learning control of quantum cartpoles},\ }\href {https://doi.org/10.1103/PhysRevLett.125.100401} {\bibfield  {journal} {\bibinfo  {journal} {Phys. Rev. Lett.}\ }\textbf {\bibinfo {volume} {125}},\ \bibinfo {pages} {100401} (\bibinfo {year} {2020})}\BibitemShut {NoStop}%
\bibitem [{\citenamefont {Porotti}\ \emph {et~al.}(2022)\citenamefont {Porotti}, \citenamefont {Essig}, \citenamefont {Huard},\ and\ \citenamefont {Marquardt}}]{Porotti2022}%
  \BibitemOpen
  \bibfield  {author} {\bibinfo {author} {\bibfnamefont {R.}~\bibnamefont {Porotti}}, \bibinfo {author} {\bibfnamefont {A.}~\bibnamefont {Essig}}, \bibinfo {author} {\bibfnamefont {B.}~\bibnamefont {Huard}},\ and\ \bibinfo {author} {\bibfnamefont {F.}~\bibnamefont {Marquardt}},\ }\bibfield  {title} {\bibinfo {title} {Deep {R}einforcement {L}earning for {Q}uantum {S}tate {P}reparation with {W}eak {N}onlinear {M}easurements},\ }\href {https://doi.org/10.22331/q-2022-06-28-747} {\bibfield  {journal} {\bibinfo  {journal} {{Quantum}}\ }\textbf {\bibinfo {volume} {6}},\ \bibinfo {pages} {747} (\bibinfo {year} {2022})}\BibitemShut {NoStop}%
\bibitem [{\citenamefont {Metz}\ and\ \citenamefont {Bukov}(2023)}]{Metz2023}%
  \BibitemOpen
  \bibfield  {author} {\bibinfo {author} {\bibfnamefont {F.}~\bibnamefont {Metz}}\ and\ \bibinfo {author} {\bibfnamefont {M.}~\bibnamefont {Bukov}},\ }\bibfield  {title} {\bibinfo {title} {Self-correcting quantum many-body control using reinforcement learning with tensor networks},\ }\href {https://doi.org/10.1038/s42256-023-00687-5} {\bibfield  {journal} {\bibinfo  {journal} {Nature Machine Intelligence}\ }\textbf {\bibinfo {volume} {5}},\ \bibinfo {pages} {780} (\bibinfo {year} {2023})}\BibitemShut {NoStop}%
\bibitem [{\citenamefont {Vent}(1975)}]{Rechenberg1975}%
  \BibitemOpen
  \bibfield  {author} {\bibinfo {author} {\bibfnamefont {W.}~\bibnamefont {Vent}},\ }\bibfield  {title} {\bibinfo {title} {Rechenberg, ingo, evolutionsstrategie — optimierung technischer systeme nach prinzipien der biologischen evolution. 170 s. mit 36 abb. frommann-holzboog-verlag. stuttgart 1973. broschiert},\ }\href {https://doi.org/https://doi.org/10.1002/fedr.19750860506} {\bibfield  {journal} {\bibinfo  {journal} {Feddes Repertorium}\ }\textbf {\bibinfo {volume} {86}},\ \bibinfo {pages} {337} (\bibinfo {year} {1975})}\BibitemShut {NoStop}%
\bibitem [{\citenamefont {Hansen}(2016)}]{hansen2016cma}%
  \BibitemOpen
  \bibfield  {author} {\bibinfo {author} {\bibfnamefont {N.}~\bibnamefont {Hansen}},\ }\bibfield  {title} {\bibinfo {title} {The cma evolution strategy: A tutorial},\ }\href@noop {} {\bibfield  {journal} {\bibinfo  {journal} {arXiv preprint arXiv:1604.00772}\ } (\bibinfo {year} {2016})}\BibitemShut {NoStop}%
\bibitem [{\citenamefont {Mitarai}\ \emph {et~al.}(2018)\citenamefont {Mitarai}, \citenamefont {Negoro}, \citenamefont {Kitagawa},\ and\ \citenamefont {Fujii}}]{Mitarai2018}%
  \BibitemOpen
  \bibfield  {author} {\bibinfo {author} {\bibfnamefont {K.}~\bibnamefont {Mitarai}}, \bibinfo {author} {\bibfnamefont {M.}~\bibnamefont {Negoro}}, \bibinfo {author} {\bibfnamefont {M.}~\bibnamefont {Kitagawa}},\ and\ \bibinfo {author} {\bibfnamefont {K.}~\bibnamefont {Fujii}},\ }\bibfield  {title} {\bibinfo {title} {Quantum circuit learning},\ }\href {https://doi.org/10.1103/PhysRevA.98.032309} {\bibfield  {journal} {\bibinfo  {journal} {Phys. Rev. A}\ }\textbf {\bibinfo {volume} {98}},\ \bibinfo {pages} {032309} (\bibinfo {year} {2018})}\BibitemShut {NoStop}%
\bibitem [{\citenamefont {Schuld}\ \emph {et~al.}(2019)\citenamefont {Schuld}, \citenamefont {Bergholm}, \citenamefont {Gogolin}, \citenamefont {Izaac},\ and\ \citenamefont {Killoran}}]{Schuld2019grad}%
  \BibitemOpen
  \bibfield  {author} {\bibinfo {author} {\bibfnamefont {M.}~\bibnamefont {Schuld}}, \bibinfo {author} {\bibfnamefont {V.}~\bibnamefont {Bergholm}}, \bibinfo {author} {\bibfnamefont {C.}~\bibnamefont {Gogolin}}, \bibinfo {author} {\bibfnamefont {J.}~\bibnamefont {Izaac}},\ and\ \bibinfo {author} {\bibfnamefont {N.}~\bibnamefont {Killoran}},\ }\bibfield  {title} {\bibinfo {title} {Evaluating analytic gradients on quantum hardware},\ }\href {https://doi.org/10.1103/PhysRevA.99.032331} {\bibfield  {journal} {\bibinfo  {journal} {Phys. Rev. A}\ }\textbf {\bibinfo {volume} {99}},\ \bibinfo {pages} {032331} (\bibinfo {year} {2019})}\BibitemShut {NoStop}%
\bibitem [{\citenamefont {McClean}\ \emph {et~al.}(2018)\citenamefont {McClean}, \citenamefont {Boixo}, \citenamefont {Smelyanskiy}, \citenamefont {Babbush},\ and\ \citenamefont {Neven}}]{McClean2018}%
  \BibitemOpen
  \bibfield  {author} {\bibinfo {author} {\bibfnamefont {J.~R.}\ \bibnamefont {McClean}}, \bibinfo {author} {\bibfnamefont {S.}~\bibnamefont {Boixo}}, \bibinfo {author} {\bibfnamefont {V.~N.}\ \bibnamefont {Smelyanskiy}}, \bibinfo {author} {\bibfnamefont {R.}~\bibnamefont {Babbush}},\ and\ \bibinfo {author} {\bibfnamefont {H.}~\bibnamefont {Neven}},\ }\bibfield  {title} {\bibinfo {title} {Barren plateaus in quantum neural network training landscapes},\ }\href {https://doi.org/10.1038/s41467-018-07090-4} {\bibfield  {journal} {\bibinfo  {journal} {Nature Communications}\ }\textbf {\bibinfo {volume} {9}},\ \bibinfo {pages} {4812} (\bibinfo {year} {2018})}\BibitemShut {NoStop}%
\bibitem [{\citenamefont {Cerezo}\ and\ \citenamefont {Coles}(2021)}]{Cerezo2021}%
  \BibitemOpen
  \bibfield  {author} {\bibinfo {author} {\bibfnamefont {M.}~\bibnamefont {Cerezo}}\ and\ \bibinfo {author} {\bibfnamefont {P.~J.}\ \bibnamefont {Coles}},\ }\bibfield  {title} {\bibinfo {title} {Higher order derivatives of quantum neural networks with barren plateaus},\ }\href {https://doi.org/10.1088/2058-9565/abf51a} {\bibfield  {journal} {\bibinfo  {journal} {Quantum Science and Technology}\ }\textbf {\bibinfo {volume} {6}},\ \bibinfo {pages} {035006} (\bibinfo {year} {2021})}\BibitemShut {NoStop}%
\bibitem [{\citenamefont {Arrasmith}\ \emph {et~al.}(2021)\citenamefont {Arrasmith}, \citenamefont {Cerezo}, \citenamefont {Czarnik}, \citenamefont {Cincio},\ and\ \citenamefont {Coles}}]{Arrasmith2021}%
  \BibitemOpen
  \bibfield  {author} {\bibinfo {author} {\bibfnamefont {A.}~\bibnamefont {Arrasmith}}, \bibinfo {author} {\bibfnamefont {M.}~\bibnamefont {Cerezo}}, \bibinfo {author} {\bibfnamefont {P.}~\bibnamefont {Czarnik}}, \bibinfo {author} {\bibfnamefont {L.}~\bibnamefont {Cincio}},\ and\ \bibinfo {author} {\bibfnamefont {P.~J.}\ \bibnamefont {Coles}},\ }\bibfield  {title} {\bibinfo {title} {Effect of barren plateaus on gradient-free optimization},\ }\href {https://doi.org/10.22331/q-2021-10-05-558} {\bibfield  {journal} {\bibinfo  {journal} {{Quantum}}\ }\textbf {\bibinfo {volume} {5}},\ \bibinfo {pages} {558} (\bibinfo {year} {2021})}\BibitemShut {NoStop}%
\bibitem [{\citenamefont {Anand}\ \emph {et~al.}(2021)\citenamefont {Anand}, \citenamefont {Degroote},\ and\ \citenamefont {Aspuru-Guzik}}]{Anand2021}%
  \BibitemOpen
  \bibfield  {author} {\bibinfo {author} {\bibfnamefont {A.}~\bibnamefont {Anand}}, \bibinfo {author} {\bibfnamefont {M.}~\bibnamefont {Degroote}},\ and\ \bibinfo {author} {\bibfnamefont {A.}~\bibnamefont {Aspuru-Guzik}},\ }\bibfield  {title} {\bibinfo {title} {Natural evolutionary strategies for variational quantum computation},\ }\href {https://doi.org/10.1088/2632-2153/abf3ac} {\bibfield  {journal} {\bibinfo  {journal} {Machine Learning: Science and Technology}\ }\textbf {\bibinfo {volume} {2}},\ \bibinfo {pages} {045012} (\bibinfo {year} {2021})}\BibitemShut {NoStop}%
\bibitem [{\citenamefont {Xie}\ \emph {et~al.}(2023)\citenamefont {Xie}, \citenamefont {Xu}, \citenamefont {Yin}, \citenamefont {Dong},\ and\ \citenamefont {Zhang}}]{Jianshe2023}%
  \BibitemOpen
  \bibfield  {author} {\bibinfo {author} {\bibfnamefont {J.}~\bibnamefont {Xie}}, \bibinfo {author} {\bibfnamefont {C.}~\bibnamefont {Xu}}, \bibinfo {author} {\bibfnamefont {C.}~\bibnamefont {Yin}}, \bibinfo {author} {\bibfnamefont {Y.}~\bibnamefont {Dong}},\ and\ \bibinfo {author} {\bibfnamefont {Z.}~\bibnamefont {Zhang}},\ }\bibfield  {title} {\bibinfo {title} {Natural evolutionary gradient descent strategy for variational quantum algorithms},\ }\href {https://doi.org/10.34133/icomputing.0042} {\bibfield  {journal} {\bibinfo  {journal} {Intelligent Computing}\ }\textbf {\bibinfo {volume} {2}},\ \bibinfo {pages} {0042} (\bibinfo {year} {2023})}\BibitemShut {NoStop}%
\bibitem [{\citenamefont {Mezzadri}(2007)}]{mezzadri2007}%
  \BibitemOpen
  \bibfield  {author} {\bibinfo {author} {\bibfnamefont {F.}~\bibnamefont {Mezzadri}},\ }\href {https://arxiv.org/abs/math-ph/0609050} {\bibinfo {title} {How to generate random matrices from the classical compact groups}} (\bibinfo {year} {2007}),\ \Eprint {https://arxiv.org/abs/math-ph/0609050} {arXiv:math-ph/0609050 [math-ph]} \BibitemShut {NoStop}%
\bibitem [{\citenamefont {Paszke}\ \emph {et~al.}(2019)\citenamefont {Paszke}, \citenamefont {Gross}, \citenamefont {Massa}, \citenamefont {Lerer}, \citenamefont {Bradbury}, \citenamefont {Chanan}, \citenamefont {Killeen}, \citenamefont {Lin}, \citenamefont {Gimelshein}, \citenamefont {Antiga} \emph {et~al.}}]{paszke2019pytorch}%
  \BibitemOpen
  \bibfield  {author} {\bibinfo {author} {\bibfnamefont {A.}~\bibnamefont {Paszke}}, \bibinfo {author} {\bibfnamefont {S.}~\bibnamefont {Gross}}, \bibinfo {author} {\bibfnamefont {F.}~\bibnamefont {Massa}}, \bibinfo {author} {\bibfnamefont {A.}~\bibnamefont {Lerer}}, \bibinfo {author} {\bibfnamefont {J.}~\bibnamefont {Bradbury}}, \bibinfo {author} {\bibfnamefont {G.}~\bibnamefont {Chanan}}, \bibinfo {author} {\bibfnamefont {T.}~\bibnamefont {Killeen}}, \bibinfo {author} {\bibfnamefont {Z.}~\bibnamefont {Lin}}, \bibinfo {author} {\bibfnamefont {N.}~\bibnamefont {Gimelshein}}, \bibinfo {author} {\bibfnamefont {L.}~\bibnamefont {Antiga}}, \emph {et~al.},\ }\bibfield  {title} {\bibinfo {title} {Pytorch: An imperative style, high-performance deep learning library},\ }\href@noop {} {\bibfield  {journal} {\bibinfo  {journal} {Advances in neural information processing systems}\ }\textbf {\bibinfo {volume} {32}} (\bibinfo {year} {2019})}\BibitemShut {NoStop}%
\bibitem [{cod()}]{code}%
  \BibitemOpen
  \bibinfo {note} {See https://github.com/quantum-jwjae/RL2LQS}\BibitemShut {NoStop}%
\bibitem [{\citenamefont {Pan}\ and\ \citenamefont {Yang}(2010)}]{Pan2010}%
  \BibitemOpen
  \bibfield  {author} {\bibinfo {author} {\bibfnamefont {S.~J.}\ \bibnamefont {Pan}}\ and\ \bibinfo {author} {\bibfnamefont {Q.}~\bibnamefont {Yang}},\ }\bibfield  {title} {\bibinfo {title} {A survey on transfer learning},\ }\href {https://doi.org/10.1109/TKDE.2009.191} {\bibfield  {journal} {\bibinfo  {journal} {IEEE Transactions on Knowledge and Data Engineering}\ }\textbf {\bibinfo {volume} {22}},\ \bibinfo {pages} {1345} (\bibinfo {year} {2010})}\BibitemShut {NoStop}%
\bibitem [{\citenamefont {Zen}\ \emph {et~al.}(2020)\citenamefont {Zen}, \citenamefont {My}, \citenamefont {Tan}, \citenamefont {H\'ebert}, \citenamefont {Gattobigio}, \citenamefont {Miniatura}, \citenamefont {Poletti},\ and\ \citenamefont {Bressan}}]{Zen2020}%
  \BibitemOpen
  \bibfield  {author} {\bibinfo {author} {\bibfnamefont {R.}~\bibnamefont {Zen}}, \bibinfo {author} {\bibfnamefont {L.}~\bibnamefont {My}}, \bibinfo {author} {\bibfnamefont {R.}~\bibnamefont {Tan}}, \bibinfo {author} {\bibfnamefont {F.}~\bibnamefont {H\'ebert}}, \bibinfo {author} {\bibfnamefont {M.}~\bibnamefont {Gattobigio}}, \bibinfo {author} {\bibfnamefont {C.}~\bibnamefont {Miniatura}}, \bibinfo {author} {\bibfnamefont {D.}~\bibnamefont {Poletti}},\ and\ \bibinfo {author} {\bibfnamefont {S.}~\bibnamefont {Bressan}},\ }\bibfield  {title} {\bibinfo {title} {Transfer learning for scalability of neural-network quantum states},\ }\href {https://doi.org/10.1103/PhysRevE.101.053301} {\bibfield  {journal} {\bibinfo  {journal} {Phys. Rev. E}\ }\textbf {\bibinfo {volume} {101}},\ \bibinfo {pages} {053301} (\bibinfo {year} {2020})}\BibitemShut {NoStop}%
\bibitem [{\citenamefont {Mari}\ \emph {et~al.}(2020)\citenamefont {Mari}, \citenamefont {Bromley}, \citenamefont {Izaac}, \citenamefont {Schuld},\ and\ \citenamefont {Killoran}}]{Mari2020transfer}%
  \BibitemOpen
  \bibfield  {author} {\bibinfo {author} {\bibfnamefont {A.}~\bibnamefont {Mari}}, \bibinfo {author} {\bibfnamefont {T.~R.}\ \bibnamefont {Bromley}}, \bibinfo {author} {\bibfnamefont {J.}~\bibnamefont {Izaac}}, \bibinfo {author} {\bibfnamefont {M.}~\bibnamefont {Schuld}},\ and\ \bibinfo {author} {\bibfnamefont {N.}~\bibnamefont {Killoran}},\ }\bibfield  {title} {\bibinfo {title} {Transfer learning in hybrid classical-quantum neural networks},\ }\href {https://doi.org/10.22331/q-2020-10-09-340} {\bibfield  {journal} {\bibinfo  {journal} {{Quantum}}\ }\textbf {\bibinfo {volume} {4}},\ \bibinfo {pages} {340} (\bibinfo {year} {2020})}\BibitemShut {NoStop}%
\bibitem [{\citenamefont {Bagan}\ \emph {et~al.}(2004)\citenamefont {Bagan}, \citenamefont {Baig}, \citenamefont {Mu\~noz Tapia},\ and\ \citenamefont {Rodriguez}}]{Bagan2004}%
  \BibitemOpen
  \bibfield  {author} {\bibinfo {author} {\bibfnamefont {E.}~\bibnamefont {Bagan}}, \bibinfo {author} {\bibfnamefont {M.}~\bibnamefont {Baig}}, \bibinfo {author} {\bibfnamefont {R.}~\bibnamefont {Mu\~noz Tapia}},\ and\ \bibinfo {author} {\bibfnamefont {A.}~\bibnamefont {Rodriguez}},\ }\bibfield  {title} {\bibinfo {title} {Collective versus local measurements in a qubit mixed-state estimation},\ }\href {https://doi.org/10.1103/PhysRevA.69.010304} {\bibfield  {journal} {\bibinfo  {journal} {Phys. Rev. A}\ }\textbf {\bibinfo {volume} {69}},\ \bibinfo {pages} {010304} (\bibinfo {year} {2004})}\BibitemShut {NoStop}%
\bibitem [{\citenamefont {Mahler}\ \emph {et~al.}(2013)\citenamefont {Mahler}, \citenamefont {Rozema}, \citenamefont {Darabi}, \citenamefont {Ferrie}, \citenamefont {Blume-Kohout},\ and\ \citenamefont {Steinberg}}]{Mahler2013}%
  \BibitemOpen
  \bibfield  {author} {\bibinfo {author} {\bibfnamefont {D.~H.}\ \bibnamefont {Mahler}}, \bibinfo {author} {\bibfnamefont {L.~A.}\ \bibnamefont {Rozema}}, \bibinfo {author} {\bibfnamefont {A.}~\bibnamefont {Darabi}}, \bibinfo {author} {\bibfnamefont {C.}~\bibnamefont {Ferrie}}, \bibinfo {author} {\bibfnamefont {R.}~\bibnamefont {Blume-Kohout}},\ and\ \bibinfo {author} {\bibfnamefont {A.~M.}\ \bibnamefont {Steinberg}},\ }\bibfield  {title} {\bibinfo {title} {Adaptive quantum state tomography improves accuracy quadratically},\ }\href {https://doi.org/10.1103/PhysRevLett.111.183601} {\bibfield  {journal} {\bibinfo  {journal} {Phys. Rev. Lett.}\ }\textbf {\bibinfo {volume} {111}},\ \bibinfo {pages} {183601} (\bibinfo {year} {2013})}\BibitemShut {NoStop}%
\bibitem [{\citenamefont {Kravtsov}\ \emph {et~al.}(2013)\citenamefont {Kravtsov}, \citenamefont {Straupe}, \citenamefont {Radchenko}, \citenamefont {Houlsby}, \citenamefont {Husz\'ar},\ and\ \citenamefont {Kulik}}]{Kravtsov2013}%
  \BibitemOpen
  \bibfield  {author} {\bibinfo {author} {\bibfnamefont {K.~S.}\ \bibnamefont {Kravtsov}}, \bibinfo {author} {\bibfnamefont {S.~S.}\ \bibnamefont {Straupe}}, \bibinfo {author} {\bibfnamefont {I.~V.}\ \bibnamefont {Radchenko}}, \bibinfo {author} {\bibfnamefont {N.~M.~T.}\ \bibnamefont {Houlsby}}, \bibinfo {author} {\bibfnamefont {F.}~\bibnamefont {Husz\'ar}},\ and\ \bibinfo {author} {\bibfnamefont {S.~P.}\ \bibnamefont {Kulik}},\ }\bibfield  {title} {\bibinfo {title} {Experimental adaptive bayesian tomography},\ }\href {https://doi.org/10.1103/PhysRevA.87.062122} {\bibfield  {journal} {\bibinfo  {journal} {Phys. Rev. A}\ }\textbf {\bibinfo {volume} {87}},\ \bibinfo {pages} {062122} (\bibinfo {year} {2013})}\BibitemShut {NoStop}%
\bibitem [{\citenamefont {Ferrie}(2014)}]{Ferrie2014SGTQT}%
  \BibitemOpen
  \bibfield  {author} {\bibinfo {author} {\bibfnamefont {C.}~\bibnamefont {Ferrie}},\ }\bibfield  {title} {\bibinfo {title} {Self-guided quantum tomography},\ }\href {https://doi.org/10.1103/PhysRevLett.113.190404} {\bibfield  {journal} {\bibinfo  {journal} {Phys. Rev. Lett.}\ }\textbf {\bibinfo {volume} {113}},\ \bibinfo {pages} {190404} (\bibinfo {year} {2014})}\BibitemShut {NoStop}%
\bibitem [{\citenamefont {Shen}\ and\ \citenamefont {Castan}(1992)}]{SHEN1992112}%
  \BibitemOpen
  \bibfield  {author} {\bibinfo {author} {\bibfnamefont {J.}~\bibnamefont {Shen}}\ and\ \bibinfo {author} {\bibfnamefont {S.}~\bibnamefont {Castan}},\ }\bibfield  {title} {\bibinfo {title} {An optimal linear operator for step edge detection},\ }\href {https://doi.org/https://doi.org/10.1016/1049-9652(92)90060-B} {\bibfield  {journal} {\bibinfo  {journal} {CVGIP: Graphical Models and Image Processing}\ }\textbf {\bibinfo {volume} {54}},\ \bibinfo {pages} {112} (\bibinfo {year} {1992})}\BibitemShut {NoStop}%
\bibitem [{\citenamefont {Page}(1993)}]{Page1993}%
  \BibitemOpen
  \bibfield  {author} {\bibinfo {author} {\bibfnamefont {D.~N.}\ \bibnamefont {Page}},\ }\bibfield  {title} {\bibinfo {title} {Average entropy of a subsystem},\ }\href {https://doi.org/10.1103/PhysRevLett.71.1291} {\bibfield  {journal} {\bibinfo  {journal} {Phys. Rev. Lett.}\ }\textbf {\bibinfo {volume} {71}},\ \bibinfo {pages} {1291} (\bibinfo {year} {1993})}\BibitemShut {NoStop}%
\bibitem [{\citenamefont {Sharma}\ \emph {et~al.}(2017)\citenamefont {Sharma}, \citenamefont {Lakshminarayanan},\ and\ \citenamefont {Ravindran}}]{sharma2017learning}%
  \BibitemOpen
  \bibfield  {author} {\bibinfo {author} {\bibfnamefont {S.}~\bibnamefont {Sharma}}, \bibinfo {author} {\bibfnamefont {A.~S.}\ \bibnamefont {Lakshminarayanan}},\ and\ \bibinfo {author} {\bibfnamefont {B.}~\bibnamefont {Ravindran}},\ }\bibfield  {title} {\bibinfo {title} {Learning to repeat: Fine grained action repetition for deep reinforcement learning},\ }in\ \href {https://openreview.net/forum?id=B1GOWV5eg} {\emph {\bibinfo {booktitle} {International Conference on Learning Representations}}}\ (\bibinfo {year} {2017})\BibitemShut {NoStop}%
\bibitem [{\citenamefont {Silver}\ \emph {et~al.}(2016)\citenamefont {Silver}, \citenamefont {Huang}, \citenamefont {Maddison}, \citenamefont {Guez}, \citenamefont {Sifre}, \citenamefont {van~den Driessche}, \citenamefont {Schrittwieser}, \citenamefont {Antonoglou}, \citenamefont {Panneershelvam}, \citenamefont {Lanctot}, \citenamefont {Dieleman}, \citenamefont {Grewe}, \citenamefont {Nham}, \citenamefont {Kalchbrenner}, \citenamefont {Sutskever}, \citenamefont {Lillicrap}, \citenamefont {Leach}, \citenamefont {Kavukcuoglu}, \citenamefont {Graepel},\ and\ \citenamefont {Hassabis}}]{Silver2016}%
  \BibitemOpen
  \bibfield  {author} {\bibinfo {author} {\bibfnamefont {D.}~\bibnamefont {Silver}}, \bibinfo {author} {\bibfnamefont {A.}~\bibnamefont {Huang}}, \bibinfo {author} {\bibfnamefont {C.~J.}\ \bibnamefont {Maddison}}, \bibinfo {author} {\bibfnamefont {A.}~\bibnamefont {Guez}}, \bibinfo {author} {\bibfnamefont {L.}~\bibnamefont {Sifre}}, \bibinfo {author} {\bibfnamefont {G.}~\bibnamefont {van~den Driessche}}, \bibinfo {author} {\bibfnamefont {J.}~\bibnamefont {Schrittwieser}}, \bibinfo {author} {\bibfnamefont {I.}~\bibnamefont {Antonoglou}}, \bibinfo {author} {\bibfnamefont {V.}~\bibnamefont {Panneershelvam}}, \bibinfo {author} {\bibfnamefont {M.}~\bibnamefont {Lanctot}}, \bibinfo {author} {\bibfnamefont {S.}~\bibnamefont {Dieleman}}, \bibinfo {author} {\bibfnamefont {D.}~\bibnamefont {Grewe}}, \bibinfo {author} {\bibfnamefont {J.}~\bibnamefont {Nham}}, \bibinfo {author} {\bibfnamefont {N.}~\bibnamefont {Kalchbrenner}}, \bibinfo {author} {\bibfnamefont {I.}~\bibnamefont {Sutskever}}, \bibinfo {author}
  {\bibfnamefont {T.}~\bibnamefont {Lillicrap}}, \bibinfo {author} {\bibfnamefont {M.}~\bibnamefont {Leach}}, \bibinfo {author} {\bibfnamefont {K.}~\bibnamefont {Kavukcuoglu}}, \bibinfo {author} {\bibfnamefont {T.}~\bibnamefont {Graepel}},\ and\ \bibinfo {author} {\bibfnamefont {D.}~\bibnamefont {Hassabis}},\ }\bibfield  {title} {\bibinfo {title} {Mastering the game of go with deep neural networks and tree search},\ }\href {https://doi.org/10.1038/nature16961} {\bibfield  {journal} {\bibinfo  {journal} {Nature}\ }\textbf {\bibinfo {volume} {529}},\ \bibinfo {pages} {484} (\bibinfo {year} {2016})}\BibitemShut {NoStop}%
\bibitem [{\citenamefont {Porotti}\ \emph {et~al.}(2023)\citenamefont {Porotti}, \citenamefont {Peano},\ and\ \citenamefont {Marquardt}}]{Porotti2023}%
  \BibitemOpen
  \bibfield  {author} {\bibinfo {author} {\bibfnamefont {R.}~\bibnamefont {Porotti}}, \bibinfo {author} {\bibfnamefont {V.}~\bibnamefont {Peano}},\ and\ \bibinfo {author} {\bibfnamefont {F.}~\bibnamefont {Marquardt}},\ }\bibfield  {title} {\bibinfo {title} {Gradient-ascent pulse engineering with feedback},\ }\href {https://doi.org/10.1103/PRXQuantum.4.030305} {\bibfield  {journal} {\bibinfo  {journal} {PRX Quantum}\ }\textbf {\bibinfo {volume} {4}},\ \bibinfo {pages} {030305} (\bibinfo {year} {2023})}\BibitemShut {NoStop}%
\bibitem [{\citenamefont {Maria~Schuld}\ and\ \citenamefont {Petruccione}(2015)}]{Maria2015Intro}%
  \BibitemOpen
  \bibfield  {author} {\bibinfo {author} {\bibfnamefont {I.~S.}\ \bibnamefont {Maria~Schuld}}\ and\ \bibinfo {author} {\bibfnamefont {F.}~\bibnamefont {Petruccione}},\ }\bibfield  {title} {\bibinfo {title} {An introduction to quantum machine learning},\ }\href {https://doi.org/10.1080/00107514.2014.964942} {\bibfield  {journal} {\bibinfo  {journal} {Contemporary Physics}\ }\textbf {\bibinfo {volume} {56}},\ \bibinfo {pages} {172} (\bibinfo {year} {2015})},\ \Eprint {https://arxiv.org/abs/https://doi.org/10.1080/00107514.2014.964942} {https://doi.org/10.1080/00107514.2014.964942} \BibitemShut {NoStop}%
\bibitem [{\citenamefont {Abbas}\ \emph {et~al.}(2024)\citenamefont {Abbas}, \citenamefont {Ambainis}, \citenamefont {Augustino}, \citenamefont {B{\"a}rtschi}, \citenamefont {Buhrman}, \citenamefont {Coffrin}, \citenamefont {Cortiana}, \citenamefont {Dunjko}, \citenamefont {Egger}, \citenamefont {Elmegreen}, \citenamefont {Franco}, \citenamefont {Fratini}, \citenamefont {Fuller}, \citenamefont {Gacon}, \citenamefont {Gonciulea}, \citenamefont {Gribling}, \citenamefont {Gupta}, \citenamefont {Hadfield}, \citenamefont {Heese}, \citenamefont {Kircher}, \citenamefont {Kleinert}, \citenamefont {Koch}, \citenamefont {Korpas}, \citenamefont {Lenk}, \citenamefont {Marecek}, \citenamefont {Markov}, \citenamefont {Mazzola}, \citenamefont {Mensa}, \citenamefont {Mohseni}, \citenamefont {Nannicini}, \citenamefont {O'Meara}, \citenamefont {Tapia}, \citenamefont {Pokutta}, \citenamefont {Proissl}, \citenamefont {Rebentrost}, \citenamefont {Sahin}, \citenamefont {Symons}, \citenamefont {Tornow}, \citenamefont {Valls},
  \citenamefont {Woerner}, \citenamefont {Wolf-Bauwens}, \citenamefont {Yard}, \citenamefont {Yarkoni}, \citenamefont {Zechiel}, \citenamefont {Zhuk},\ and\ \citenamefont {Zoufal}}]{Abbas2024}%
  \BibitemOpen
  \bibfield  {author} {\bibinfo {author} {\bibfnamefont {A.}~\bibnamefont {Abbas}}, \bibinfo {author} {\bibfnamefont {A.}~\bibnamefont {Ambainis}}, \bibinfo {author} {\bibfnamefont {B.}~\bibnamefont {Augustino}}, \bibinfo {author} {\bibfnamefont {A.}~\bibnamefont {B{\"a}rtschi}}, \bibinfo {author} {\bibfnamefont {H.}~\bibnamefont {Buhrman}}, \bibinfo {author} {\bibfnamefont {C.}~\bibnamefont {Coffrin}}, \bibinfo {author} {\bibfnamefont {G.}~\bibnamefont {Cortiana}}, \bibinfo {author} {\bibfnamefont {V.}~\bibnamefont {Dunjko}}, \bibinfo {author} {\bibfnamefont {D.~J.}\ \bibnamefont {Egger}}, \bibinfo {author} {\bibfnamefont {B.~G.}\ \bibnamefont {Elmegreen}}, \bibinfo {author} {\bibfnamefont {N.}~\bibnamefont {Franco}}, \bibinfo {author} {\bibfnamefont {F.}~\bibnamefont {Fratini}}, \bibinfo {author} {\bibfnamefont {B.}~\bibnamefont {Fuller}}, \bibinfo {author} {\bibfnamefont {J.}~\bibnamefont {Gacon}}, \bibinfo {author} {\bibfnamefont {C.}~\bibnamefont {Gonciulea}}, \bibinfo {author} {\bibfnamefont
  {S.}~\bibnamefont {Gribling}}, \bibinfo {author} {\bibfnamefont {S.}~\bibnamefont {Gupta}}, \bibinfo {author} {\bibfnamefont {S.}~\bibnamefont {Hadfield}}, \bibinfo {author} {\bibfnamefont {R.}~\bibnamefont {Heese}}, \bibinfo {author} {\bibfnamefont {G.}~\bibnamefont {Kircher}}, \bibinfo {author} {\bibfnamefont {T.}~\bibnamefont {Kleinert}}, \bibinfo {author} {\bibfnamefont {T.}~\bibnamefont {Koch}}, \bibinfo {author} {\bibfnamefont {G.}~\bibnamefont {Korpas}}, \bibinfo {author} {\bibfnamefont {S.}~\bibnamefont {Lenk}}, \bibinfo {author} {\bibfnamefont {J.}~\bibnamefont {Marecek}}, \bibinfo {author} {\bibfnamefont {V.}~\bibnamefont {Markov}}, \bibinfo {author} {\bibfnamefont {G.}~\bibnamefont {Mazzola}}, \bibinfo {author} {\bibfnamefont {S.}~\bibnamefont {Mensa}}, \bibinfo {author} {\bibfnamefont {N.}~\bibnamefont {Mohseni}}, \bibinfo {author} {\bibfnamefont {G.}~\bibnamefont {Nannicini}}, \bibinfo {author} {\bibfnamefont {C.}~\bibnamefont {O'Meara}}, \bibinfo {author} {\bibfnamefont {E.~P.}\ \bibnamefont
  {Tapia}}, \bibinfo {author} {\bibfnamefont {S.}~\bibnamefont {Pokutta}}, \bibinfo {author} {\bibfnamefont {M.}~\bibnamefont {Proissl}}, \bibinfo {author} {\bibfnamefont {P.}~\bibnamefont {Rebentrost}}, \bibinfo {author} {\bibfnamefont {E.}~\bibnamefont {Sahin}}, \bibinfo {author} {\bibfnamefont {B.~C.~B.}\ \bibnamefont {Symons}}, \bibinfo {author} {\bibfnamefont {S.}~\bibnamefont {Tornow}}, \bibinfo {author} {\bibfnamefont {V.}~\bibnamefont {Valls}}, \bibinfo {author} {\bibfnamefont {S.}~\bibnamefont {Woerner}}, \bibinfo {author} {\bibfnamefont {M.~L.}\ \bibnamefont {Wolf-Bauwens}}, \bibinfo {author} {\bibfnamefont {J.}~\bibnamefont {Yard}}, \bibinfo {author} {\bibfnamefont {S.}~\bibnamefont {Yarkoni}}, \bibinfo {author} {\bibfnamefont {D.}~\bibnamefont {Zechiel}}, \bibinfo {author} {\bibfnamefont {S.}~\bibnamefont {Zhuk}},\ and\ \bibinfo {author} {\bibfnamefont {C.}~\bibnamefont {Zoufal}},\ }\bibfield  {title} {\bibinfo {title} {Challenges and opportunities in quantum optimization},\ }\bibfield  {journal}
  {\bibinfo  {journal} {Nature Reviews Physics}\ }\href {https://doi.org/10.1038/s42254-024-00770-9} {10.1038/s42254-024-00770-9} (\bibinfo {year} {2024})\BibitemShut {NoStop}%
\bibitem [{\citenamefont {Chen}\ \emph {et~al.}(2017{\natexlab{b}})\citenamefont {Chen}, \citenamefont {Hoffman}, \citenamefont {Colmenarejo}, \citenamefont {Denil}, \citenamefont {Lillicrap}, \citenamefont {Botvinick},\ and\ \citenamefont {de~Freitas}}]{chen17e}%
  \BibitemOpen
  \bibfield  {author} {\bibinfo {author} {\bibfnamefont {Y.}~\bibnamefont {Chen}}, \bibinfo {author} {\bibfnamefont {M.~W.}\ \bibnamefont {Hoffman}}, \bibinfo {author} {\bibfnamefont {S.~G.}\ \bibnamefont {Colmenarejo}}, \bibinfo {author} {\bibfnamefont {M.}~\bibnamefont {Denil}}, \bibinfo {author} {\bibfnamefont {T.~P.}\ \bibnamefont {Lillicrap}}, \bibinfo {author} {\bibfnamefont {M.}~\bibnamefont {Botvinick}},\ and\ \bibinfo {author} {\bibfnamefont {N.}~\bibnamefont {de~Freitas}},\ }\bibfield  {title} {\bibinfo {title} {Learning to learn without gradient descent by gradient descent},\ }in\ \href {https://proceedings.mlr.press/v70/chen17e.html} {\emph {\bibinfo {booktitle} {Proceedings of the 34th International Conference on Machine Learning}}},\ \bibinfo {series} {Proceedings of Machine Learning Research}, Vol.~\bibinfo {volume} {70},\ \bibinfo {editor} {edited by\ \bibinfo {editor} {\bibfnamefont {D.}~\bibnamefont {Precup}}\ and\ \bibinfo {editor} {\bibfnamefont {Y.~W.}\ \bibnamefont {Teh}}}\ (\bibinfo
  {publisher} {PMLR},\ \bibinfo {year} {2017})\ pp.\ \bibinfo {pages} {748--756}\BibitemShut {NoStop}%
\bibitem [{\citenamefont {TV}\ \emph {et~al.}(2019)\citenamefont {TV}, \citenamefont {Malhotra}, \citenamefont {Narwariya}, \citenamefont {Vig},\ and\ \citenamefont {Shroff}}]{tv2019meta}%
  \BibitemOpen
  \bibfield  {author} {\bibinfo {author} {\bibfnamefont {V.}~\bibnamefont {TV}}, \bibinfo {author} {\bibfnamefont {P.}~\bibnamefont {Malhotra}}, \bibinfo {author} {\bibfnamefont {J.}~\bibnamefont {Narwariya}}, \bibinfo {author} {\bibfnamefont {L.}~\bibnamefont {Vig}},\ and\ \bibinfo {author} {\bibfnamefont {G.}~\bibnamefont {Shroff}},\ }\bibfield  {title} {\bibinfo {title} {Meta-learning for black-box optimization},\ }in\ \href@noop {} {\emph {\bibinfo {booktitle} {Joint European Conference on Machine Learning and Knowledge Discovery in Databases}}}\ (\bibinfo {organization} {Springer},\ \bibinfo {year} {2019})\ pp.\ \bibinfo {pages} {366--381}\BibitemShut {NoStop}%
\bibitem [{\citenamefont {Shala}\ \emph {et~al.}(2020)\citenamefont {Shala}, \citenamefont {Biedenkapp}, \citenamefont {Awad}, \citenamefont {Adriaensen}, \citenamefont {Lindauer},\ and\ \citenamefont {Hutter}}]{Shala2020}%
  \BibitemOpen
  \bibfield  {author} {\bibinfo {author} {\bibfnamefont {G.}~\bibnamefont {Shala}}, \bibinfo {author} {\bibfnamefont {A.}~\bibnamefont {Biedenkapp}}, \bibinfo {author} {\bibfnamefont {N.}~\bibnamefont {Awad}}, \bibinfo {author} {\bibfnamefont {S.}~\bibnamefont {Adriaensen}}, \bibinfo {author} {\bibfnamefont {M.}~\bibnamefont {Lindauer}},\ and\ \bibinfo {author} {\bibfnamefont {F.}~\bibnamefont {Hutter}},\ }\bibfield  {title} {\bibinfo {title} {Learning step-size adaptation in cma-es},\ }in\ \href@noop {} {\emph {\bibinfo {booktitle} {Parallel Problem Solving from Nature -- PPSN XVI}}},\ \bibinfo {editor} {edited by\ \bibinfo {editor} {\bibfnamefont {T.}~\bibnamefont {B{\"a}ck}}, \bibinfo {editor} {\bibfnamefont {M.}~\bibnamefont {Preuss}}, \bibinfo {editor} {\bibfnamefont {A.}~\bibnamefont {Deutz}}, \bibinfo {editor} {\bibfnamefont {H.}~\bibnamefont {Wang}}, \bibinfo {editor} {\bibfnamefont {C.}~\bibnamefont {Doerr}}, \bibinfo {editor} {\bibfnamefont {M.}~\bibnamefont {Emmerich}},\ and\ \bibinfo {editor}
  {\bibfnamefont {H.}~\bibnamefont {Trautmann}}}\ (\bibinfo  {publisher} {Springer International Publishing},\ \bibinfo {address} {Cham},\ \bibinfo {year} {2020})\ pp.\ \bibinfo {pages} {691--706}\BibitemShut {NoStop}%
\bibitem [{\citenamefont {Lange}\ \emph {et~al.}(2023{\natexlab{b}})\citenamefont {Lange}, \citenamefont {Schaul}, \citenamefont {Chen}, \citenamefont {Zahavy}, \citenamefont {Dalibard}, \citenamefont {Lu}, \citenamefont {Singh},\ and\ \citenamefont {Flennerhag}}]{lange2023discovering}%
  \BibitemOpen
  \bibfield  {author} {\bibinfo {author} {\bibfnamefont {R.~T.}\ \bibnamefont {Lange}}, \bibinfo {author} {\bibfnamefont {T.}~\bibnamefont {Schaul}}, \bibinfo {author} {\bibfnamefont {Y.}~\bibnamefont {Chen}}, \bibinfo {author} {\bibfnamefont {T.}~\bibnamefont {Zahavy}}, \bibinfo {author} {\bibfnamefont {V.}~\bibnamefont {Dalibard}}, \bibinfo {author} {\bibfnamefont {C.}~\bibnamefont {Lu}}, \bibinfo {author} {\bibfnamefont {S.}~\bibnamefont {Singh}},\ and\ \bibinfo {author} {\bibfnamefont {S.}~\bibnamefont {Flennerhag}},\ }\bibfield  {title} {\bibinfo {title} {Discovering evolution strategies via meta-black-box optimization},\ }in\ \href {https://openreview.net/forum?id=mFDU0fP3EQH} {\emph {\bibinfo {booktitle} {The Eleventh International Conference on Learning Representations}}}\ (\bibinfo {year} {2023})\BibitemShut {NoStop}%
\bibitem [{\citenamefont {Konda}\ and\ \citenamefont {Tsitsiklis}(1999)}]{konda1999actor}%
  \BibitemOpen
  \bibfield  {author} {\bibinfo {author} {\bibfnamefont {V.}~\bibnamefont {Konda}}\ and\ \bibinfo {author} {\bibfnamefont {J.}~\bibnamefont {Tsitsiklis}},\ }\bibfield  {title} {\bibinfo {title} {Actor-critic algorithms},\ }\href@noop {} {\bibfield  {journal} {\bibinfo  {journal} {Advances in Neural Information Processing Systems}\ }\textbf {\bibinfo {volume} {12}} (\bibinfo {year} {1999})}\BibitemShut {NoStop}%
\bibitem [{\citenamefont {Wang}\ \emph {et~al.}(2017)\citenamefont {Wang}, \citenamefont {Bapst}, \citenamefont {Heess}, \citenamefont {Mnih}, \citenamefont {Munos}, \citenamefont {Kavukcuoglu},\ and\ \citenamefont {De~Freitas}}]{wang2016sample}%
  \BibitemOpen
  \bibfield  {author} {\bibinfo {author} {\bibfnamefont {Z.}~\bibnamefont {Wang}}, \bibinfo {author} {\bibfnamefont {V.}~\bibnamefont {Bapst}}, \bibinfo {author} {\bibfnamefont {N.}~\bibnamefont {Heess}}, \bibinfo {author} {\bibfnamefont {V.}~\bibnamefont {Mnih}}, \bibinfo {author} {\bibfnamefont {R.}~\bibnamefont {Munos}}, \bibinfo {author} {\bibfnamefont {K.}~\bibnamefont {Kavukcuoglu}},\ and\ \bibinfo {author} {\bibfnamefont {N.}~\bibnamefont {De~Freitas}},\ }\bibfield  {title} {\bibinfo {title} {Sample efficient actor-critic with experience replay},\ }\href {https://openreview.net/pdf?id=HyM25Mqel} {\bibfield  {journal} {\bibinfo  {journal} {Proceedings of the International Conference on Learning Representations (ICLR)}\ } (\bibinfo {year} {2017})}\BibitemShut {NoStop}%
\bibitem [{\citenamefont {Mnih}\ \emph {et~al.}(2015)\citenamefont {Mnih}, \citenamefont {Kavukcuoglu}, \citenamefont {Silver}, \citenamefont {Rusu}, \citenamefont {Veness}, \citenamefont {Bellemare}, \citenamefont {Graves}, \citenamefont {Riedmiller}, \citenamefont {Fidjeland}, \citenamefont {Ostrovski}, \citenamefont {Petersen}, \citenamefont {Beattie}, \citenamefont {Sadik}, \citenamefont {Antonoglou}, \citenamefont {King}, \citenamefont {Kumaran}, \citenamefont {Wierstra}, \citenamefont {Legg},\ and\ \citenamefont {Hassabis}}]{Mnih2015}%
  \BibitemOpen
  \bibfield  {author} {\bibinfo {author} {\bibfnamefont {V.}~\bibnamefont {Mnih}}, \bibinfo {author} {\bibfnamefont {K.}~\bibnamefont {Kavukcuoglu}}, \bibinfo {author} {\bibfnamefont {D.}~\bibnamefont {Silver}}, \bibinfo {author} {\bibfnamefont {A.~A.}\ \bibnamefont {Rusu}}, \bibinfo {author} {\bibfnamefont {J.}~\bibnamefont {Veness}}, \bibinfo {author} {\bibfnamefont {M.~G.}\ \bibnamefont {Bellemare}}, \bibinfo {author} {\bibfnamefont {A.}~\bibnamefont {Graves}}, \bibinfo {author} {\bibfnamefont {M.}~\bibnamefont {Riedmiller}}, \bibinfo {author} {\bibfnamefont {A.~K.}\ \bibnamefont {Fidjeland}}, \bibinfo {author} {\bibfnamefont {G.}~\bibnamefont {Ostrovski}}, \bibinfo {author} {\bibfnamefont {S.}~\bibnamefont {Petersen}}, \bibinfo {author} {\bibfnamefont {C.}~\bibnamefont {Beattie}}, \bibinfo {author} {\bibfnamefont {A.}~\bibnamefont {Sadik}}, \bibinfo {author} {\bibfnamefont {I.}~\bibnamefont {Antonoglou}}, \bibinfo {author} {\bibfnamefont {H.}~\bibnamefont {King}}, \bibinfo {author} {\bibfnamefont
  {D.}~\bibnamefont {Kumaran}}, \bibinfo {author} {\bibfnamefont {D.}~\bibnamefont {Wierstra}}, \bibinfo {author} {\bibfnamefont {S.}~\bibnamefont {Legg}},\ and\ \bibinfo {author} {\bibfnamefont {D.}~\bibnamefont {Hassabis}},\ }\bibfield  {title} {\bibinfo {title} {Human-level control through deep reinforcement learning},\ }\href {https://doi.org/10.1038/nature14236} {\bibfield  {journal} {\bibinfo  {journal} {Nature}\ }\textbf {\bibinfo {volume} {518}},\ \bibinfo {pages} {529} (\bibinfo {year} {2015})}\BibitemShut {NoStop}%
\bibitem [{\citenamefont {Mnih}\ \emph {et~al.}(2016)\citenamefont {Mnih}, \citenamefont {Badia}, \citenamefont {Mirza}, \citenamefont {Graves}, \citenamefont {Lillicrap}, \citenamefont {Harley}, \citenamefont {Silver},\ and\ \citenamefont {Kavukcuoglu}}]{pmlr-v48-mniha16}%
  \BibitemOpen
  \bibfield  {author} {\bibinfo {author} {\bibfnamefont {V.}~\bibnamefont {Mnih}}, \bibinfo {author} {\bibfnamefont {A.~P.}\ \bibnamefont {Badia}}, \bibinfo {author} {\bibfnamefont {M.}~\bibnamefont {Mirza}}, \bibinfo {author} {\bibfnamefont {A.}~\bibnamefont {Graves}}, \bibinfo {author} {\bibfnamefont {T.}~\bibnamefont {Lillicrap}}, \bibinfo {author} {\bibfnamefont {T.}~\bibnamefont {Harley}}, \bibinfo {author} {\bibfnamefont {D.}~\bibnamefont {Silver}},\ and\ \bibinfo {author} {\bibfnamefont {K.}~\bibnamefont {Kavukcuoglu}},\ }\bibfield  {title} {\bibinfo {title} {Asynchronous methods for deep reinforcement learning},\ }in\ \href {https://proceedings.mlr.press/v48/mniha16.html} {\emph {\bibinfo {booktitle} {Proceedings of The 33rd International Conference on Machine Learning}}},\ \bibinfo {series} {Proceedings of Machine Learning Research}, Vol.~\bibinfo {volume} {48},\ \bibinfo {editor} {edited by\ \bibinfo {editor} {\bibfnamefont {M.~F.}\ \bibnamefont {Balcan}}\ and\ \bibinfo {editor} {\bibfnamefont
  {K.~Q.}\ \bibnamefont {Weinberger}}}\ (\bibinfo  {publisher} {PMLR},\ \bibinfo {address} {New York, New York, USA},\ \bibinfo {year} {2016})\ pp.\ \bibinfo {pages} {1928--1937}\BibitemShut {NoStop}%
\bibitem [{\citenamefont {Kocsis}\ and\ \citenamefont {Szepesv{\'a}ri}(2006)}]{Kocsis2006}%
  \BibitemOpen
  \bibfield  {author} {\bibinfo {author} {\bibfnamefont {L.}~\bibnamefont {Kocsis}}\ and\ \bibinfo {author} {\bibfnamefont {C.}~\bibnamefont {Szepesv{\'a}ri}},\ }\bibfield  {title} {\bibinfo {title} {Bandit based monte-carlo planning},\ }in\ \href@noop {} {\emph {\bibinfo {booktitle} {Machine Learning: ECML 2006}}},\ \bibinfo {editor} {edited by\ \bibinfo {editor} {\bibfnamefont {J.}~\bibnamefont {F{\"u}rnkranz}}, \bibinfo {editor} {\bibfnamefont {T.}~\bibnamefont {Scheffer}},\ and\ \bibinfo {editor} {\bibfnamefont {M.}~\bibnamefont {Spiliopoulou}}}\ (\bibinfo  {publisher} {Springer Berlin Heidelberg},\ \bibinfo {address} {Berlin, Heidelberg},\ \bibinfo {year} {2006})\ pp.\ \bibinfo {pages} {282--293}\BibitemShut {NoStop}%
\bibitem [{\citenamefont {Kingma}\ and\ \citenamefont {Ba}(2014)}]{kingma2014adam}%
  \BibitemOpen
  \bibfield  {author} {\bibinfo {author} {\bibfnamefont {D.~P.}\ \bibnamefont {Kingma}}\ and\ \bibinfo {author} {\bibfnamefont {J.}~\bibnamefont {Ba}},\ }\bibfield  {title} {\bibinfo {title} {Adam: A method for stochastic optimization},\ }\href@noop {} {\bibfield  {journal} {\bibinfo  {journal} {arXiv preprint arXiv:1412.6980}\ } (\bibinfo {year} {2014})}\BibitemShut {NoStop}%
\bibitem [{\citenamefont {\ifmmode \check{R}\else \v{R}\fi{}eh\'a\ifmmode~\check{c}\else \v{c}\fi{}ek}\ \emph {et~al.}(2007)\citenamefont {\ifmmode \check{R}\else \v{R}\fi{}eh\'a\ifmmode~\check{c}\else \v{c}\fi{}ek}, \citenamefont {Hradil}, \citenamefont {Knill},\ and\ \citenamefont {Lvovsky}}]{Jaroslav2007}%
  \BibitemOpen
  \bibfield  {author} {\bibinfo {author} {\bibfnamefont {J.}~\bibnamefont {\ifmmode \check{R}\else \v{R}\fi{}eh\'a\ifmmode~\check{c}\else \v{c}\fi{}ek}}, \bibinfo {author} {\bibfnamefont {Z.~c.~v.}\ \bibnamefont {Hradil}}, \bibinfo {author} {\bibfnamefont {E.}~\bibnamefont {Knill}},\ and\ \bibinfo {author} {\bibfnamefont {A.~I.}\ \bibnamefont {Lvovsky}},\ }\bibfield  {title} {\bibinfo {title} {Diluted maximum-likelihood algorithm for quantum tomography},\ }\href {https://doi.org/10.1103/PhysRevA.75.042108} {\bibfield  {journal} {\bibinfo  {journal} {Phys. Rev. A}\ }\textbf {\bibinfo {volume} {75}},\ \bibinfo {pages} {042108} (\bibinfo {year} {2007})}\BibitemShut {NoStop}%
\end{thebibliography}

%

\end{document}